\documentclass[a4paper,11pt]{article}
\usepackage{amsmath}
\usepackage{amsthm}
\usepackage{amsfonts}
\usepackage{mathrsfs}
\usepackage{amssymb}
\usepackage{mathpazo}
\usepackage[dvipsnames]{xcolor}
\usepackage{bm}
\usepackage[no-weekday]{eukdate}
\usepackage[bb=boondox]{mathalfa}
\usepackage{paralist}
\usepackage{natbib}
\usepackage[title]{appendix}%
\usepackage{url}
\usepackage[textwidth=6.1in,textheight = 10in]{geometry}
\usepackage{placeins}
\usepackage[hidelinks]{hyperref}
\hypersetup{
	pdftitle = {{Point and probabilistic forecast reconciliation for general linearly constrained multiple time series}}, 
	pdfauthor = {{Daniele Girolimetto and Tommaso Di Fonzo}}
}

\usepackage{multirow, float}
\usepackage{makecell}
\usepackage{calc}
\usepackage{graphicx}
\usepackage{enumitem}
\usepackage{microtype}
\usepackage{todonotes}
\usepackage{caption}
\DeclareCaptionStyle{italic}[justification=centering]
 {labelfont={bf},textfont={it},labelsep=colon}
\usepackage[compact]{titlesec}
\titlespacing{\section}{0pt}{2ex}{1ex}
\titlespacing{\subsection}{0pt}{1ex}{0ex}
\titlespacing{\subsubsection}{0pt}{0.5ex}{0ex}
\usepackage{titling}\pretitle{\begin{center}\LARGE\bfseries}
\usepackage[bottom,hang,flushmargin]{footmisc}
\usepackage{adjustbox}

\allowdisplaybreaks
\usepackage{caption}
\usepackage{subcaption}

\usepackage{longtable}
\usepackage{booktabs}
\usepackage{array}
\newcolumntype{M}[1]{>{\centering\arraybackslash}m{#1}}
\newcolumntype{L}[1]{>{\raggedright\arraybackslash}m{#1}}
\newcolumntype{R}[1]{>{\raggedleft\arraybackslash}m{#1}}

\newcommand{\avet}{\bm{a}}
\newcommand{\bvet}{\bm{b}}
\newcommand{\cvet}{\bm{c}}

\newcommand{\evet}{\bm{e}}

\newcommand{\uvet}{\bm{u}}

\newcommand{\xvet}{\bm{x}}
\newcommand{\yvet}{\bm{y}}
\newcommand{\zvet}{\bm{z}}
\newcommand{\Avet}{\bm{A}}

\newcommand{\Cvet}{\bm{C}}

\newcommand{\Evet}{\bm{E}}

\newcommand{\Gvet}{\bm{G}}

\newcommand{\Ivet}{\bm{I}}

\newcommand{\Kvet}{\bm{K}}

\newcommand{\Mvet}{\bm{M}}

\newcommand{\Pvet}{\bm{P}}
\newcommand{\Qvet}{\bm{Q}}
\newcommand{\Rvet}{\bm{R}}
\newcommand{\Svet}{\bm{S}}

\newcommand{\Wvet}{\bm{W}}

\newcommand{\Zvet}{\bm{Z}}
\newcommand{\Zerovet}{\bm{0}}

\newcommand{\Gammavet}{\bm{\Gamma}}

\makeatletter
\def\@endtheorem{\endtrivlist}
\makeatother

\usepackage{tikz}
\usetikzlibrary{arrows,shapes,positioning,shadows,trees}
\usetikzlibrary{matrix, decorations.pathreplacing, arrows, calc, fit, arrows.meta, decorations.pathmorphing, decorations.markings, decorations,calligraphy}

\tikzset{
  basic/.style  = {draw, text width=2cm, drop shadow, font=\sffamily, rectangle},
  root/.style   = {basic, rounded corners=2pt, thin, align=center,
                   fill=green!30},
  level 2/.style = {basic, rounded corners=6pt, thin,align=center, fill=green!60,
                   text width=4em},
  level 3/.style = {basic, thin, align=left, fill=pink!60, text width=1.5em}
}
\newcommand{\relation}[3]
{
	\draw (#3.south) -- +(0,-#1) -| ($ (#2.north) $)
}

\newcommand{\relationD}[3]
{
	\draw (#3.east) -- +(#1,0) |- (#2.west)
}

\theoremstyle{definition}
\newtheorem{definition}{Definition}[section]

\usepackage[affil-it]{authblk}
\makeatletter
\renewenvironment{abstract}{%
    \if@twocolumn
      \section*{\abstractname}%
    \else %
      \begin{center}%
        {\bfseries \large\abstractname\vspace{\z@}}%
      \end{center}%
      \quotation
    \fi}
    {\if@twocolumn\else\endquotation\fi}
\makeatother

\title{\normalfont \textbf{Point and probabilistic forecast reconciliation for general linearly constrained multiple time series}}

\author{Daniele Girolimetto}
\author{Tommaso Di Fonzo}
\affil{\small Department of Statistical Sciences, University of Padua, Padova 35121, Italy}

\date{21 December 2023}

\begin{document}

\def\spacingset#1{\renewcommand{\baselinestretch}{#1}\small\normalsize}
\spacingset{1.1}

\thispagestyle{empty} \clearpage\maketitle

\begingroup
\let\thefootnote\relax\footnotetext{\raggedright Email: \href{mailto:daniele.girolimetto@phd.unipd.it}{daniele.girolimetto@unipd.it} (DG) and \href{mailto:difonzo@stat.unipd.it}{difonzo@stat.unipd.it} (TDF)}
\endgroup

\begin{abstract}
\noindent Forecast reconciliation is the post-forecasting process aimed to revise a set of incoherent base forecasts into coherent forecasts in line with given data structures. Most of the point and probabilistic regression-based forecast reconciliation results ground on the so called ``structural representation'' and on the related unconstrained generalized least squares reconciliation formula. However, the structural representation naturally applies to genuine hierarchical/grouped time series, where the top- and bottom-level variables are uniquely identified. When a general linearly constrained multiple time series is considered, the forecast reconciliation is naturally expressed according to a projection approach. While it is well known that the classic structural reconciliation formula is equivalent to its projection approach counterpart, so far it is not completely understood if and how a structural-like reconciliation formula may be derived for a general linearly constrained multiple time series. Such an expression would permit to extend reconciliation definitions, theorems and results in a straightforward manner.
In this paper, we show that for general linearly constrained multiple time series it is possible to express the reconciliation formula according to a ``structural-like'' approach that keeps distinct free and constrained, instead of bottom and upper (aggregated), variables, establish the probabilistic forecast reconciliation framework, and apply these findings to obtain fully reconciled point and probabilistic forecasts for the aggregates of the Australian $GDP$ from income and expenditure sides, and for the European Area $GDP$ disaggregated by income, expenditure and output sides and by 19 countries.
\end{abstract}

\begin{itemize}[nosep, align=left, leftmargin = !]
	\item[\textbf{Keywords}] \textit{Linearly constrained multiple time series, Hierarchical/grouped time series, \\ Point and probabilistic forecast reconciliation, Quarterly National Accounts, $GDP$}
\end{itemize}
\vfill

\newpage
\spacingset{1.3}


\section{Introduction}\label{sec:intro}
Starting from \cite{hyndman2011}, regression-based forecast reconciliation has become an hot topic in the forecasting literature \citep{vanerven2015, wickramasuriya2019, wickramasuriya2021a, panagiotelis2021, jeon2019, bentaieb2019, bentaieb2021, panagiotelis2020prob, wickramasuriya2021b}. 
By forecast reconciliation, we mean a post-forecasting procedure \citep{difonzo2021a} in which previously (and however) generated incoherent predictions (called base forecasts) for all the components of a multiple time series are adjusted in order to fulfill some externally given linear constraints. These reconciled forecasts are said to be coherent.
In many real world applications, forecasts of a large collection of time series have a natural organization according to a hierarchical structure. More precisely, a system is classified as hierarchical when series are created by aggregating others in a tree shape. When the system is formed by a unique tree, then the collection is called ``hierarchical time series'' \citep{hyndman2011}. On the other side, when various hierarchies share the same series at both the most aggregated and disaggregated levels (top and bottom level, respectively), we face a ``grouped time series'' \citep{hyndman2016}.

In the field of forecast reconciliation, the bottom-up and top-down approaches are among the earliest and best known ones. Bottom-up forecasting \citep{dunn1976} uses only forecasts at the bottom level to obtain all the reconciled forecasts. In contrast, top-down forecasting \citep{gross1990} uses only the forecasts at the highest aggregated level. Having observed that neither forecasting approach uses all the available information, \cite{hyndman2011} developed a regression-based reconciliation approach consisting in (i) forecasting all the series with no regard for the constraints, and (ii) using then a regression model to optimally combining these  (base) forecasts in order to produce coherent forecasts. This approach has witnessed a continuous growth of the related literature \citep{hyndman2016, athanasopoulos2020, wickramasuriya2021a, panagiotelis2020prob}, that in most cases grounds on the so-called structural representation of a hierarchical/grouped time series \citep{athanasopoulos2009}, in which the variables are classified either bottom if they belong to the most disaggregated level, or upper if they are obtained by summing the lower levels' variables. In mathematical terms, upper and bottom variables are linked by an aggregation matrix, which describes how the upper series are obtained from the bottom ones \citep{hyndman2011}. This representation is directly related  to the hierarchical structure, where the series are naturally classifiable and an aggregation matrix may be obtained with little effort. Theoretical aspects for point and probabilistic forecast reconciliation have been developed using a structural representation by \cite{panagiotelis2021} and \cite{panagiotelis2020prob}, respectively (see also \citealp{wickramasuriya2021a, wickramasuriya2021b}).

However, it can be shown \citep{vanerven2015, wickramasuriya2019, bisaglia2020, difonzo2021a} that reconciled forecasts may be obtained as the solution to a linearly constrained quadratic optimization problem\footnote{This approach dates back to the seminal paper by \cite{stone1942} on the least squares adjustment of noisy data with accounting constraints \citep{byron1978, byron1979}. Recent applications to the reconciliation of systems of seasonally adjusted time series are provided by \cite{difonzomarini2011, difonzomarini2015}.}, that does not require any ``upper and bottom'' classification of the involved variables. This approach grounds on a zero-constrained representation of the linearly constrained multiple time series (\citealp{difonzo2021a}) describing the relationships linking all the individual series in the system. For a genuine hierarchical/grouped time series, where the top and bottom level variables are uniquely identified, it is easy to express the relationship between structural and zero-constrained representations. \cite{wickramasuriya2019} show that the structural  and the corresponding projection reconciliation approaches produce the same final reconciled forecasts. Unfortunately, given a zero-constrained representation of a linearly constrained multiple time series, finding the corresponding structural representation is not trivial \citep{difonzo2021a}, and one of the objective of the present paper is to fill this gap.

Most of the forecast reconciliation approaches proposed in the literature refer to genuine hierarchical/grouped time series, that do not take into account the full spectrum of possible cases encountered in real-life situations. As pointed out by \cite{panagiotelis2021}, “\textit{concepts such as coherence and reconciliation (...) require the data to have only two important characteristics: the first is that they are multivariate, and the second is that they adhere to linear constraints}". Using a novel geometric interpretation, \cite{panagiotelis2021} develop definitions and a formulation for linearly constrained multiple time series within a general framework and not just for simple summation hierarchical relationships. However, their results still ground on the structural representation valid only for genuine hierarchical/grouped time series, and this also holds for the probabilistic forecast reconciliation approach developed by \cite{panagiotelis2020prob}. Nevertheless, the point we wish to stress here is that when working with general linear constraints and many variables, the interchangeability between structural and zero-constrained representations, easy to recover for a genuine hierarchical/grouped time series, is no more always straightforward. 

In this paper we address a number of open issues in point and probabilistic cross-sectional forecast reconciliation for general linearly constrained multiple time series. First, we introduce a general linearly constrained multiple time series by exploiting its analogies with an homogeneous linear system. Second, we show that the classical hierarchical representation for a multiple time series is a simple special case of the general representation. Third, we show that if it is possible to express a general linearly constrained multiple time series according to a ``structural-like'' representation, we can easily achieve the formulation for point and probabilistic regression-based reconciled forecasts using a linear combination matrix, with elements in $\mathbb{R}$, that is the natural extension of the aggregation matrix used in the structural reconciliation approach, with elements only in $\{0,1\}$. When the distinction between bottom and upper variables is no longer meaningful, we adopt a classification involving free and constrained variables, respectively, and show how to obtain a structural-like representation, possibly using well known linear algebra techniques, such as the Reduced Row Echelon Form and the QR decomposition (\citealp{golub1996matrix}, \citealp{meyer2000}).

The remainder of the paper is structured as follows. In \autoref{sec:consTS}, we present the notation and the main results about the point forecast reconciliation for a genuine hierarchical/grouped time series and in the general case. We define the structural-like representation for a general linearly constrained multiple time series by distinguishing between free and constrained variables, instead of bottom and upper, and use this result in the probabilistic forecast reconciliation framework set out by \cite{panagiotelis2020prob}. In \autoref{sec:Cbar}, we show how to obtain the linear combination matrix for the structural-like representation in a computationally efficient way\footnote{The procedures used in this paper are implemented in the \textsf{R} package \texttt{FoReco} \citep{girolimetto2022}. A complete set of results is available at the GitHub repository: \url{https://github.com/danigiro/mtsreco}.}. Two empirical applications are presented in \autoref{sec:appl}. First, we extend the forecast reconciliation experiment for the Australian $GDP$ originally developed by \cite{athanasopoulos2020} and \cite{bisaglia2020}, to obtain both point and probabilistic $GDP$ forecasts simultaneously coherent with their disaggregate counterpart forecasts from income and expenditure sides. Second, point and probabilistic forecasts are obtained for the European Area $GDP$ from income, expenditure and output sides, geographically disaggregated by 19 component countries, where the large number of series and constraints makes a full row rank zero-constraints matrix difficult to build. \autoref{sec:conclusions} contains conclusions. 

\section{Cross-sectional forecast reconciliation of a linearly constrained multiple time series}\label{sec:consTS}

Let $\xvet_t$ be a $n$-dimensional linearly constrained multiple time series such that all the values for $t=1,...$ (either observed or not) lie in the \textit{coherent} linear subspace $\mathcal{S}\in \mathbb{R}^n$, $\xvet_t \in \mathcal{S}\; \forall t>0$ \citep{panagiotelis2021}, which means that all linear constraints are satisfied at time $t$. These constraints can be represented through linear equations and grouped as a (rectangular) linear system. Following \cite{meyer2000} and \cite{leon2015}, let $x_{1,t}, \dots, x_{n,t}$ be the observations of $n$ individual time series at a given time $t = 1, ..., T$. An homogeneous linear system of $p$ equations in the $n$ variables present in $\xvet_t$ can be written as
$$
\left\{\begin{array}{ccccccccc}
	\gamma_{1,1} \; x_{1,t} & + & \gamma_{1,2} \; x_{2,t} & + & \dots & + & \gamma_{1,n} \; x_{n,t} & = & 0\\
	\vdots & \vdots & \vdots & \vdots & & \vdots & \vdots & \vdots & \vdots \\
	\gamma_{p,1} \; x_{1,t} & + & \gamma_{p,2} \; x_{2,t} & + & \dots & + & \gamma_{p,n} \; x_{n,t} & = & 0\\
\end{array}\right. ,
$$
where the $\gamma_{ij}$’s are real-valued coefficients. This system can be expressed in matrix form as
\begin{equation}
\label{eq:Uy0}
	\Gammavet \xvet_t = \Zerovet_{(p\times 1)} ,
\end{equation}
where $\Gammavet\in\mathbb{R}^{(p \times n)}$ is the coefficient matrix 
$$
\Gammavet = \begin{bmatrix}
	\gamma_{1,1} & \gamma_{1,2} & \dots & \gamma_{1,n} \\
	\vdots & \vdots & \ddots & \vdots \\
	\gamma_{p,1} & \gamma_{p,2} & \dots & \gamma_{p,n} \\
\end{bmatrix}.
$$
Expression (\ref{eq:Uy0}) is called ``zero-constrained representation'' of a linearly constrained multiple time series (\citealp{difonzo2021a}).

\subsection{Forecast reconciliation of a genuine hierarchical time series}\label{sec:hts}

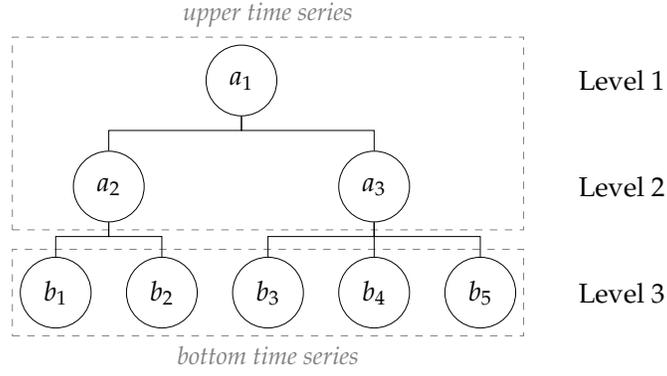
\begin{figure}[!t]
	\centering
\begin{tikzpicture}[baseline=(current  bounding  box.center),
			rel/.append style={shape=circle,
				draw=black,
			minimum width=1cm,
			minimum height=1cm},
			connection/.style ={inner sep =0, outer sep =0}]
				
			\node[rel] at (0, 0) (AA){$b_1$};
			\node[rel] at (1.5, 0) (AB){$b_2$};
			\node[rel] at (3, 0) (BA){$b_3$};
			\node[rel] at (4.5, 0) (BB){$b_4$};
			\node[rel] at (6, 0) (BC){$b_5$};
			
			\node[rel] at (0.75, 1.5) (A){$a_2$};
			\node[rel] at (4.5, 1.5) (B){$a_3$};
			\node[rel] at (2.625, 3) (T){$a_1$};
			
			\node at (8, 3) (L1){Level 1};
			\node at (8, 1.5) (L2){Level 2};
			\node at (8, 0) (L3){Level 3};
			
			\relation{0.2}{AA}{A};
			\relation{0.2}{AB}{A};
			\relation{0.2}{BA}{B};
			\relation{0.2}{BB}{B};
			\relation{0.2}{BC}{B};

			\relation{0.2}{A}{T};
			\relation{0.2}{B}{T};
			
			\node[minimum width=1cm,
			minimum height=1cm] at (0, 1.5) (con0){};
			\node[minimum width=1cm,
			minimum height=1cm] at (6, 3) (con1){};
			
			\node[fit=(con0)(con1), rectangle, dashed, inner sep = 3, draw, opacity = 0.5, label= {[font=\small,text=black!50]above:{\textit{upper time series}}}] {};
			\node[fit=(AA)(BC), rectangle, dashed, inner sep = 3, draw, opacity = 0.5, label= {[font=\small,text=black!50]below:{\textit{bottom time series}}}] {};
		\end{tikzpicture}
	\caption{A simple three-level hierarchical structure for a linearly constrained multiple time series}
	\label{fig:hierS}
\end{figure}

In \autoref{fig:hierS} it is shown a simple, genuine hierarchical time series \citep{athanasopoulos2009, hyndman2011, athanasopoulos2020, hyndman2021b}, that can be seen as a particular case of a linearly constrained multiple time series. This hierarchical structure is defined only by simple summation constraints,
\begin{equation}
\begin{array}{lcl}
	a_1 & = & b_1 + b_2 + b_3 + b_4 + b_5 \\
	a_2 & = & b_1 + b_2 \\
	a_3 & = & b_3 + b_4 + b_5 \\
\end{array} ,
\end{equation}
that can be easily transformed into a zero-constrained representation $\Gammavet \xvet = \Zerovet_{(8 \times 1)}$, with $\xvet=\left[a_1 \; a_2 \; a_3 \; b_1 \; b_2 \;b_3 \;b_4 \;b_5 \;\right]' = \left[\avet ' \quad \bvet'\right]'$, and
$$
\Gammavet = \left[\begin{array}{cccccccc}
1 & 0 & 0 & -1 & -1 & -1 & -1 & -1 \\
0 & 1 & 0 & -1 & -1 &  0 &  0 &  0 \\
0 & 0 & 1 &  0 &  0 & -1 & -1 & -1 \\
\end{array}\right] = \left[\Ivet_3 \quad -\Avet\right],
$$
such as the $\gamma_{i,j}$'s coefficients are only -1, 0 and 1, and $\avet = \Avet \bvet$.

In general, let
$
\bvet_t = \begin{bmatrix}
	b_{1,t} &
	\dots &
	b_{n_b,t}
\end{bmatrix}'\in \mathbb{R}^{(n_b \times 1)}$ and $\avet_t = \begin{bmatrix}
	a_{1,t}&
	\dots &
	a_{n_a,t}
\end{bmatrix}'\in \mathbb{R}^{(n_a \times 1)}$, $t = 1, ..., T$, with $n = n_a + n_b$,
be the $T$ vectors containing the bottom and the aggregated series, respectively, of a hierarchy. All series can be collected in the $T$ vectors 
$$
\yvet_t = \Pvet\xvet_t = \begin{bmatrix}\avet_t \\\bvet_t \end{bmatrix}\in\mathbb{R}^{(n \times 1)},
$$ 
where $\Pvet\in\{0,1\}^{(n \times n)}$ is a permutation matrix used to appropriately re-order the original vector $\xvet_t$. If the classification as upper or bottom of the single time series in $\xvet_t$ is known in advance, we assume $\xvet_t = \yvet_t = \left[\avet_t' \quad \bvet_t'\right]'$, i.e. $\Pvet = \Ivet_n$ (no permutation of the original vector is needed). Moreover, also the linear combination (aggregation) matrix $\Avet\in\{0,1\}^{(n_a \times n_b)}$ describing the summation constraints linking the upper time series to the bottom ones, $\avet_t = \Avet\bvet_t$, is assumed known in advance. Thus the “structural representation" is simply given by (\citealp{athanasopoulos2009})
\begin{equation}
\label{eq:Smat}
	\yvet_t = \Svet \bvet_t\qquad\mbox{with}\qquad\Svet = \begin{bmatrix}
	\Avet\\
	\Ivet_{n_b}
\end{bmatrix},
\end{equation}
where $\Svet\in\{0,1\}^{(n\times n_a)}$ is the structural (summation) matrix.

Suppose now we have the vector $\widehat{\yvet}_h = \begin{bmatrix}\widehat{\avet}_h' & \widehat{\bvet}_h' \end{bmatrix}'\in \mathbb{R}^{(n\times 1)}$ of unbiased and incoherent (i.e., $\widehat{\yvet}_h \neq \Svet \widehat{\bvet}_h$) base forecasts for the $n$ variables of the linearly constrained series $\yvet_t$ for the forecast horizon $h$. \cite{hyndman2011} use the structural representation (\ref{eq:Smat}) to obtain the reconciled forecasts $\widetilde{\yvet}_h$ as
\begin{equation}
	\label{eq:Seq}
\widetilde{\yvet}_h = \Svet\Gvet\widehat{\yvet}_h, \quad \Gvet = \left(\Svet'\Wvet^{-1}\Svet\right)^{-1}\Svet'\Wvet^{-1},
\end{equation}
where $\Wvet$ is a $(n\times n)$ p.d. matrix assumed known and $\widehat{\yvet}_h$ ($\widetilde{\yvet}_h$) is the vector containing the base (reconciled) forecasts at forecast horizon $h$. Some alternative matrices $\Wvet$ have been proposed in the literature for the cross-sectional forecast reconciliation case:\footnote{Dealing with the uncertainty in base forecasts, and their error covariance matrices, is of primary interest to establish characteristics and properties of the reconciled forecasts. This topic is left to future research.}
\begin{itemize}
	\item identity (ols): $\widehat{\Wvet}_{\mathrm{ols}} = \Ivet_n$ \citep{hyndman2011},
	\item series variance (wls): $\widehat{\Wvet}_{\mathrm{wls}} = \Ivet_{n} \odot \widehat{\Wvet}_{\mathrm{sam}}$ \citep{hyndman2016},
	\item MinT-shr (shr): $\widehat{\Wvet}_{\mathrm{shr}}=\widehat{\lambda} \widehat{\Wvet}_{\mathrm{wls}} +(1-$ $\widehat{\lambda}) \widehat{\Wvet}_{\mathrm{sam}}$ \citep{wickramasuriya2019},
	\item MinT-sam (sam): $\widehat{\Wvet}_{\mathrm{sam}}=\frac{1}{T} \sum_{t=1}^{T} \widehat{\evet}_{t} \widehat{\evet}_{t}^{\prime}$ is the covariance matrix of the one-step ahead in-sample forecast errors $\widehat{\mathbf{e}}_{t}$ \citep{wickramasuriya2019},
\end{itemize}
where the symbol $\odot$ denotes the Hadamard product, and $\widehat{\lambda}$ is an estimated shrinkage coefficient \citep{ledoit2004a}.

The structural representation (\ref{eq:Smat}) may be transformed into the equivalent zero-constrained representation $\avet_t - \Avet\bvet_t = \Zerovet_{(n_a \times 1)}$, that is \citep{wickramasuriya2019}
\begin{equation}
	\label{eq:zeroconstraints}
\Cvet\yvet_t = \Zerovet_{(n_a \times 1)}
\qquad
\text{with} \qquad
	\Cvet = \begin{bmatrix}
		\Ivet_{n_a} & -\Avet
	\end{bmatrix}.
\end{equation}
$\Cvet \in\{-1,0,1\}^{(n_a \times n)}$ is a full row-rank zero constraints matrix used to obtain the point reconciled forecasts according to the \textit{projection approach}
\citep{vanerven2015, wickramasuriya2019, difonzo2021a}:
\begin{equation}
\label{eq:Meq}
	\widetilde{\yvet}_h = \Mvet\widehat{\yvet}_h, \quad \Mvet = \Ivet_n - \Wvet\Cvet'\left(\Cvet\Wvet\Cvet'\right)^{-1}\Cvet .
\end{equation}
Structural and zero-constrained representations can be interchangeable depending on which of the two is more convenient to use, allowing for greater flexibility in the calculation of the reconciled forecasts.
For, the zero-constrained representation appears to be less computational intensive: in equation (\ref{eq:Seq}) two matrices must be inverted, one of size ($n\times n$) and the other ($n_b \times n_b$), whereas only the inversion of an ($n_a \times n_a$) matrix is required in formula (\ref{eq:Meq}).

\subsection{The general case: zero-constrained and structural-like representations}\label{sec:struc}
In the hierarchical/grouped cross-sectional forecast reconciliation there is a natural distinction between upper and bottom time series that leads to the construction of the matrix $\Cvet$ as in (\ref{eq:zeroconstraints}), where the first $n_a$ columns refer to the upper and the remaining $n_b$ ($=n-n_a$) to the bottom variables, respectively. The time series in these two sets are categorized logically: all those related to the last level belong to the second group, the rest to the first. 
Most of the forecast reconciliation approaches proposed in the literature refer to genuine hierarchical/grouped time series and its structural representation. However, these do not take into account the full spectrum of possible cases encountered in real life situations.
For a general linearly constrained multiple time series
$\left( \xvet_t, \; t=1,\ldots,T\right)$, the classification between upper and bottom variables
might not be meaningful,
prompting us rather to look for two new sets: the \textbf{constrained} variables, denoted as $\cvet_t\in\mathbb{R}^{(n_c \times 1)}$, and the \textbf{free} (unconstrained) variables, denoted as $\uvet_t\in\mathbb{R}^{(n_u \times 1)}$,
with $n = n_c + n_u$, 
such that
$\yvet_t = \Pvet\xvet_t = \left[\cvet_t' \quad \uvet_t'\right]'$, and
\begin{equation}
\label{eq:genC}
	\cvet_t =  \Avet \uvet_t.
\end{equation}
In this general framework, $\Avet\in \mathbb{R}^{(n_c \times n_u)}$ is the \textit{linear combination} matrix associated to the linearly constrained multiple time series $\xvet_t = \Pvet'\yvet_t$, that can be thus expressed \textit{via} the ``structural-like representation''
\begin{equation}
\label{eq:barSmat}
	\yvet_t = \Pvet \xvet_t = \begin{bmatrix}
	\cvet_t\\
	\uvet_t
 \end{bmatrix} = \Svet \uvet_t \quad \mathrm{with} \quad \Svet = \begin{bmatrix}
	\Avet\\
	\Ivet_{n_u}
 \end{bmatrix},
\end{equation}
where $\Svet \in \mathbb{R}^{(n_c \times n_u)}$ is the structural-like matrix\footnote{Unlike the structural (summation) matrix in (\ref{eq:Smat}),  that describes the simple summation relationships valid for genuine hierarchical/grouped time series, and has only elements in $\{0,1\}$, in expression (\ref{eq:barSmat}) $\Svet$ consists of real coefficients, appropriately organized to highlight the links between constrained and free variables.}. It is worth noting that a full-rank zero-constraints matrix $\Cvet\in\mathbb{R}^{(n_c \times n)}$ like in expression (\ref{eq:zeroconstraints}) may be easily obtained by expression (\ref{eq:barSmat}) and used for the full-rank zero constrained representation $\Cvet\yvet_t = \Cvet\Pvet\xvet_t = \Zerovet_{(n_c\times 1)}$. Therefore, even for a general, possibly not genuine hierarchical/grouped, linearly constrained multiple time series, either the structural (\ref{eq:Seq}) or the projection (\ref{eq:Meq}) approaches may be used to perform the point forecast reconciliation of incoherent base forecasts.

\begin{figure}[t]
\centering
\resizebox{0.9\linewidth}{!}{
\begin{tikzpicture}[baseline=(current  bounding  box.center),
			rel/.append style={shape=circle,
				draw=black,
			minimum width=1cm,
			minimum height=1cm},
			connection/.style ={inner sep =0, outer sep =0}]
				
			\node[rel] at (0, 0) (A1){A1};
			\node[rel] at (1.5, 0) (A2){A2};
			\node[rel] at (3, 0) (B){B};
			\node[rel] at (0.75, 1.75) (A){A};
			\node[rel] at (1.9, 3.5) (F){X};
			
			\relation{0.2}{A1}{A};
			\relation{0.2}{A2}{A};
			\relation{0.2}{A}{F};
			\relation{0.2}{B}{F};
		\end{tikzpicture}$\;\;\bigcup\;$
\begin{tikzpicture}[baseline=(current  bounding  box.center),
			rel/.append style={shape=circle,
				draw=black,
			minimum width=1cm,
			minimum height=1cm},
			connection/.style ={inner sep =0, outer sep =0}]
				
			\node[rel] at (0, 0) (C){C};
			\node[rel] at (1.5, 0) (D){D};
			\node[rel] at (0.75, 3.5) (F){X};
			
			\relation{0.1}{C}{F};
			\relation{0.1}{D}{F};
		\end{tikzpicture}$\qquad\longrightarrow\qquad$
\begin{tikzpicture}[baseline=(current  bounding  box.center),
			rel/.append style={shape=circle,
				draw=black,
			minimum width=1cm,
			minimum height=1cm},
			connection/.style ={inner sep =0, outer sep =0}]
				
			\node[rel] at (0, 0) (A1){A1};
			\node[rel] at (0, 1.5) (A2){A2};
			\node[rel] at (0, 3) (B){B};
			\node[rel] at (5.25, 0.75) (C){C};
			\node[rel] at (5.25, 2.25) (D){D};		
			\node[rel] at (1.75, 0.75) (A){A};
			
			\node[rel, fill = black!20] at (3.5, 1.5) (X){X};
			
			\relationD{0.5}{A}{A1};
			\relationD{0.5}{A}{A2};
			\relationD{0.2}{C}{X};
			\relationD{0.2}{D}{X};
			\relationD{0.5}{X}{A};
			\relationD{2.25}{X}{B};
		\end{tikzpicture}}
		\vskip0.15cm
		\caption{A general linearly constrained structure: two hierarchies sharing only the same top-level series.}
\label{fig:hierpap}
\end{figure}
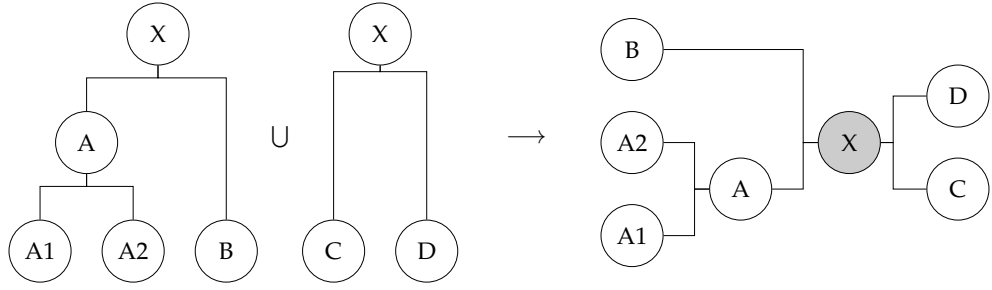

A simple example of a linearly constrained multiple time series that cannot be expressed as a genuine hierarchical/grouped structure is shown in \autoref{fig:hierpap}. In this case the variable X is at the top of two distinct hierarchies, that do not share the same bottom-level variables. The aggregation relationships between the upper variables X and A, and the bottom ones A1, A2, B, C, and D are given by:
\begin{equation}\label{sys:mg}
	\begin{aligned}
		X &= A1+A2+B \\
		X &= C+D\\
		A &= A1+A2
	\end{aligned} .
\end{equation}
In this case, both the zero-constrained and the structural-like representations are found in a rather straightforward manner. We consider $\left\{A2, B, C, D\right\}$ as free variables, such that $\yvet_t = \xvet_t$ (i.e., $\Pvet = \Ivet_4$), with $\cvet_t = \begin{bmatrix}
	x_{{X},t} & x_{{A},t} & x_{{A1},t}
\end{bmatrix}' $ 
and $\uvet_t = \begin{bmatrix}
	x_{{A2},t}&
	x_{{B},t}&
	x_{{C},t} & 
	x_{{D},t}
\end{bmatrix}'$.
Thus, the coefficient matrix of the zero-constrained representation (\ref{eq:Uy0}) is
$$
\Gammavet = \begin{bmatrix}
	1 & 0 & -1 & -1 & -1 & 0 & 0 \\
	1 & 0 & 0 & 0 & 0 & -1 & -1 \\
	0 & 1 & -1 & -1 & 0 & 0 & 0 \\
\end{bmatrix} .
$$
It is immediate to check that the system of linear constraints (\ref{sys:mg}) may be re-written as
\begin{align*}
	X &= C+D \\
	A &= -B+C+D\\
	A1 &= -A2-B+C+D
\end{align*}
that is $\Cvet \xvet_t = \Zerovet_{(3 \times 1)}$, where
$$
\Cvet = \begin{bmatrix} \Ivet_3 & -\Avet \end{bmatrix} = \begin{bmatrix}
	1 & 0 & 0 & 0 & 0 & -1 & -1 \\
	0 & 1 & 0 & 0 & 1 & -1 & -1 \\
	0 & 0 & 1 & 1 & 1 & -1 & -1 \\
\end{bmatrix} \quad \mathrm{and}\quad  
\Avet = \begin{bmatrix}
	0 & 0 & 1 & 1 \\
	0 & -1 & 1 & 1 \\
	-1 & -1 & 1 & 1 \\
\end{bmatrix}.
$$
The structural-like representation of the general linearly constrained multiple time series in \autoref{fig:hierpap} is then $\yvet_t = \Svet  \uvet_t$, with $\Svet = \left[\Avet' \quad \Ivet_3 \right]'$. 

It should be mentioned, however, that for medium/large systems (with many constraints and/or variables), manually operating on the constraints could be time-consuming and challenging. In \autoref{sec:Cbar} we will show a general technique to derive the linear combination matrix $\Avet$ from a general zero-constraints matrix $\Gammavet$.

\subsection{Probabilistic forecast reconciliation for general linearly constrained multiple time series}\label{sec:prob}

So far we have dealt with only point forecasting, but if one wishes to account for forecast uncertainty, probabilistic forecasts should be considered, as they - in the form of probability distributions over future quantities or events - measure the uncertainty in forecasts and are an important component of optimal decision making \citep{gneiting2014}. 

Representation (\ref{eq:barSmat}) states that $\yvet_t$ lies in an $n$-dimensional subspace of $\mathbb{R}^n$ spanned by the columns of $\Svet$, called “coherent subspace” and denoted by $\mathcal{S}$ \citep{panagiotelis2020prob}. Now, let $\mathscr{F}_{\mathbb{R}^{n_u}}$ be the Borel $\sigma$-algebra on $\mathbb{R}^{n_u}$, $\left(\mathbb{R}^{n_u}, \mathscr{F}_{\mathbb{R}^{n_u}}, \nu\right)$ a probability space for the free variables, and ${s}: \mathbb{R}^{n_u} \rightarrow \mathbb{R}^n$ a continuous mapping matrix. Then a $\sigma$-algebra $\mathscr{F}_{\mathcal{S}}$ can be constructed from the collection of sets ${s}(\mathcal{B})$ for all $\mathcal{B} \in \mathscr{F}_{\mathbb{R}^{n}}$.
\begin{definition}[\textit{Coherent probabilistic forecast for a linearly constrained multiple time series}]
Given the triple $\left(\mathbb{R}^{n_u}, \mathscr{F}_{\mathbb{R}^{n_u}}, \nu\right)$, we define a coherent probability triple $\left(\mathcal{S}, \mathscr{F}_{\mathcal{S}}, \breve{\nu}\right)$ such that $\breve{\nu}({s}(\mathcal{B}))=\nu(\mathcal{B})$, $\forall \mathcal{B} \in \mathscr{F}_{\mathbb{R}^{n_u}}$.
\end{definition}

In order to extend forecast reconciliation to the probabilistic setting, let $\left(\mathbb{R}^{n}, \mathscr{F}_{\mathbb{R}^{n}}, \widehat{\nu}\right)$ be a probability triple characterizing base (incoherent) probabilistic forecasts for all $n$ series, and let $\psi: \mathbb{R}^{n_u} \rightarrow \mathbb{R}^n$ be a continuous mapping function defined by \cite{panagiotelis2020prob} as the composition of two transformations, ${s} \circ {g}$, where ${g} : \mathbb{R}^n \rightarrow \mathbb{R}^{n_u}$ is a continuous function corresponding to matrix ${\Gvet}$ in equation (\ref{eq:Seq}).
\begin{definition}[\textit{Probabilistic forecast reconciliation for a linearly constrained multiple time series}] The reconciled probability measure of $\widehat{\nu}$ with respect to $\psi$ is a probability measure $\widetilde{\nu}$ on $\mathcal{S}$ with $\sigma$-algebra $\mathscr{F}_{\mathcal{S}}$ such that
	\begin{equation}\label{def:vrec}
	\widetilde{\nu}(\mathcal{A})=\widehat{\nu}\left(\psi^{-1}(\mathcal{A})\right), \quad \forall \mathcal{A} \in \mathscr{F}_{\mathcal{S}},
	\end{equation}
	where $\psi^{-1}(\mathcal{A})=\left\{\boldsymbol{x} \in \mathbb{R}^{n}: \psi(\boldsymbol{x}) \in \mathcal{A}\right\}$ is the pre-image of $\mathcal{A}$. 
\end{definition}
We consider two alternative approaches to deal with probabilistic forecast reconciliation according to the above definitions: a parametric framework, where probabilistic forecasts are produced under the assumption that the density function of the forecast errors is known, and a non-parametric framework, where no distributional assumption is made.

\subsubsection{Parametric framework: Gaussian reconciliation}

A reconciled probabilistic forecast may be obtained analytically for some parametric distributions, such as the multivariate normal \citep{yagli2020, corani2021, eckert2021, panagiotelis2020prob, wickramasuriya2021b}. 
In particular, if the base forecasts distribution is $\mathcal{N}(\widehat{\yvet}_h, \Wvet_h)$, then the reconciled forecasts distribution is $\mathcal{N}(\widetilde{\yvet}_h, \widetilde{\Wvet}_h)$, with
$$
\widetilde{\yvet}_h = {\Svet} \,{\Gvet}\widehat{\yvet}_h \quad \mbox{and} \quad \widetilde{\Wvet}_h = {\Svet}\, {\Gvet} \Wvet_h {\Gvet}'\, {\Svet}'.
$$
The covariance matrix $\widetilde{\Wvet}_h$ deserves special attention. In the simple case assumed by \cite{wickramasuriya2019}, $\Wvet_h = k_h\Wvet$, where $k_h$ is a proportionality constant, and the reconciled covariance matrix reduces to (see Appendix~\ref{app:covh}):
\begin{equation}
\label{eq:covh}
	\widetilde{\Wvet}_h = k_h {\Svet}\, {\Gvet} \Wvet .
\end{equation}
However, the proportionality assumption along the forecast horizons $h$ may be too restrictive, and computing $k_h$ cannot be an easy task. Thus, three alternative formulations of ${\Wvet}_h$, already shown in \autoref{sec:hts}, have been proposed in the forecast reconciliation literature:
\begin{itemize}[nosep]
	\item diagonal covariance matrix: $\Wvet_h = \widehat{\Wvet}_{\mathrm{wls}}$ \citep{corani2021, panagiotelis2020prob};
	\item sample covariance matrix: $\Wvet_h = \widehat{\Wvet}_{\mathrm{sam}}$ \citep{panagiotelis2020prob};
	\item shrinkage covariance matrix: $\Wvet_h = \widehat{\Wvet}_{\mathrm{shr}}$ \citep{athanasopoulos2020}.
\end{itemize}

\subsubsection{Non-parametric framework: joint bootstrap-based reconciliation}\label{sec:jointboot}
When an analytical expression of the forecast distribution is either unavailable, or relies on unrealistic parametric assumptions, the empirical evaluation of the results may be grounded on reconciled samples \citep{jeon2019, yang2020, panagiotelis2020prob}. At this end, we extend theorem 4.5 in \cite{panagiotelis2020prob}, originally formulated for genuine hierarchical/grouped time series, to the case of a general linearly constrained multiple time series. This theorem states that, if $\left(\widehat{\yvet}^{[1]}, \ldots, \widehat{\yvet}^{[L]}\right)$ is a sample drawn from an incoherent probability measure $\widehat{\nu}$, then $\left(\widetilde{\yvet}^{[1]}, \ldots, \widetilde{\yvet}^{[L]}\right)$, where $\widetilde{\yvet}^{[\ell]}:=\psi\left(\widehat{\yvet}^{[\ell]}\right)$ for $\ell=$ $1, \ldots, L\;$, is a sample drawn from the reconciled probability measure $\widetilde{\nu}$ as defined in (\ref{def:vrec}). According to this result, reconciling each member of a sample obtained from an incoherent distribution yields a sample from the reconciled distribution. As a consequence, coherent probabilistic forecasts may be developed through a post-forecasting mechanism analogous to the point forecast reconciliation setting. 
For this purpose, the bootstrap procedure by \cite{athanasopoulos2020} is applied:
\begin{enumerate}
    \item appropriate univariate models $M_i$ for each series in the system are fitted based on the training data $\{y_{i,t}\}_{t=1}^T$, $i = 1,\dots,n$, and the one-step-ahead in-sample forecast errors are stacked in an $(n \times T)$ matrix, $\widehat{\Evet}=\left\{\widehat{e}_{i, t}\right\}$;
    \item $\widehat{\yvet}^{[l]}_{i,h} = f_i\left(M_i, \widehat{e}_{i, h}^{[l]}\right)$ is computed for $h = 1,\dots,H$ and $l = 1,\dots, L$, where $f(\cdot)$ is a function of the fitted univariate model and its associated error, $\widehat{\yvet}^{[l]}_{i,h}$ is a sample path simulated for the $i-$th series, and $\widehat{e}_{i, h}^{[l]}$ is the $(i, h)-$th element of an $(n\times H)$ block bootstrap matrix containing $H$ consecutive columns randomly drawn from $\widehat{\Evet}$;
	\item the optimal reconciliation formula, either according to the structural approach (\ref{eq:Seq}) or the projection approach (\ref{eq:Meq}), is applied to each $\widehat{\yvet}^{[l]}_{h}$.  
\end{enumerate}

\section{Building the linear combination matrix $\mathbf{A}$}\label{sec:Cbar}
In the previous section, we limited ourselves to introduce the linear combination matrix $\Avet$ in expression (\ref{eq:genC}), in line with the novel classification distinguishing between free (unconstrained) and constrained variables. In this section we propose two ways of building such a matrix in practice. 

First, consider the simple case where there are no redundant constraints ($n_c = p$) and the first $n_c$ columns of $\Gammavet$ are linearly independent, so that $\yvet_t = \xvet_t =  \Big[\cvet_t' \quad \uvet_t'\Big]'$ and $\Gammavet\yvet_t = \Zerovet_{(n_c\times 1)}$. This homogeneous linear system can be written as
$$
\Gammavet \yvet_t = \Zerovet_{(n_c\times 1)} \quad \Longleftrightarrow \quad \begin{bmatrix}
	\Gammavet_c & \Gammavet_u
\end{bmatrix}\begin{bmatrix}
	\cvet_t\\
	\uvet_t
\end{bmatrix} = \Zerovet_{(n_c\times 1)} ,
$$
where $\Gammavet_c\in\mathbb{R}^{(n_c \times n_c)}$ contains the coefficients for the constrained variables, and $\Gammavet_u\in\mathbb{R}^{(n_c \times n_u)}$  those for the free ones. Thanks to its non-singularity, $\Gammavet_c$ can be used to derive the equivalent zero-constrained representation:
\begin{equation}
\label{eq:meth1}
\begin{bmatrix}\Ivet_{n_c} & \left(\Gammavet_c\right)^{-1}\Gammavet_u\end{bmatrix}\begin{bmatrix}
	\cvet_t\\
	\uvet_t
\end{bmatrix} = \Cvet \yvet_t= \Zerovet_{(n_c\times 1)} ,
\end{equation}
where
\begin{equation}
\label{eq:easyC}
	\Cvet = \begin{bmatrix}\Ivet_{n_c} & -\Avet\end{bmatrix} \qquad \mathrm{and} \qquad \Avet = -\left(\Gammavet_c\right)^{-1}\Gammavet_u .
\end{equation}
In practical situations, mostly if many variables and/or constraints are involved, categorizing variables as either constrained or free may be a challenging task\footnote{The issue of defining a valid set of free variables is studied by \cite{zhang2022} in the framework of hierarchical/grouped reconciliation with immutable forecasts.}: the goal is to identify a valid set of free variables with invertible coefficient matrix $\Gammavet_c$.

\subsection{General (redundant) linear constraints framework} \label{sec:genconst}

When $n$ is large, it is not always immediate to find a set of non-redundant constraints, so the method shown in expression (\ref{eq:meth1}) may be hardly applied in several real-life situations. We propose to overcome these issues by employing standard linear algebra tools, like the Reduced Row Echelon Form or the QR decomposition (\citealp{golub1996matrix}, \citealp{meyer2000}), that are able to deal with redundant constraints and do not request any a priori classification of the single variables entering the multiple time series.

\subsubsection*{Reduced Row Echelon Form (rref)}

\noindent A matrix is said to be in rref \citep{meyer2000} if and only if the following three conditions hold:
\begin{itemize}[nosep]
	\item it is in row echelon form; 
	\item the pivot in each row is 1;
	\item all entries above each pivot are 0.
\end{itemize}
The idea is then very simple: classify as constrained the variables corresponding to the pivot positions of the rref representation coefficient matrix, while the remaining columns form the linear combination matrix $\Avet$. 
Usually a rref form is obtained through a Gauss-Jordan elimination (more details in \citealp{meyer2000}). So, let $\Zvet\in\mathbb{R}^{(n_c\times n)}$ be the rref of $\Gammavet$ deprived of any possible null rows, then a permutation matrix $\Pvet$ can be obtained starting from the pivot columns of $\Zvet$, such that
$$
\yvet_t = \Pvet \xvet_t = \begin{bmatrix}
	x_{\pi_c(1), t} & \dots & x_{\pi_c(n_c), t} & x_{\pi_u(1), t} & \dots &  x_{\pi_u(n_u), t}
\end{bmatrix} ,
$$
where $\pi_c(i)$, $i = 1,..., n_c$, is the position of the $i$-th pivot column (i.e., one of the columuns that identify the constrained variables) and $\pi_u(j)$, $j = 1,..., n_u$, is the position of the $j$-th no-pivot column (i.e., one of the columns associated to the free variables). 
Then, the linear combination matrix $\Avet$ can be extracted from the expression
\begin{equation*}
	\Cvet = \Zvet \Pvet' = \begin{bmatrix}\Ivet_{n_c} & -\Avet\end{bmatrix}.
\end{equation*}
Additional examples can be found in the online appendix.

\subsubsection*{QR decomposition}
Given the $(p \times n)$ coefficient matrix $\Gammavet$ of the zero-constrained representation (\ref{eq:Uy0}), $\Gammavet = \Qvet \Rvet\Pvet$ is a QR decomposition with column pivoting \citep{lyche2020}, where $\Qvet\in\mathbb{R}^{(p \times p)}$ is a square and orthonormal matrix ($\Qvet'\Qvet = \Qvet\Qvet'=\Ivet_p$), $\Pvet\in\{0,1\}^{(n \times n)}$ is a permutation matrix, and $\Rvet\in\mathbb{R}^{(p \times n)}$ is an upper trapezoidal matrix \citep{Anderson1992,anderson1999lapack} such that
$$
\Rvet = \begin{cases}
\big[ \Rvet_c \quad \Rvet_u\big] & \mbox{if}\;\Gammavet\;\mbox{is full-rank}\\[0.1cm]
\begin{bmatrix}     \Rvet_c & \Rvet_u \\
    \Zerovet_{[(p-n_c) \times n_c]} & \Zerovet_{[(p-n_c) \times n_u]}
\end{bmatrix}& \mbox{if}\;\Gammavet\;\mbox{is not full-rank}
\end{cases},
$$
where $\Rvet_{c}\in \mathbb{R}^{(n_c \times n_c)}$ is upper triangular, and $\Rvet_{u}\in \mathbb{R}^{(n_c \times n_u)}$ is nonsingular \citep{golub1996matrix}. Applying this decomposition to the homogenous system (\ref{eq:Uy0}), we obtain
$$
\Gammavet\xvet_t = \Qvet \Rvet\Pvet\xvet_t = \Qvet \Rvet\yvet_t = \Zerovet_{(p\times 1)} ,
$$
that is equivalent to \citep{meyer2000}
$$
\begin{cases}
	\Qvet\zvet = \Zerovet_{(p \times 1)}\\
	\Rvet\yvet_t = \zvet = \Zerovet_{(p \times 1)}
\end{cases} .
$$
Due to the non-singularity of $\Qvet$, $\zvet = \Zerovet_{(p \times 1)}$ is the unique solution to the homogenous system $\Qvet\xvet = \Zerovet_{(p \times 1)}$ \citep{lyche2020}. Then, the homogeneous system (\ref{eq:Uy0}) representing a general linearly constrained time series may be re-written as\footnote{Possible null rows, present if $\Gammavet$ is not full-rank, are removed.} $\big[ \Rvet_c \quad \Rvet_u\big]\yvet_t = \Zerovet_{(n_c\times 1)}$. Finally, from (\ref{eq:meth1}) we obtain
$$
\Cvet = \big[\Ivet_{n_c}\quad -\Avet\big] \qquad \mathrm{and} \qquad \Avet = -\Rvet_c^{-1}\Rvet_u ,
$$
where $\Rvet_c$ is invertible by construction \citep{golub1996matrix}. It is worth noting that the Pivoted QR decomposition generates a permutation matrix $\Pvet$ that ``moves'' the free variables in $\xvet_t$ to the bottom part of the re-ordered vector $\yvet_t$, that is $\Pvet\xvet_t = \yvet_t = \big[\cvet_t' \quad \uvet_t'\big]'$.

It should be noted that in both cases (QR and rref), $\Pvet = \Ivet_n$ if the first $n_c$ columns of $\Gammavet$ are linearly independent. This means that both algorithms start by assuming as constrained and free the variables as they appear in $\xvet$, whose ordering is then changed only if it is not feasible for the constraints operating on the multivariate time series in equation (\ref{eq:Uy0}).

\section{Empirical applications}\label{sec:appl}
In this section we present two macroeconomics applications involving general linearly constrained multiple time series which do not have a genuine hierarchical/grouped structure. In the first case, in the wake of \cite{athanasopoulos2020}, we forecast the Australian $GDP$ from income and expenditure sides, for which \cite{bisaglia2020} already provided a full row-rank $\Cvet$ matrix. The second application concerns the European $GDP$ disaggregated by three sides (income, expenditure and output) and 19 member countries. In this case, building a full row-rank zero constraints matrix is not an easy task, so we use a QR decomposition (\autoref{sec:genconst}). Detailed informations on the variables in either dataset are reported in the online appendix.

\subsection{Reconciled probabilistic forecasts of the Australian $GDP$ from income and expenditure sides}\label{sec:ausGDP}

\cite{athanasopoulos2020} first considered the reconciliation of point and probabilistic forecasts of the 95 Australian Quarterly National Accounts (QNA) variables that describe the Gross Domestic Product ($GDP$) at current prices from the income and expenditure sides,  interpreted as two distinct hierarchies. In the former case (income), $GDP$ is the top level aggregate of a hierarchy of 15 lower level aggregates with $n_a^I = 6$ and $n_b^I = 10$, whereas in the latter (expenditure), $GDP$ is the top level aggregate of a hierarchy of 79 time series, with $n_a^E = 27$ and $n_b^E = 53$ (for details, \citealp{athanasopoulos2020, bisaglia2020, difonzo2022c, difonzo2021a}). 

\begin{figure}[!t]
	\includegraphics[width = \linewidth]{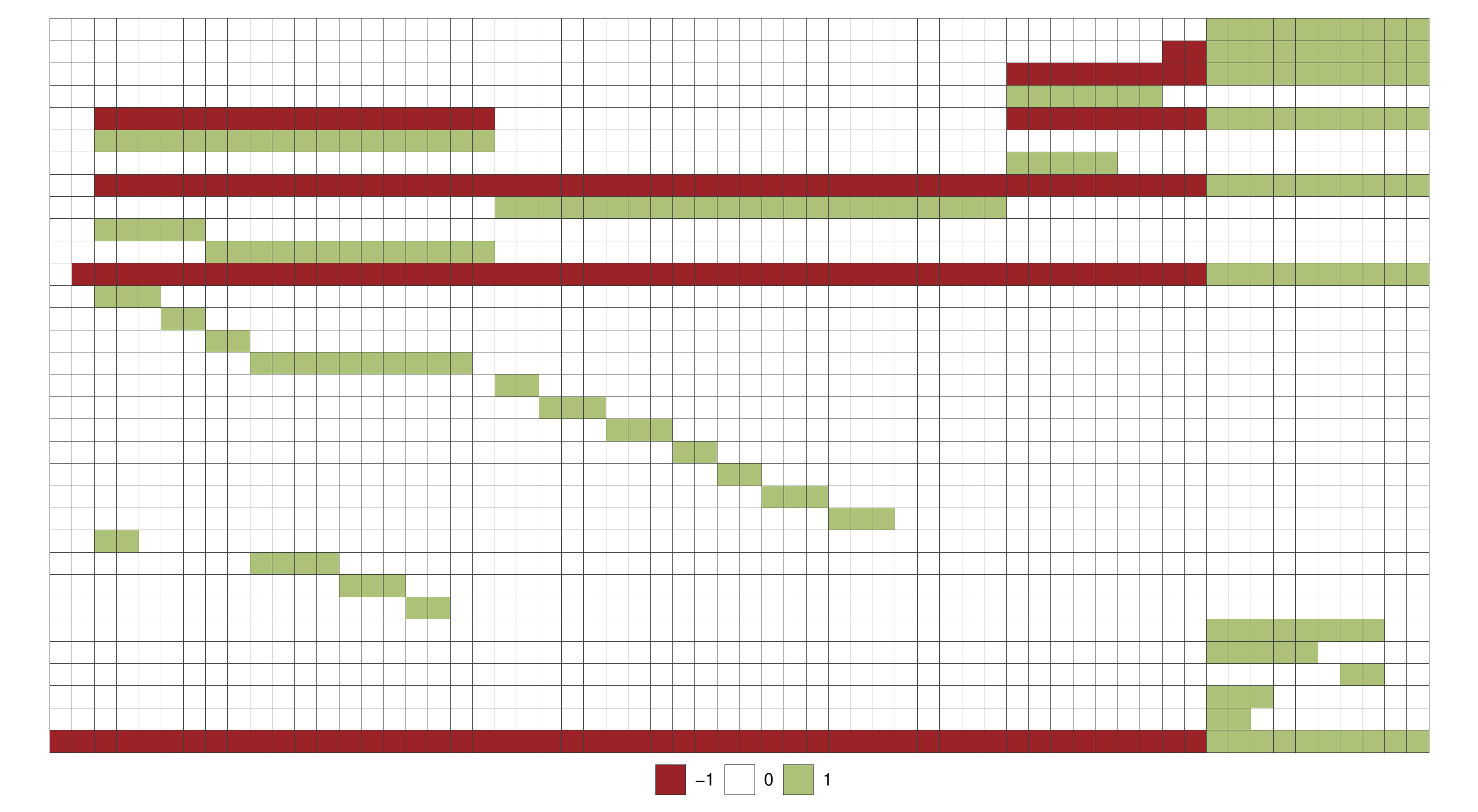}
	\caption{Linear combination matrix $\Avet$ for the Australian $GDP$ from income and expenditure sides.}
	\label{fig:aus_MAT}
	
\end{figure}

Considering these two hierarchies as distinct yields different $GDP$ forecasts depending on the considered disaggregation (either by income or expenditure). The fact that the two hierarchical structures describing the National Accounts share only the same top-level series ($GDP$), prevents the adoption for the whole set of $n=95$ distinct variables of the original structural reconciliation approach proposed by \cite{hyndman2011}. However, it is possible to use the results shown so far for a general linearly constrained multiple time series. The homogeneous constraints valid for the variables are described by the following $(33 \times 95)$ matrix $\Gammavet$ \citep{bisaglia2020}:
\begin{equation}
\label{eq:Ut}
\Gammavet = \begin{bmatrix}
 1 & \Zerovet_5' & -{\bf 1}_{10}' & \Zerovet_{26}' &  \Zerovet_{53}' \\
 1 & \Zerovet_5' &  \Zerovet_{10}' & \Zerovet_{26}' & -{\bf 1}_{53}' \\
 \Zerovet_5 & {\bf I}_5 & -\Avet^I & \Zerovet_{5 \times 26} & \Zerovet_{5 \times 53} \\
 \Zerovet_{26} & \Zerovet_{26 \times 5} & \Zerovet_{26 \times 10} & {\bf I}_{26}          & -\Avet^E 
\end{bmatrix},
\end{equation}
where $\Avet^I \in \{0,1\}^{(5 \times 10)}$ and $\Avet^E \in \{0,1\}^{(26 \times 53)}$ are the aggregation matrices for the income and the expenditure sides, respectively, and  $\Gammavet$ has already full row-rank. A structural-like representation of the multiple time series that incorporates both sides' accounting constraints may be obtained by transforming $\Gammavet$ through, for example, the QR technique described in \autoref{sec:genconst}. This operation results in a $(33 \times 95)$ matrix $\Cvet = \left[\Ivet_{33} \; -\Avet\right]$, where $\Avet$ is the $(33 \times 62)$ linear combination matrix shown in \autoref{fig:aus_MAT}, and $\Svet = \left[\Avet' \quad \Ivet_{62}\right]'$ is the structural-like matrix (see \autoref{sec:struc}).

We perform a forecasting experiment as the one designed by \cite{athanasopoulos2020}. Base forecasts from $h = 1$ quarter ahead up to $h = 4$ quarters ahead for all the 95 separate time series have been obtained through simple univariate ARIMA models selected using the {\tt auto.arima} function of the R-package {\tt forecast} \citep{hyndman2008a}. The first training sample is set from 1984:Q4 to 1994:Q3, and a recursive training sample with expanding window length is used, for a total of 94 forecast origins. Finally the reconciled forecasts are obtained using three reconciliation approaches (ols, wls and shr, see \autoref{sec:hts}) through the \textsf{R} package \texttt{Foreco} \citep{girolimetto2022}.

\begin{figure}[!t]
\centering
\includegraphics[width=\linewidth]{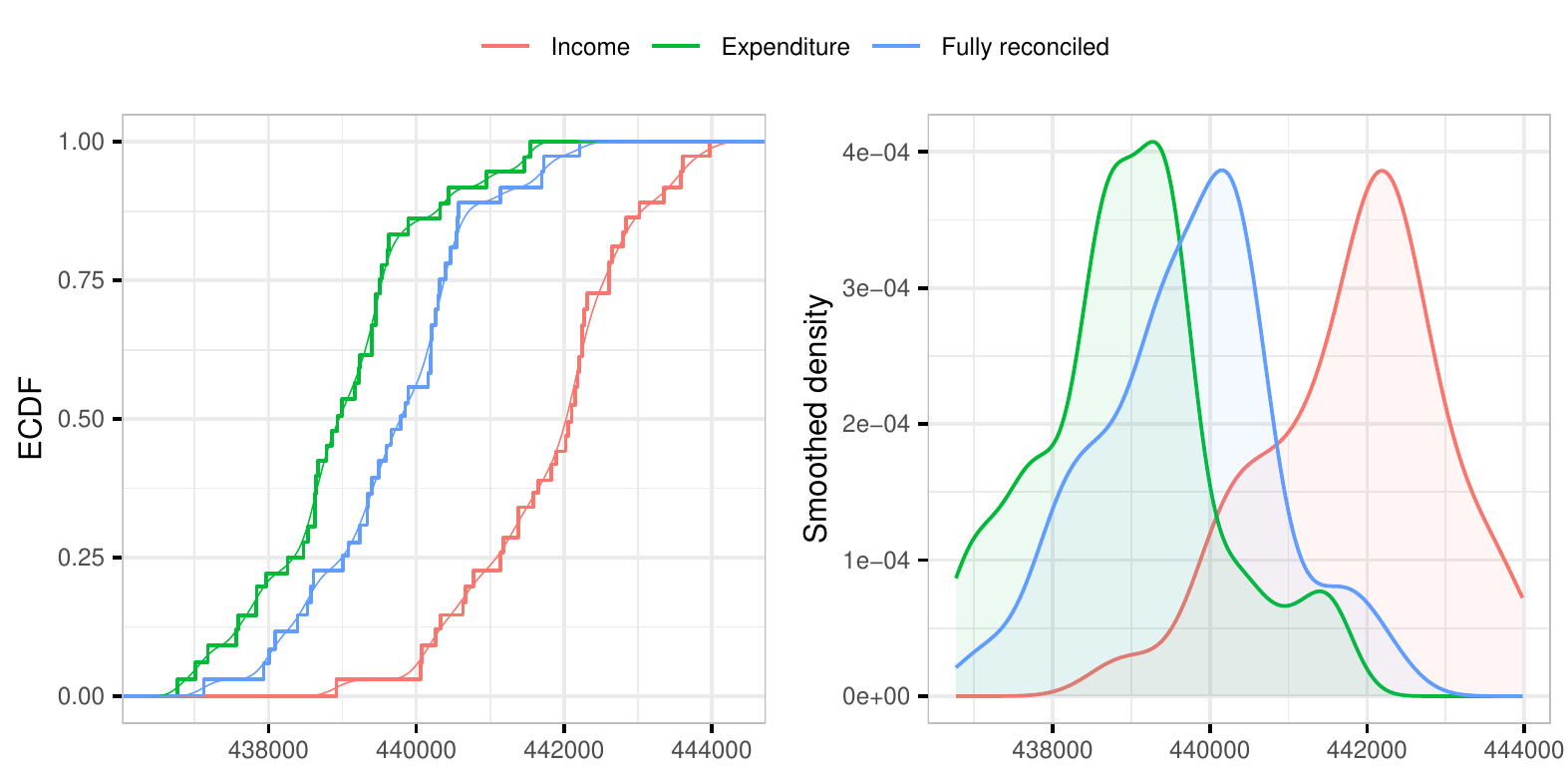}
 \vspace{-.55cm}
 \caption{Australian $GDP$ empirical one-step-ahead forecast distributions for 2018:Q1, shr joint bootstrap-based reconciliation approach. Empirical Cumulative Distribution Function (left), and Smoothed density (right).}
 \label{fig:diffdistr}
 
\end{figure}

In \cite{athanasopoulos2020} the probabilistic forecasts of the Australian quarterly $GDP$ aggregates are separately reconciled from income $\big(\widetilde{X}_{GDP}^I\big)$ and expenditure $\big(\widetilde{X}_{GDP}^E\big)$ sides. This means that the empirical forecast distributions $\widetilde{X}_{GDP}^I$ and $\widetilde{X}_{GDP}^E$ are each coherent (see \autoref{sec:prob}) within its own pertaining side with the other empirical forecast distributions, but in general $\widetilde{X}_{GDP}^I \ne \widetilde{X}_{GDP}^E$ at any forecast horizon. This circumstance could confuse the user, mostly when the difference between the empirical forecast distributions is not negligible, as shown in \autoref{fig:diffdistr}, where the $GDP$ empirical forecast distributions from income and expenditure sides for 2018:Q1 are presented along with their fully reconciled counterparts through the shr joint bootstrap-based reconciliation approach (see \autoref{sec:jointboot})\footnote{Note that the naive practice of averaging $GDP$ forecasts from different sides yields a single forecast, that is though inconsistent with the component variables from both sides.}.

\subsubsection*{Point and probabilistic forecasting accuracy}
To evaluate the accuracy of the point forecasts we use the Mean Square Error (MSE)\footnote{The Mean Absolute Scaled Error (MASE) leads to the same conclusions (see the online appendix).}:
$$
MSE_{j,h} = \frac{1}{n T_h}\sum_{i = 1}^{n}\sum_{t = 1}^{T_h} \tilde{e}_{i,j,t+h}^{\;2},
$$
where $\tilde{e}_{i,j,t+h} = \left(\tilde{y}_{i,j,t+h} - {y}_{i,j,t+h}\right)$ is the the $h$-step-ahead forecast error using the approach $j$ to forecast the $i$-th  series, $j = 0$ denotes the base forecast (i.e., $\tilde{y}_{i,0,t+h} = \widehat{y}_{i,t+h}$), and $t$ is the forecast origin. To assess any improvement in the reconciled forecasts compared to the base ones, we use the $MSE$-skill score: 
$$
\left(1 - \frac{MSE_{j,h}}{MSE_{0,h}}\right) \times 100.
$$
The accuracy of the probabilistic forecasts is evaluated using the Cumulative Rank Probability Score (CRPS, \citealp{gneiting2014}): 
$$
\operatorname{CRPS}(\hat{P}_i, z_i)=\frac{1}{L} \sum_{l=1}^{L}\left|x_{i,l}-z_i\right|-\frac{1}{2 L^{2}} \sum_{l=1}^{L} \sum_{j=1}^{L}\left|x_{i,l}-x_{i,j}\right|, \quad i = 1,\dots,n,
$$
where $\hat{P}_i(\omega)=\displaystyle\frac{1}{L} \displaystyle\sum_{l=1}^{L} \mathbf{1}\left(x_{i,l} \leq \omega\right)$, $\xvet_{1}, \xvet_{2}, \ldots, \xvet_{L}\in \mathbb{R}^{n}$ is a collection of $L$ random draws taken from the predictive distribution and $\zvet \in \mathbb{R}^{n}$ is the observation vector. In addition, to evaluate the forecasting accuracy for the whole system, we employ the Energy Score (ES), that is the CRPS extension to the multivariate case\footnote{An alternative to the Energy Score is the Variogram Score \citep{scheuerer2015}, considered in the online appendix, that leads to similar conclusions.}: 
$$
\operatorname{ES}(\hat{P}, \zvet)=\frac{1}{L} \sum_{l=1}^{L}\left\|\xvet_{l}-\zvet\right\|_{2}-\frac{1}{2(L-1)} \sum_{i=1}^{L-1}\left\|\xvet_{l}-\xvet_{l+1}\right\|_{2} .
$$

\subsubsection*{Results}

\begin{table}[!t]
\centering
\caption{MSE and CRPS-skill scores (relative to base forecasts) for the point and probabilistic Australian $GDP$ forecasts from alternative reconciliation approaches. Negative values are highlighted in red, the best for each row is marked in bold.}
\label{tab:AUS_gdp}
\begingroup
	\spacingset{1.1}
\begin{tabular}[t]{c|rrr|rrr|rrr}
\toprule
& \multicolumn{3}{c|}{Point forecasts}  & \multicolumn{6}{c}{Probabilistic forecasts - CRPS(\%)} \\
& \multicolumn{3}{c|}{MSE (\%)} & \multicolumn{3}{c|}{Bootstrap} & \multicolumn{3}{c}{Gaussian}\\
\hspace{1em}$h$ & \multicolumn{1}{c}{ols} & \multicolumn{1}{c}{wls} & \multicolumn{1}{c}{shr} & \multicolumn{1}{|c}{ols} & \multicolumn{1}{c}{wls} & \multicolumn{1}{c}{shr} & \multicolumn{1}{|c}{ols} & \multicolumn{1}{c}{wls} & \multicolumn{1}{c}{shr}\\
\midrule
\addlinespace[0.3em]
\multicolumn{10}{l}{\textbf{Income}}\\
\hspace{1em}1 & \textcolor{black}{1.63} & \textcolor{black}{1.07} & \textcolor{black}{\textbf{5.41}} & \textcolor{black}{\textbf{0.57}} & \textcolor{red}{-1.26} & \textcolor{black}{0.52} & \textcolor{black}{\textbf{0.13}} & \textcolor{red}{-1.87} & \textcolor{red}{-0.45}\\
\hspace{1em}2 & \textcolor{black}{2.54} & \textcolor{black}{5.68} & \textcolor{black}{\textbf{6.10}} & \textcolor{black}{\textbf{1.46}} & \textcolor{black}{0.68} & \textcolor{red}{-0.50} & \textcolor{black}{1.11} & \textcolor{black}{\textbf{1.26}} & \textcolor{red}{-0.34}\\
\hspace{1em}3 & \textcolor{black}{2.28} & \textcolor{black}{\textbf{7.81}} & \textcolor{black}{4.56} & \textcolor{black}{\textbf{0.18}} & \textcolor{red}{-0.64} & \textcolor{red}{-1.93} & \textcolor{black}{\textbf{0.14}} & \textcolor{red}{-0.24} & \textcolor{red}{-1.69}\\
\hspace{1em}4 & \textcolor{black}{1.98} & \textcolor{black}{\textbf{9.33}} & \textcolor{black}{7.04} & \textcolor{red}{\textbf{-0.08}} & \textcolor{red}{-1.12} & \textcolor{red}{-1.35} & \textcolor{red}{\textbf{-0.23}} & \textcolor{red}{-1.00} & \textcolor{red}{-1.39}\\
\addlinespace[0.3em]
\multicolumn{10}{l}{\textbf{Expenditure}}\\
\hspace{1em}1 & \textcolor{black}{\textbf{4.53}} & \textcolor{black}{0.07} & \textcolor{black}{2.48} & \textcolor{black}{\textbf{1.10}} & \textcolor{black}{0.78} & \textcolor{black}{0.08} & \textcolor{black}{\textbf{0.64}} & \textcolor{red}{-0.83} & \textcolor{red}{-0.70}\\
\hspace{1em}2 & \textcolor{black}{\textbf{5.09}} & \textcolor{black}{3.90} & \textcolor{black}{1.72} & \textcolor{black}{\textbf{2.34}} & \textcolor{black}{1.25} & \textcolor{red}{-1.07} & \textcolor{black}{\textbf{1.68}} & \textcolor{black}{0.58} & \textcolor{red}{-1.95}\\
\hspace{1em}3 & \textcolor{black}{6.96} & \textcolor{black}{\textbf{9.18}} & \textcolor{black}{6.24} & \textcolor{black}{\textbf{2.42}} & \textcolor{black}{1.50} & \textcolor{red}{-0.38} & \textcolor{black}{\textbf{1.69}} & \textcolor{black}{1.00} & \textcolor{red}{-1.02}\\
\hspace{1em}4 & \textcolor{black}{8.01} & \textcolor{black}{\textbf{11.76}} & \textcolor{black}{8.34} & \textcolor{black}{\textbf{3.73}} & \textcolor{black}{2.73} & \textcolor{black}{0.72} & \textcolor{black}{\textbf{3.18}} & \textcolor{black}{2.49} & \textcolor{black}{0.20}\\
\addlinespace[0.3em]
\multicolumn{10}{l}{\textbf{Fully reconciled}}\\
\hspace{1em}1 & \textcolor{black}{4.59} & \textcolor{black}{1.14} & \textcolor{black}{\textbf{4.77}} & \textcolor{black}{\textbf{1.13}} & \textcolor{red}{-0.75} & \textcolor{red}{-0.20} & \textcolor{red}{\textbf{-0.59}} & \textcolor{red}{-3.46} & \textcolor{red}{-2.85}\\
\hspace{1em}2 & \textcolor{black}{5.76} & \textcolor{black}{\textbf{6.24}} & \textcolor{black}{4.76} & \textcolor{black}{\textbf{2.81}} & \textcolor{black}{1.07} & \textcolor{red}{-1.27} & \textcolor{black}{\textbf{1.30}} & \textcolor{red}{-0.33} & \textcolor{red}{-3.02}\\
\hspace{1em}3 & \textcolor{black}{7.31} & \textcolor{black}{\textbf{10.94}} & \textcolor{black}{8.21} & \textcolor{black}{\textbf{1.99}} & \textcolor{black}{0.26} & \textcolor{red}{-1.46} & \textcolor{black}{\textbf{0.91}} & \textcolor{red}{-0.62} & \textcolor{red}{-2.80}\\
\hspace{1em}4 & \textcolor{black}{7.90} & \textcolor{black}{\textbf{13.24}} & \textcolor{black}{10.81} & \textcolor{black}{\textbf{2.83}} & \textcolor{black}{0.75} & \textcolor{red}{-0.63} & \textcolor{black}{\textbf{1.86}} & \textcolor{red}{-0.43} & \textcolor{red}{-2.05}\\
\bottomrule
\end{tabular}
	\endgroup
\end{table}

\autoref{tab:AUS_gdp} shows the MSE and CRPS-skill scores for the $GDP$ point and probabilistic reconciled forecasts. \autoref{tab:AUS_all} presents the MSE and ES-skill scores for all 95 Australian QNA variables from both income and expenditure sides. The `Income' and `Expenditure' panels, respectively, reproduce the results found by \cite{athanasopoulos2020}. The `Fully reconciled' panels show the skill scores for the simultaneously reconciled forecasts. For the point forecasts, all the reconciliation approaches improve forecast accuracy compared to the base forecasts. In detail, shr is almost always the best approach for the one-step-ahead forecasts, whereas wls is competitive for $h\geq 2$. Looking at the probabilistic reconciliation results in \autoref{tab:AUS_gdp}, it is worth noting that for $GDP$ ols outperforms both wls and shr, whatever side and framework (parametric or not) is considered. However, when all 95 variables are considered (\autoref{tab:AUS_all}), shr and wls approaches almost always show the best performance. In the Gaussian framework, these results are confirmed for the income side, whereas shr performs poorly when we look at the expenditure side (either fully reconciled or not).

\begin{table}[t]
\centering
\caption{MSE and ES-skill scores (relative to base forecasts) for the point and probabilistic forecasts from alternative reconciliation approaches (all Australian QNA variables). Negative values are highlighted in red, the best for each row is marked in bold.}
\label{tab:AUS_all}
\begingroup
	\spacingset{1.1}
\begin{tabular}[t]{c|rrr|rrr|rrr}
\toprule
& \multicolumn{3}{c|}{Point forecasts}  & \multicolumn{6}{c}{Probabilistic forecasts - ES(\%)} \\
& \multicolumn{3}{c|}{MSE (\%)} & \multicolumn{3}{c|}{Bootstrap} & \multicolumn{3}{c}{Gaussian}\\
\hspace{1em}$h$ & \multicolumn{1}{c}{ols} & \multicolumn{1}{c}{wls} & \multicolumn{1}{c}{shr} & \multicolumn{1}{|c}{ols} & \multicolumn{1}{c}{wls} & \multicolumn{1}{c}{shr} & \multicolumn{1}{|c}{ols} & \multicolumn{1}{c}{wls} & \multicolumn{1}{c}{shr}\\
\midrule
\addlinespace[0.3em]
\multicolumn{10}{l}{\textbf{Income}}\\
\hspace{1em}1 & \textcolor{black}{3.16} & \textcolor{black}{6.32} & \textcolor{black}{\textbf{10.55}} & \textcolor{black}{2.30} & \textcolor{black}{4.24} & \textcolor{black}{\textbf{6.15}} & \textcolor{black}{1.87} & \textcolor{black}{3.47} & \textcolor{black}{\textbf{5.15}}\\
\hspace{1em}2 & \textcolor{black}{2.58} & \textcolor{black}{6.07} & \textcolor{black}{\textbf{8.18}} & \textcolor{black}{2.14} & \textcolor{black}{4.08} & \textcolor{black}{\textbf{4.08}} & \textcolor{black}{1.62} & \textcolor{black}{\textbf{3.41}} & \textcolor{black}{3.08}\\
\hspace{1em}3 & \textcolor{black}{2.18} & \textcolor{black}{\textbf{5.81}} & \textcolor{black}{4.09} & \textcolor{black}{1.78} & \textcolor{black}{\textbf{3.29}} & \textcolor{black}{3.19} & \textcolor{black}{1.28} & \textcolor{black}{\textbf{2.67}} & \textcolor{black}{2.35}\\
\hspace{1em}4 & \textcolor{black}{2.18} & \textcolor{black}{\textbf{6.78}} & \textcolor{black}{5.51} & \textcolor{black}{1.86} & \textcolor{black}{3.73} & \textcolor{black}{\textbf{4.42}} & \textcolor{black}{1.42} & \textcolor{black}{3.10} & \textcolor{black}{\textbf{3.67}}\\
\addlinespace[0.3em]
\multicolumn{10}{l}{\textbf{Income - Fully reconciled}}\\
\hspace{1em}1 & \textcolor{black}{3.78} & \textcolor{black}{7.57} & \textcolor{black}{\textbf{8.85}} & \textcolor{black}{2.77} & \textcolor{black}{4.58} & \textcolor{black}{\textbf{4.87}} & \textcolor{black}{1.30} & \textcolor{black}{1.94} & \textcolor{black}{\textbf{2.25}}\\
\hspace{1em}2 & \textcolor{black}{2.91} & \textcolor{black}{6.12} & \textcolor{black}{\textbf{6.92}} & \textcolor{black}{2.59} & \textcolor{black}{\textbf{3.95}} & \textcolor{black}{3.40} & \textcolor{black}{1.20} & \textcolor{black}{\textbf{1.59}} & \textcolor{black}{0.87}\\
\hspace{1em}3 & \textcolor{black}{2.67} & \textcolor{black}{\textbf{6.23}} & \textcolor{black}{5.57} & \textcolor{black}{2.24} & \textcolor{black}{\textbf{3.70}} & \textcolor{black}{3.17} & \textcolor{black}{0.92} & \textcolor{black}{\textbf{1.60}} & \textcolor{black}{0.79}\\
\hspace{1em}4 & \textcolor{black}{2.87} & \textcolor{black}{\textbf{7.21}} & \textcolor{black}{6.07} & \textcolor{black}{2.65} & \textcolor{black}{\textbf{4.40}} & \textcolor{black}{3.95} & \textcolor{black}{1.44} & \textcolor{black}{\textbf{2.34}} & \textcolor{black}{1.59}\\
\addlinespace[0.3em]
\multicolumn{10}{l}{\textbf{Expenditure}}\\
\hspace{1em}1 & \textcolor{black}{6.50} & \textcolor{black}{6.75} & \textcolor{black}{\textbf{8.78}} & \textcolor{black}{3.71} & \textcolor{black}{4.54} & \textcolor{black}{\textbf{4.94}} & \textcolor{black}{2.20} & \textcolor{black}{1.92} & \textcolor{black}{\textbf{2.58}}\\
\hspace{1em}2 & \textcolor{black}{4.90} & \textcolor{black}{5.50} & \textcolor{black}{\textbf{5.52}} & \textcolor{black}{2.88} & \textcolor{black}{\textbf{3.04}} & \textcolor{black}{2.68} & \textcolor{black}{\textbf{1.28}} & \textcolor{black}{0.67} & \textcolor{black}{0.32}\\
\hspace{1em}3 & \textcolor{black}{4.27} & \textcolor{black}{\textbf{6.08}} & \textcolor{black}{5.65} & \textcolor{black}{2.57} & \textcolor{black}{\textbf{2.94}} & \textcolor{black}{2.29} & \textcolor{black}{\textbf{0.88}} & \textcolor{black}{0.71} & \textcolor{black}{0.09}\\
\hspace{1em}4 & \textcolor{black}{4.01} & \textcolor{black}{\textbf{6.69}} & \textcolor{black}{5.20} & \textcolor{black}{2.43} & \textcolor{black}{\textbf{2.65}} & \textcolor{black}{1.63} & \textcolor{black}{\textbf{0.77}} & \textcolor{black}{0.44} & \textcolor{red}{-0.65}\\
\addlinespace[0.3em]
\multicolumn{10}{l}{\textbf{Expenditure - Fully reconciled}}\\
\hspace{1em}1 & \textcolor{black}{6.51} & \textcolor{black}{6.82} & \textcolor{black}{\textbf{9.08}} & \textcolor{black}{3.76} & \textcolor{black}{4.57} & \textcolor{black}{\textbf{4.99}} & \textcolor{black}{2.19} & \textcolor{black}{1.79} & \textcolor{black}{\textbf{2.44}}\\
\hspace{1em}2 & \textcolor{black}{5.09} & \textcolor{black}{6.24} & \textcolor{black}{\textbf{6.54}} & \textcolor{black}{3.03} & \textcolor{black}{\textbf{3.48}} & \textcolor{black}{3.14} & \textcolor{black}{\textbf{1.18}} & \textcolor{black}{0.82} & \textcolor{black}{0.57}\\
\hspace{1em}3 & \textcolor{black}{4.38} & \textcolor{black}{\textbf{6.75}} & \textcolor{black}{5.94} & \textcolor{black}{2.58} & \textcolor{black}{\textbf{3.11}} & \textcolor{black}{2.29} & \textcolor{black}{\textbf{0.81}} & \textcolor{black}{0.69} & \textcolor{red}{-0.12}\\
\hspace{1em}4 & \textcolor{black}{3.98} & \textcolor{black}{\textbf{7.33}} & \textcolor{black}{5.82} & \textcolor{black}{2.36} & \textcolor{black}{\textbf{2.82}} & \textcolor{black}{1.81} & \textcolor{black}{\textbf{0.60}} & \textcolor{black}{0.32} & \textcolor{red}{-0.70}\\
\bottomrule
\end{tabular}
	\endgroup
\end{table}

\begin{figure}[!t]
	\centering
	\begin{subfigure}[b]{\textwidth}
         \centering
         \includegraphics[width = \linewidth]{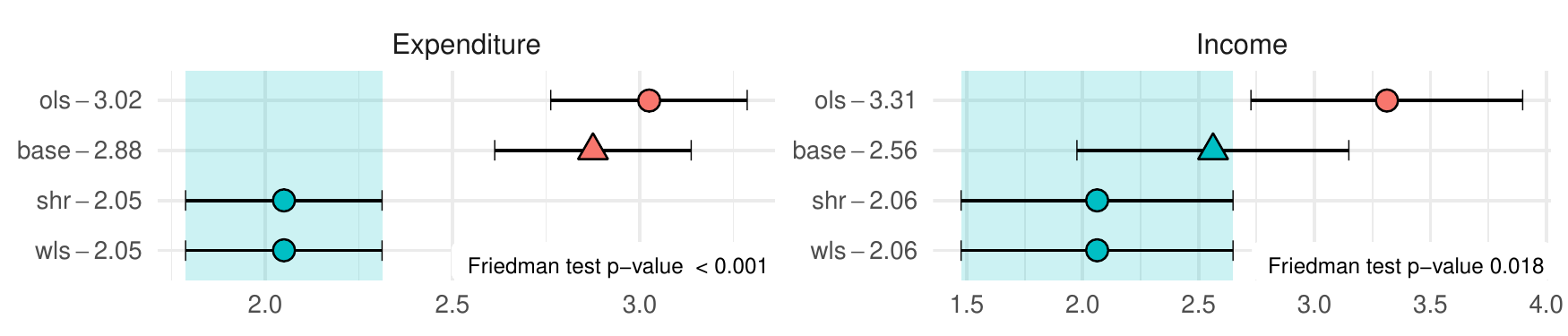}
         \caption{Point forecasts (MSE)}
         \label{img:AUS_es_point_mcb}
         \vspace{-0.25cm}
     \end{subfigure}
	\begin{subfigure}[b]{\textwidth}
         \centering
         \includegraphics[width = \linewidth]{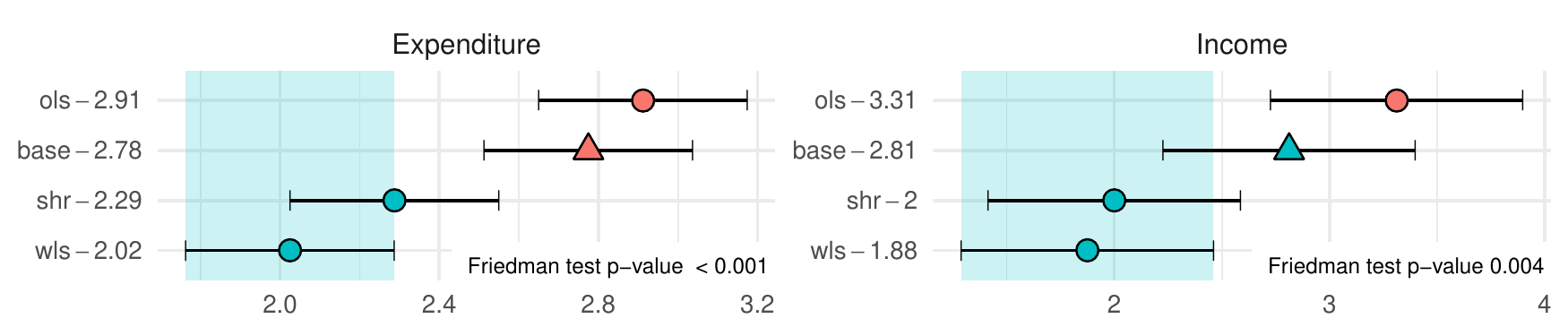}
         \caption{Non parametric joint bootstrap probabilistic forecasts (CRPS)}
         \label{img:AUS_es_boot_mcb}
         \vspace{-0.25cm}
     \end{subfigure}
	\begin{subfigure}[b]{\textwidth}
         \centering
         \includegraphics[width = \linewidth]{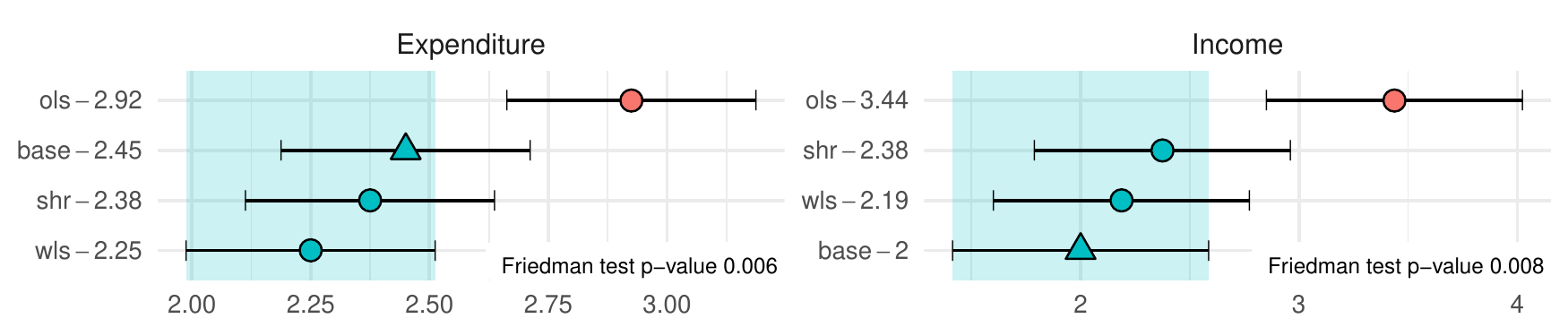}
         \caption{Gaussian probabilistic forecasts (CRPS)}
         \label{img:AUS_es_gauss_mcb}
         \vspace{-0.15cm}
     \end{subfigure}
	\caption{MCB Nemenyi test for the fully reconciled forecasts of the Australian QNA variables at any forecast horizon. In each panel, the Friedman test $p$-value is reported in the lower right corner. The mean rank of each approach is shown to the right of its name.
	Statistical differences in performance are indicated if the intervals of two forecast reconciliation approaches do not overlap. Thus, approaches that do not overlap with the blue interval are considered significantly worse than the best, and vice-versa.}
	\label{img:AUS_es_mcb}
	
\end{figure}

In \autoref{img:AUS_es_mcb} are shown the results obtained by the non-parametric Friedman test and the post hoc “Multiple Comparison with the Best” (MCB) Nemenyi test \citep{koning2005, kourentzes2019, makridakis2022} to determine if the forecasting performances of the different techniques are significantly different from one another. In general, wls always falls in the set of the best performing approaches. For the bootstrap-based probabilistic reconciled forecasts of the expenditure side variables, wls and shr significantly improves in terms of MSE and CRPS compared to the base forecasts. This result is confirmed in the remaining cases as well.

In conclusion, when income and expenditure sides are simultaneously considered for both point and probabilistic forecasts, forecast reconciliation succeeds in improving the base forecasts of $GDP$ and its component aggregates, while preserving the full coherence with the National Accounts constraints.

\subsection{Reconciled probabilistic forecasts of the European Area $GDP$ from output, income and expenditure sides}\label{sec:eaGDP}

In this section, we consider the system of European QNA for the $GDP$ at current prices (in euro), with time series spanning the period 2000:Q1–2019:Q4. This system has many variables linked by several, possibly redundant, accounting constraints, such that it is difficult to manually build a system of non-redundant constraints. 

\begin{figure}[!t]
\centering
  \includegraphics[width = 0.95\linewidth]{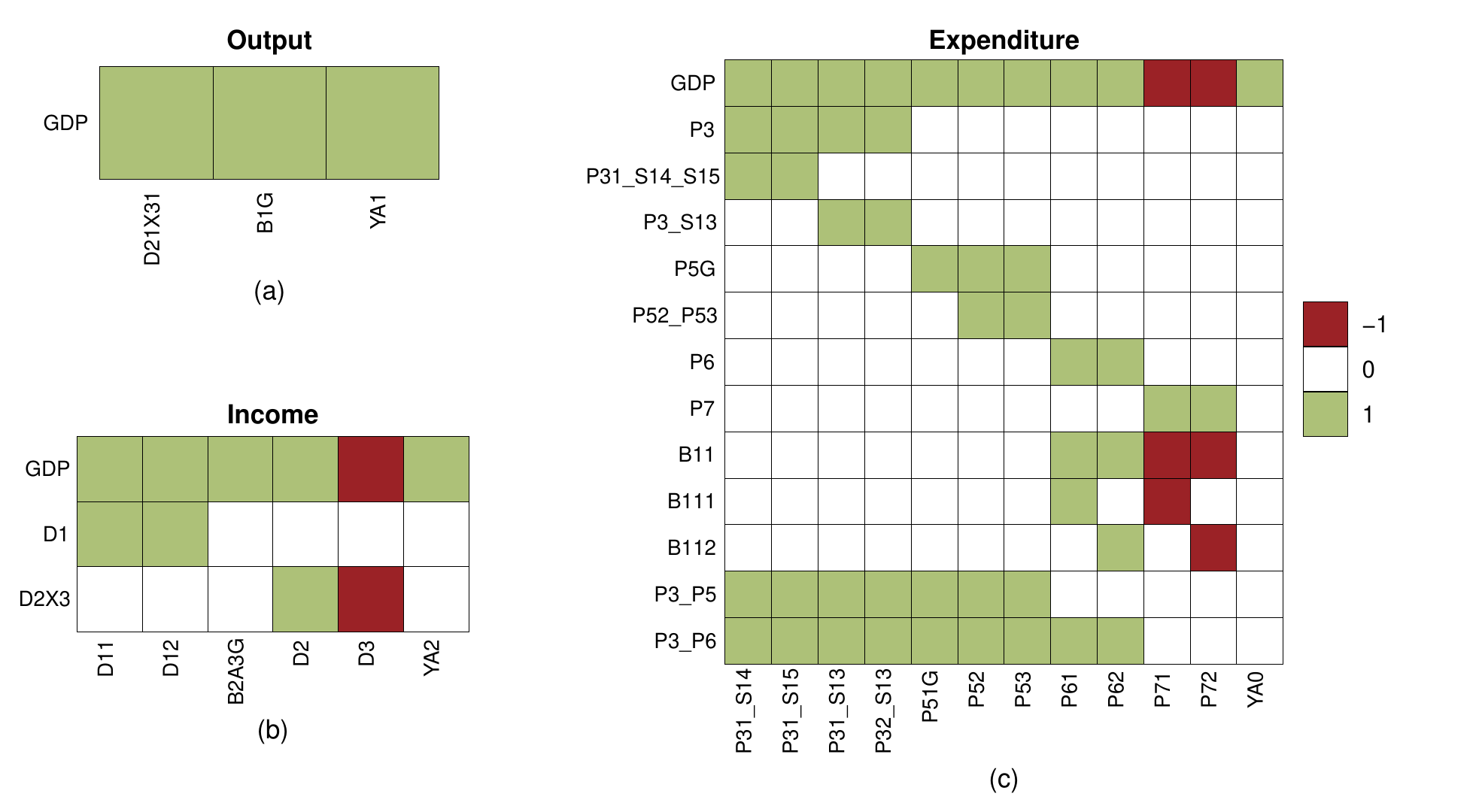}
  \caption{Linear combination matrices $\Avet$ for the European Area $GDP$: output side in panel (a), income side in panel (b), expenditure side in panel (c).}
  \label{fig:CMAT}
	
\end{figure}

The National Accounts are a coherent and consistent set of macroeconomic indicators that are used mostly for economic research and forecasting, policy design, and coordination mechanisms. In this dataset, $GDP$ is a key macroeconomic quantity that is measured using three main approaches, namely output (or production), income and expenditure. These parallel systems internally present a well-defined hierarchical structure of variables with relevant economic significance, such as Final consumption, on the expenditure side, Gross operating surplus and mixed income on the income side, and Total gross value added on the output side. In the EU countries, the data is processed on the basis of the ESA 2010 classification and are released by Eurostat\footnote{Further information can be found at \url{https://ec.europa.eu/eurostat/esa2010/} and \url{https://ec.europa.eu/eurostat/web/national-accounts/data/database}.}. We consider the 19 Euro Area member countries (Austria, Belgium, Finland, France, Germany, Ireland, Italy, Luxembourg, Netherlands, Portugal, Spain, Greece, Slovenia, Cyprus, Malta, Slovakia, Estonia, Latvia, and Lithuania) that have been using the euro since 2015. In \autoref{fig:CMAT} we have represented the aggregation matrices describing output, income, and expenditure constraints, respectively: in panel (a), matrix $\Avet_{\mathrm{O}}$ for the output side, in panel (b) matrix $\Avet_{\mathrm{I}}$ for the income side, and in panel (c) matrix $\Avet_{\mathrm{E}}$ for the expenditure side. The zero-constraints coefficient matrix describing the QNA variables for a single country can thus be written as
$$
\Gammavet_{\mathrm{GDP}} = \begin{bmatrix}
   \Kvet_{\mathrm{E}}  & - \Avet_{\mathrm{E}} & \Zerovet_{(13 \times 6)} & \Zerovet_{(13 \times 3)} \\
	\Kvet_{\mathrm{I}} & \Zerovet_{(3\times 12)} & - \Avet_{\mathrm{I}} & \Zerovet_{(3\times 3)} \\
    \Kvet_{\mathrm{O}} & \Zerovet_{(1\times 12)} & \Zerovet_{(1\times 6)} & - \Avet_{\mathrm{O}}\\
\end{bmatrix} ,
$$
where $\Kvet_{\mathrm{E}}$, $\Kvet_{\mathrm{I}}$ and $\Kvet_{\mathrm{O}}$, respectively, are  the following $(13 \times 15)$, $(3 \times 15)$ and $(1 \times 15)$ matrices:
$$
    \Kvet_{\mathrm{E}}  = \begin{bmatrix}
		1 & \Zerovet_{(1\times 12)}' & \Zerovet_{(1\times 2)}' \\
		\Zerovet_{(12\times 1)} & \Ivet_{12} & \Zerovet_{(12\times 2)} \\
	\end{bmatrix}, \; \Kvet_{\mathrm{I}} = \begin{bmatrix}
		1 & \Zerovet_{(1\times 12)}' & \Zerovet_{(1 \times 2)}' \\
		\Zerovet_{(2\times 1)} & \Zerovet_{(2 \times 12)} & \Ivet_{2} \\
	\end{bmatrix}, \;
\Kvet_{\mathrm{O}} = \begin{bmatrix}
    1 & \Zerovet_{(1\times 14)}'
    \end{bmatrix}.
$$
This disaggregation is common for almost all European countries, the only differences being related to the presence/absence of an aggregate measuring the statistical discrepancy in each accounting side\footnote{For 14 countries (Belgium, France, Germany, Italy, Luxembourg, Netherlands, Spain, Greece, Slovenia, Cyprus, Malta, Slovakia, Latvia, and Lithuania), the expenditure, income and output statistical discrepancies are not present. A statistical discrepancy aggregate is present in the output QNA of Portugal and in the expenditure QNA of Finland, Estonia and Austria. Ireland is the only country where a statistical discrepancy aggregate is present in all accounting sides. \label{footnote:statdiscr}}. The $(361 \times 720)$ matrix describing the accounting relationships for the whole EA19 QNA by countries and accounting sides can be written as follows:
$$
\Gammavet = \begin{bmatrix}
	\Gammavet_{\mathrm{GDP}} & \Zerovet_{(17 \times 684)} \\
	\left[\Zerovet_{(21 \times 15)} \quad \Ivet_{21}\right] & - \mathbf{1}_{19}' \otimes \left[\Zerovet_{(21 \times 15)} \quad \Ivet_{21}\right] \\
	\Zerovet_{(323 \times 36)} & \Ivet_{19} \otimes \Gammavet_{\mathrm{GDP}}\\	
\end{bmatrix},
$$
where $\otimes$ is the Kronecker product, and the top-left portion of $\Gammavet$ refers to the European Area aggregates as a whole. In order to proceed with the calculations, it is necessary to eliminate the columns related to null variables (e.g., the statistical discrepancy aggregate for the countries/sides where it is not contemplated, see footnote \ref{footnote:statdiscr}). Then, to eliminate possible remaining redundant constraints, we apply the QR decomposition (\autoref{sec:genconst}). Finally, we obtain the linear combination matrix $\Avet$, which refers to 311 free and 358 constrained time series, and the full rank zero-constraints matrix $\Cvet = \begin{bmatrix} \Ivet_{358} & -\Avet \end{bmatrix}$.

A rolling forecast experiment with expanding window is performed using ARIMA models to produce the individual series' base forecasts. The first training set is set from 2000:Q1 to 2009:Q4, which gives 40 one-step-ahead, 39 two-step-ahead, 38 three-step-ahead and 37 four-step-ahead ARIMA forecasts, respectively. The used reconciliation approaches are ols, wls and shr, and the forecast accuracy is evaluated through MSE, CRPS and ES indices, as described in \autoref{sec:ausGDP}.

\begin{table}[!t]
\centering
\caption{MSE and ES-skill scores (relative to base forecasts) for the point and probabilistic forecasts from alternative reconciliation approaches (European Area QNA). Negative values are highlighted in red, the best for each row is marked in bold.}
\label{tab:EA_es}
	\begingroup
	\spacingset{1.1}
\begin{tabular}[t]{l|rrr|rrr|rrr}
\toprule
& \multicolumn{3}{c|}{Point forecasts}  & \multicolumn{6}{c}{Probabilistic forecasts - ES(\%)} \\
& \multicolumn{3}{c|}{MSE (\%)} & \multicolumn{3}{c|}{Bootstrap} & \multicolumn{3}{c}{Gaussian}\\
\hspace{1em} & \multicolumn{1}{c}{ols} & \multicolumn{1}{c}{wls} & \multicolumn{1}{c}{shr} & \multicolumn{1}{|c}{ols} & \multicolumn{1}{c}{wls} & \multicolumn{1}{c}{shr} & \multicolumn{1}{|c}{ols} & \multicolumn{1}{c}{wls} & \multicolumn{1}{c}{shr}\\
\midrule
\addlinespace[0.3em]
\hspace{1em}GDP & \textcolor{black}{14.0} & \textcolor{black}{\textbf{41.1}} & \textcolor{black}{21.6} & \textcolor{black}{9.7} & \textcolor{black}{\textbf{25.8}} & \textcolor{black}{15.7} & \textcolor{black}{3.4} & \textcolor{black}{\textbf{20.8}} & \textcolor{black}{3.7}\\
\addlinespace[0.3em]
\multicolumn{10}{l}{\textbf{Sides}}\\
\hspace{1em}Expenditure & \textcolor{black}{15.8} & \textcolor{black}{\textbf{31.0}} & \textcolor{black}{28.3} & \textcolor{black}{9.1} & \textcolor{black}{\textbf{20.2}} & \textcolor{black}{16.3} & \textcolor{black}{6.8} & \textcolor{black}{\textbf{15.7}} & \textcolor{black}{8.7}\\
\hspace{1em}Income & \textcolor{red}{-2.2} & \textcolor{black}{\textbf{27.3}} & \textcolor{black}{9.2} & \textcolor{red}{-0.3} & \textcolor{black}{\textbf{15.3}} & \textcolor{black}{5.6} & \textcolor{red}{-2.7} & \textcolor{black}{\textbf{11.6}} & \textcolor{red}{-3.2}\\
\hspace{1em}Output & \textcolor{black}{0.3} & \textcolor{black}{\textbf{35.0}} & \textcolor{black}{15.0} & \textcolor{black}{1.2} & \textcolor{black}{\textbf{21.6}} & \textcolor{black}{12.9} & \textcolor{red}{-1.5} & \textcolor{black}{\textbf{17.1}} & \textcolor{black}{1.3}\\
\addlinespace[0.3em]
\multicolumn{10}{l}{\textbf{Countries}}\\
\hspace{1em}EA19 & \textcolor{black}{17.6} & \textcolor{black}{\textbf{37.1}} & \textcolor{black}{31.6} & \textcolor{black}{10.1} & \textcolor{black}{\textbf{23.5}} & \textcolor{black}{18.1} & \textcolor{black}{7.8} & \textcolor{black}{\textbf{19.3}} & \textcolor{black}{10.0}\\
\hspace{1em}Austria & \textcolor{red}{$<$-30} & \textcolor{black}{18.5} & \textcolor{black}{\textbf{21.5}} & \textcolor{red}{-26.8} & \textcolor{black}{\textbf{12.1}} & \textcolor{black}{10.0} & \textcolor{red}{-21.6} & \textcolor{black}{\textbf{9.1}} & \textcolor{black}{5.4}\\
\hspace{1em}Belgium & \textcolor{red}{-5.9} & \textcolor{black}{13.5} & \textcolor{black}{\textbf{21.0}} & \textcolor{red}{-2.7} & \textcolor{black}{10.7} & \textcolor{black}{\textbf{10.8}} & \textcolor{red}{-3.1} & \textcolor{black}{\textbf{8.3}} & \textcolor{black}{5.9}\\
\hspace{1em}Cyprus & \textcolor{red}{$<$-30} & \textcolor{black}{\textbf{11.7}} & \textcolor{black}{8.6} & \textcolor{red}{$<$-30} & \textcolor{black}{\textbf{7.1}} & \textcolor{black}{1.7} & \textcolor{red}{$<$-30} & \textcolor{black}{\textbf{3.6}} & \textcolor{red}{-2.8}\\
\hspace{1em}Estonia & \textcolor{red}{$<$-30} & \textcolor{black}{20.9} & \textcolor{black}{\textbf{26.8}} & \textcolor{red}{$<$-30} & \textcolor{black}{\textbf{12.7}} & \textcolor{black}{11.2} & \textcolor{red}{$<$-30} & \textcolor{black}{\textbf{9.7}} & \textcolor{black}{6.7}\\
\hspace{1em}Finland & \textcolor{red}{$<$-30} & \textcolor{black}{\textbf{24.8}} & \textcolor{black}{14.7} & \textcolor{red}{$<$-30} & \textcolor{black}{\textbf{15.3}} & \textcolor{black}{9.5} & \textcolor{red}{$<$-30} & \textcolor{black}{\textbf{13.2}} & \textcolor{black}{3.4}\\
\hspace{1em}France & \textcolor{black}{4.4} & \textcolor{black}{\textbf{19.1}} & \textcolor{red}{-0.1} & \textcolor{black}{3.2} & \textcolor{black}{\textbf{11.7}} & \textcolor{black}{3.8} & \textcolor{black}{0.3} & \textcolor{black}{\textbf{7.9}} & \textcolor{red}{-6.8}\\
\hspace{1em}Germany & \textcolor{black}{15.0} & \textcolor{black}{20.6} & \textcolor{black}{\textbf{25.0}} & \textcolor{black}{13.5} & \textcolor{black}{\textbf{21.6}} & \textcolor{black}{18.6} & \textcolor{black}{7.5} & \textcolor{black}{\textbf{11.1}} & \textcolor{black}{6.8}\\
\hspace{1em}Greece & \textcolor{red}{$<$-30} & \textcolor{black}{\textbf{1.5}} & \textcolor{red}{$<$-30} & \textcolor{red}{-12.9} & \textcolor{black}{\textbf{3.9}} & \textcolor{red}{-7.1} & \textcolor{red}{-15.1} & \textcolor{black}{\textbf{0.8}} & \textcolor{red}{-14.1}\\
\hspace{1em}Ireland & \textcolor{black}{4.1} & \textcolor{black}{\textbf{6.6}} & \textcolor{black}{3.7} & \textcolor{black}{2.1} & \textcolor{black}{\textbf{5.9}} & \textcolor{black}{3.0} & \textcolor{black}{0.5} & \textcolor{black}{\textbf{2.9}} & \textcolor{red}{-0.7}\\
\hspace{1em}Italy & \textcolor{black}{9.8} & \textcolor{black}{12.6} & \textcolor{black}{\textbf{24.0}} & \textcolor{black}{8.4} & \textcolor{black}{10.8} & \textcolor{black}{\textbf{17.1}} & \textcolor{black}{4.4} & \textcolor{black}{4.9} & \textcolor{black}{\textbf{7.8}}\\
\hspace{1em}Latvia & \textcolor{red}{$<$-30} & \textcolor{black}{\textbf{26.2}} & \textcolor{black}{18.3} & \textcolor{red}{$<$-30} & \textcolor{black}{\textbf{13.0}} & \textcolor{black}{8.7} & \textcolor{red}{$<$-30} & \textcolor{black}{\textbf{13.7}} & \textcolor{black}{5.6}\\
\hspace{1em}Lithuania & \textcolor{red}{$<$-30} & \textcolor{black}{27.3} & \textcolor{black}{\textbf{28.5}} & \textcolor{red}{$<$-30} & \textcolor{black}{\textbf{15.0}} & \textcolor{black}{12.8} & \textcolor{red}{$<$-30} & \textcolor{black}{\textbf{12.1}} & \textcolor{black}{8.4}\\
\hspace{1em}Luxembourg & \textcolor{red}{$<$-30} & \textcolor{black}{\textbf{6.6}} & \textcolor{red}{-10.3} & \textcolor{red}{$<$-30} & \textcolor{black}{\textbf{6.1}} & \textcolor{red}{-3.0} & \textcolor{red}{$<$-30} & \textcolor{black}{\textbf{2.0}} & \textcolor{red}{-10.0}\\
\hspace{1em}Malta & \textcolor{red}{$<$-30} & \textcolor{black}{\textbf{5.4}} & \textcolor{red}{-9.9} & \textcolor{red}{$<$-30} & \textcolor{black}{\textbf{4.9}} & \textcolor{red}{-7.4} & \textcolor{red}{$<$-30} & \textcolor{black}{\textbf{0.8}} & \textcolor{red}{-12.0}\\
\hspace{1em}Netherlands & \textcolor{black}{7.4} & \textcolor{black}{\textbf{16.9}} & \textcolor{black}{11.9} & \textcolor{black}{6.1} & \textcolor{black}{\textbf{14.1}} & \textcolor{black}{7.3} & \textcolor{black}{3.5} & \textcolor{black}{\textbf{11.7}} & \textcolor{black}{4.4}\\
\hspace{1em}Portugal & \textcolor{red}{$<$-30} & \textcolor{black}{\textbf{11.4}} & \textcolor{red}{-8.1} & \textcolor{red}{$<$-30} & \textcolor{black}{\textbf{7.4}} & \textcolor{red}{-3.4} & \textcolor{red}{$<$-30} & \textcolor{black}{\textbf{4.1}} & \textcolor{red}{-10.1}\\
\hspace{1em}Slovakia & \textcolor{red}{$<$-30} & \textcolor{black}{\textbf{21.5}} & \textcolor{black}{16.4} & \textcolor{red}{$<$-30} & \textcolor{black}{\textbf{10.0}} & \textcolor{black}{6.9} & \textcolor{red}{$<$-30} & \textcolor{black}{\textbf{9.2}} & \textcolor{black}{2.0}\\
\hspace{1em}Slovenia & \textcolor{red}{$<$-30} & \textcolor{black}{\textbf{24.1}} & \textcolor{black}{15.5} & \textcolor{red}{$<$-30} & \textcolor{black}{\textbf{13.5}} & \textcolor{black}{8.3} & \textcolor{red}{$<$-30} & \textcolor{black}{\textbf{11.1}} & \textcolor{black}{1.8}\\
\hspace{1em}Spain & \textcolor{black}{6.5} & \textcolor{black}{\textbf{17.7}} & \textcolor{black}{0.9} & \textcolor{black}{5.3} & \textcolor{black}{\textbf{9.2}} & \textcolor{black}{1.4} & \textcolor{black}{1.7} & \textcolor{black}{\textbf{6.4}} & \textcolor{red}{-7.3}\\
\bottomrule
\end{tabular}
	\endgroup
\end{table}

\subsubsection*{Results}
\autoref{tab:EA_es} shows the MSE indices for point forecasts, and the ES indices for probabilistic nonparametric (bootstrap) and parametric (Gaussian) forecasts, respectively. 
The rows of the table are divided into three parts: the first row shows the results for $GDP$, the second to fourth rows the National Accounts' divisions (income, expenditure, or output sides), while the remaining rows correspond to the 19 countries and EA19.
All forecast horizons are considered\footnote{A disaggregated analysis by forecast horizon is reported in the online appendix.}.

\begin{figure}[!t]
	\centering
	\begin{subfigure}[b]{\textwidth}
         \centering
         \includegraphics[width = \linewidth]{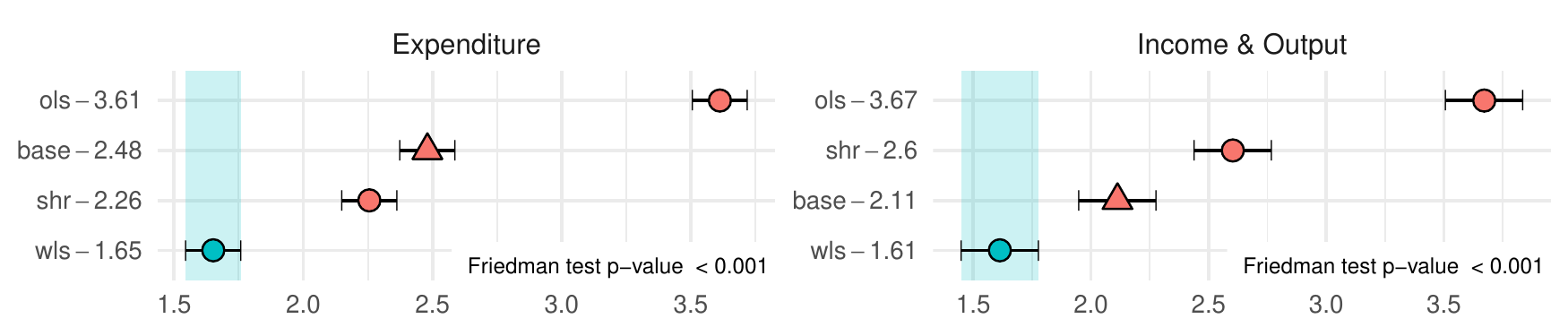}
         \caption{Point forecasts}
         \label{img:EA_es_point_mcb}
         \vspace{-0.15cm}
     \end{subfigure}
	\begin{subfigure}[b]{\textwidth}
         \centering
         \includegraphics[width = \linewidth]{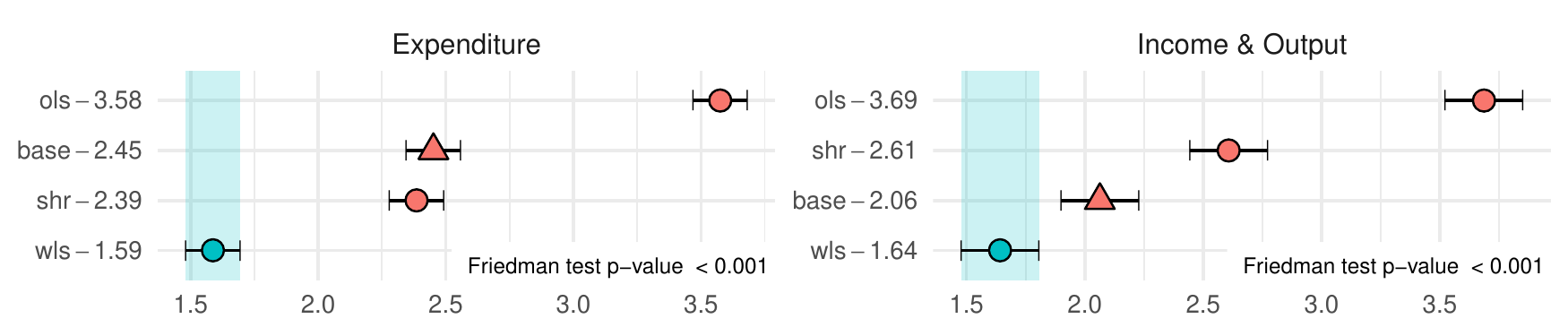}
         \caption{Non parametric joint bootstrap probabilistic forecasts}
         \label{img:EA_es_boot_mcb}
         \vspace{-0.15cm}
     \end{subfigure}
	\begin{subfigure}[b]{\textwidth}
         \centering
         \includegraphics[width = \linewidth]{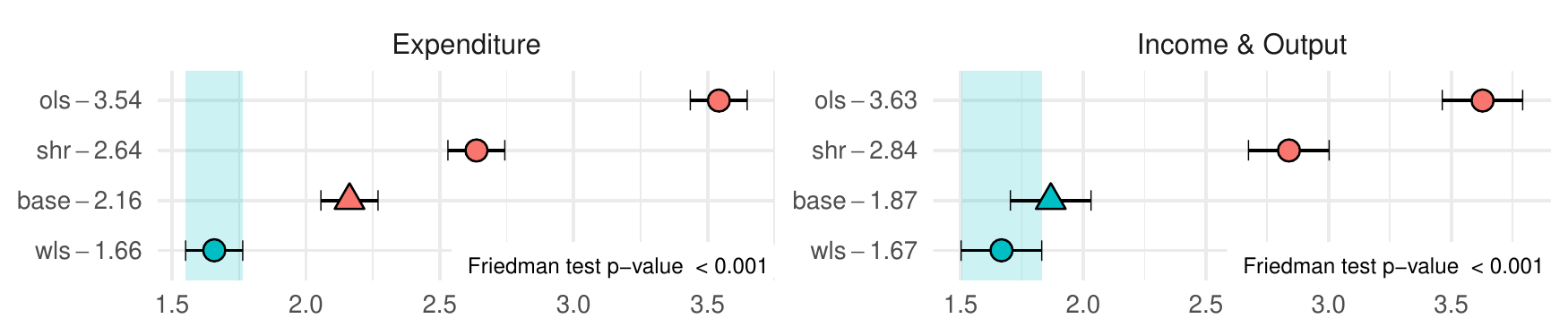}
         \caption{Gaussian probabilistic forecasts}
         \label{img:EA_es_gauss_mcb}
         \vspace{-0.15cm}
     \end{subfigure}
	\caption{MCB Nemenyi test for the fully reconciled forecasts of the European Area QNA variables at any forecast horizon. In each panel, the Friedman test p-value is reported in the lower right corner. The mean rank of each approach is shown to the right of its name.
		Statistical differences in performance are indicated if the intervals of two forecast reconciliation approaches do not overlap. Thus, approaches that do not overlap with the blue interval are considered significantly worse than the best, and vice-versa.}
	\label{img:EA_es_mcb}
	
\end{figure}

\begin{figure}[!t]
	\centering
	\includegraphics[width = \linewidth]{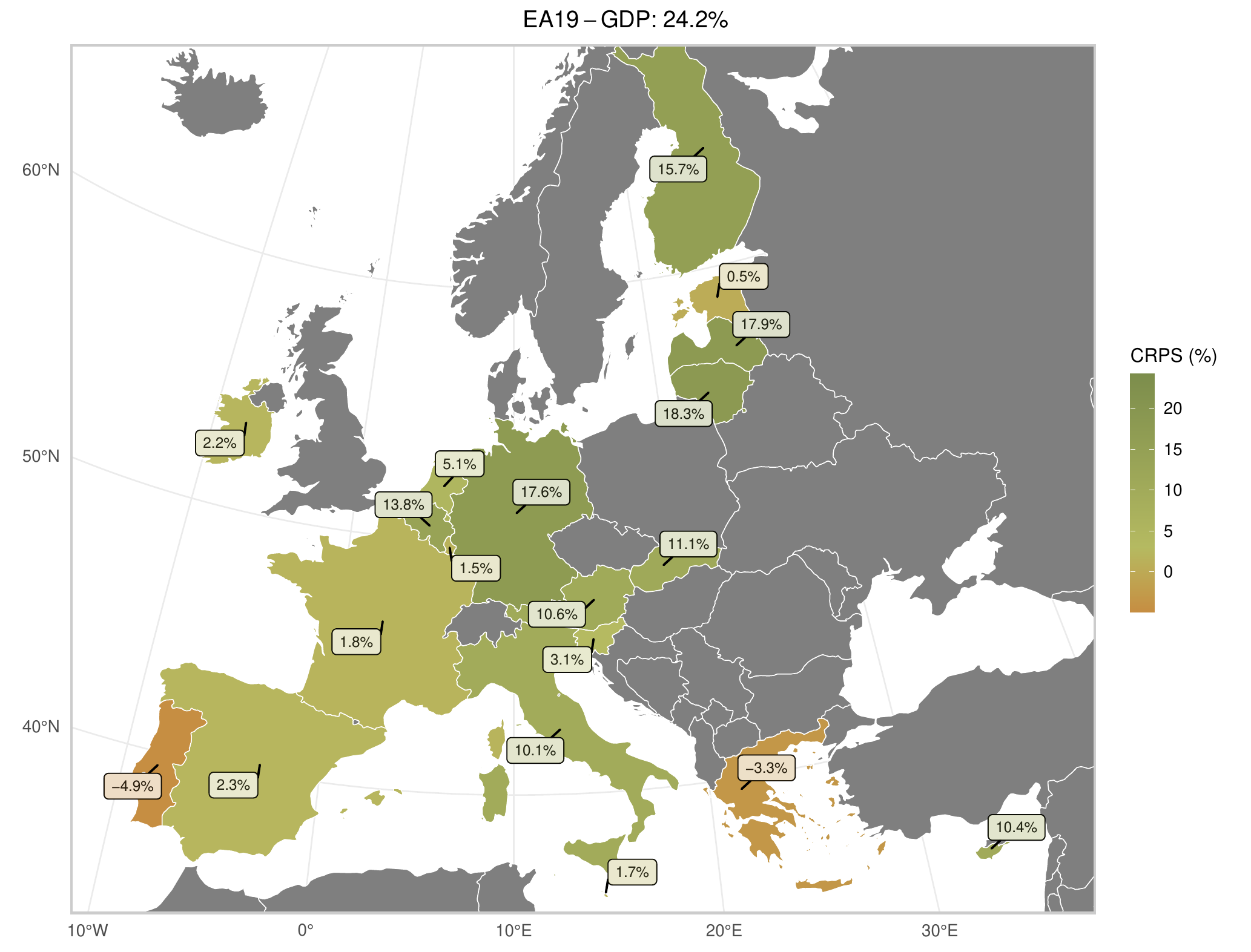}
	\caption{CRPS-skill scores of the one-step-ahead $GDP$ non-parametric joint bootstrap probabilistic reconciled (wls) forecasts for the 19 Euro Area countries.}
	\label{img:EA_crps}
	
\end{figure}

When only $GDP$ is considered, any reconciliation approach consistently outperforms the base forecasts, both in the point and probabilistic cases. The wls approach confirms a good performance when we look at the income, expenditure, and output sides, while ols shows the worst performance. When the parametric framework is considered, shr is worse than the base forecast for the income side.  At country level, ols is the approach that overall shows the worst performance, with many relative losses in the accuracy indices higher than 30\%. It is worth noting that all reconciliation approaches always perform well for the whole Euro Area (EA19). Overall, in this forecasting experiment wls appears to be the most performing reconciliation approach, showing no negative skill score, and improvements higher than 20\% and 10\% for nonparametric and parametric probabilistic frameworks, respectively. 

\autoref{img:EA_es_mcb} shows the MCB Nemenyi test at any forecast horizon, distinct by expenditure side and income and output sides, respectively. The results just seen are further confirmed by this alternative forecast assessment tool: it clearly appears that the wls approach almost always significantly improves compared to the base forecast, in terms of both point and probabilistic forecasts.

Finally, a visual evaluation of the accuracy improvement obtained through wls forecast reconciliation, although limited to a single forecast horizon, is offered by \autoref{img:EA_crps}, showing the European map with the CRPS skill scores for the one-step-ahead $GDP$ non-parametric probabilistic reconciled forecasts. It is worth noting that only for two countries (Greece and Portugal) a decrease is registered ($-3.3$\% and $-4.9$\%, respectively). In all other cases, improvements in the forecasting accuracy are obtained, with Germany and Lithuania leading the way with about 18\%. Furthermore, the improvement in the forecasting accuracy for the $EA19-GDP$ is 24.2\%, the highest throughout the whole Euro Area.

\section{Conclusions}\label{sec:conclusions}
Producing and using coherent information, disaggregated by different characteristics useful for different decision levels, is an important task for any practitioner and quantitative-based decision process. At this end, in this article we aimed to generalize the results valid for the forecast reconciliation of a genuine hierarchical/grouped time series to the case of a general linearly constrained time series, where the distinction between upper and bottom variables, which is typical in the hierarchical setting, is no longer meaningful.

Two motivating examples have been considered, both coming from the National Accounts field, namely the forecast reconciliation of quarterly $GDP$ of (i) Australia, disaggregated by income and expenditure variables, and (ii) Euro Area 19, disaggregated by the income, output and expenditure side variables of 19 component countries. In both cases, the structure of the time series involved cannot be represented according to a genuinely hierarchical/grouped scheme, so the standard forecast reconciliation techniques fail in producing a ``unique'' $GDP$ forecast, either point or probabilistic, making it necessary to solve this annoying issue. 

We have shown that using well known linear algebra tools, it is always possible to establish a formal connection between the unconstrained GLS structural approach originally developed by \cite{hyndman2011}, and the projection approach to reconciliation dating back to the work by \cite{stone1942}, and then applied to solve different reconciliation problems (\citealp{difonzomarini2011}, \citealp{vanerven2015}, \citealp{wickramasuriya2019}, \citealp{difonzo2021a}). We propose a new classification of the variables forming the multiple time series as free and constrained, respectively, that can be seen as a generalization of the standard bottom/upper variables classification used in the hierarchical setting. Furthermore, we show techniques for deriving a linear combination matrix describing the relationships between these variables, starting from the coefficient matrix summarizing the (possible redundant) constraints linking the series.

The application of these findings to both point and probabilistic reconciliation techniques proved to be easy to implement and powerful, resulting in significant improvements in the forecasting accuracy of $GDP$ and its components in both forecasting experiments.

\phantomsection\addcontentsline{toc}{section}{Acknowledgments}
\section*{Acknowledgments}
The authors acknowledge financial support from project PRIN2017 “HiDEA: Advanced Econometrics for High-frequency Data”, 2017RSMPZZ.

\begin{appendices}

\section{Derivation of equation (\ref{eq:covh})}
\label{app:covh}

Denote $\Wvet_h = k_h\Wvet$, where $k_h$ is a proportionality constant and $\Wvet$ is the p.d. covariance matrix used in the point forecast reconciliation formula (\ref{eq:Seq}). Then:
$$
\begin{aligned}
 \widetilde{\Wvet}_h & = \Svet\Gvet\Wvet_h\Gvet'\Svet' = k_h\Svet\Gvet \Wvet \Gvet'\Svet'\\
 & = k_h\Svet\left(\Svet'\Wvet^{-1}\Svet\right)^{-1}\underbrace{\Svet'\Wvet^{-1}\Wvet \Wvet^{-1}\Svet\left(\Svet'\Wvet^{-1}\Svet\right)^{-1}}_{\Ivet_n}\Svet'\\
 & = k_h\Svet\left(\Svet'\Wvet^{-1}\Svet\right)^{-1}\Svet' = k_h\Svet\underbrace{\left(\Svet'\Wvet^{-1}\Svet\right)^{-1}\Svet'\Wvet^{-1}}_{\Gvet}\Wvet = k_h\Svet\Gvet\Wvet .
\end{aligned}
$$
\end{appendices}

\phantomsection\addcontentsline{toc}{section}{References}

\bibliographystyle{agsm}
\bibliography{biblio.bib}

\end{document}


\def\spacingset#1{\renewcommand{\baselinestretch}{#1}\small\normalsize}
\spacingset{1.1}
  
\maketitleblind

\tableofcontents

\appendix
\renewcommand{\thetable}{\Alph{section}.\arabic{table}}
\renewcommand{\thefigure}{\Alph{section}.\arabic{figure}}

\clearpage
\section[Examples of derivation of the linear combination matrix A using the Reduced Row Echelon Form (rref)]{Examples of derivation of the linear combination matrix $\Avet$ using the Reduced Row Echelon Form (rref)}

\subsection{Linearly constrained multiple time series, $n = 5$ and $p = 3$}
First, consider a system of $n=5$ linearly constrained time series such that for $t = 1,\dots, T$
$$
\left\{\begin{array}{ccccccccccc}
	2x_{1,t} &-& 4x_{2,t} &-& 8x_{3,t} &+& 6x_{4,t} &+& 3x_{5,t} &=&0\\ 
   & & x_{2,t} &+& 3x_{3,t} &+& 2x_{4,t} &+& 3x_{5,t}  &=&0\\ 
  3x_{1,t} &-& 2x_{2,t} &&  &&  &+& 8x_{5,t}  &=&0\\ 
\end{array}\right.
$$
or, through a matrix representation,
$$
\Gammavet_{(3\times5)}\xvet_t = \Zerovet_{(3\times1)}, \qquad t = 1, \dots, T,
$$
where $\xvet_t = \left[x_{1,t} \; x_{2,t} \; x_{3,t} \; x_{4,t} \; x_{5,t}\right]'$ and 
$$
\Gammavet_{(3\times5)} = \begin{tikzpicture}[baseline]
\matrix[
    matrix of math nodes, 
    row sep=3pt,
    column sep=8pt,
    left delimiter={[},
    right delimiter={]},
    inner xsep=3pt
    ] (m) {
	2 & -4 & -8 & 6 & 3 \\ 
  0 & \phantom{+}1 & \phantom{+}3 & 2 & 3 \\ 
  3 & -2 & \phantom{+}0 & 0 & 8 \\ 
    };
    \node[fit=(m-1-1)(m-3-1), inner sep = 0, outer sep = 0, fill = red, opacity = 0.25] {};
    \node[fit=(m-1-2)(m-3-2), inner sep = 0, outer sep = 0, fill = red, opacity = 0.25] {};
    \node[fit=(m-1-4)(m-3-4), inner sep = 0, outer sep = 0, fill = red, opacity = 0.25] {};
    \node[fit=(m-1-3)(m-3-3), inner sep = 0, outer sep = 0, fill = blue, opacity = 0.25] {};
    \node[fit=(m-1-5)(m-3-5), inner sep = 0, outer sep = 0, fill = blue, opacity = 0.25] {};
\end{tikzpicture}.
$$
Reducing the coefficient matrix $\Gammavet_{(3\times5)}$ to $\Zvet_{(3\times5)}$ (rref) yields
$$
\Gammavet_{(3\times5)}\xrightarrow{\quad\mbox{rref}\quad} \Zvet_{(3\times5)} = 	\begin{tikzpicture}[baseline]
\matrix[
    matrix of math nodes, 
    row sep=3pt,
    column sep=8pt,
    left delimiter={[},
    right delimiter={]},
    inner xsep=3pt
    ] (m) {
1 & 0 & 2 & 0 & 4\phantom{.0} \\ 
0 & 1 & 3 & 0 & 2\phantom{.0} \\ 
0 & 0 & 0 & 1 & 0.5 \\ 
    };
    \node[fit=(m-1-1)(m-3-1), inner sep = 0, outer sep = 0, fill = red, opacity = 0.25] {};
    \node[fit=(m-1-2)(m-3-2), inner sep = 0, outer sep = 0, fill = red, opacity = 0.25] {};
    \node[fit=(m-1-4)(m-3-4), inner sep = 0, outer sep = 0, fill = red, opacity = 0.25] {};
    \node[fit=(m-1-3)(m-3-3), inner sep = 0, outer sep = 0, fill = blue, opacity = 0.25] {};
    \node[fit=(m-1-5)(m-3-5), inner sep = 0, outer sep = 0, fill = blue, opacity = 0.25] {};
\end{tikzpicture}.
$$
We observe that there are $n_c = 3$ constrained variables ({\color{red}red} background) in position $\{1,2,4\}$ and $n_u = 2$ free variables ({\color{blue}blue} background) in position $\{3,5\}$. Therefore, we can build the permutation matrix
$$
\Pvet_{(5\times 5)} = \left[\begin{array}{ccccc}
1 & 0 & 0 & 0 & 0 \\ 
0 & 1 & 0 & 0 & 0 \\ 
0 & 0 & 0 & 1 & 0 \\ 
0 & 0 & 1 & 0 & 0 \\ 
0 & 0 & 0 & 0 & 1 \\ 
\end{array}\right] 
$$
and compute
$$
\Cvet_{(3 \times 5)} = \Zvet\Pvet' =
\begin{tikzpicture}[baseline]
\matrix[
    matrix of math nodes, 
    row sep=3pt,
    column sep=8pt,
    left delimiter={[},
    right delimiter={]},
    inner xsep=3pt
    ] (m) {
1 & 0 & 0 & 2 & 4\phantom{.0} \\ 
0 & 1 & 0 & 3 & 2\phantom{.0} \\ 
0 & 0 & 1 & 0 & 0.5 \\ 
    };
\draw[dashed] ($0.5*(m-1-3.north east)+0.5*(m-1-4.north west)$) -- ($0.5*(m-3-3.south east)+0.5*(m-3-4.south west)$);
    \node[fit=(m-1-1)(m-3-1), inner sep = 0, outer sep = 0, fill = red, opacity = 0.25] {};
    \node[fit=(m-1-2)(m-3-2), inner sep = 0, outer sep = 0, fill = red, opacity = 0.25] {};
    \node[fit=(m-1-4)(m-3-4), inner sep = 0, outer sep = 0, fill = blue, opacity = 0.25] {};
    \node[fit=(m-1-3)(m-3-3), inner sep = 0, outer sep = 0, fill = red, opacity = 0.25] {};
    \node[fit=(m-1-5)(m-3-5), inner sep = 0, outer sep = 0, fill = blue, opacity = 0.25] {};
	
\node[
  fit=(m-3-4.south west) (m-3-5.south east),
  yshift = 0.1cm,
  inner xsep=0pt,inner ysep=0cm,
  below delimiter=\},
  label={[label distance=0.35cm]below:$-\Avet_{(2\times 3)}$}
] {};

\node[
  fit=(m-3-1.south west) (m-3-3.south east),
  yshift = 0.1cm,
  inner xsep=0pt,inner ysep=0cm,
  below delimiter=\},
  label={[label distance=0.35cm]below:$\Ivet_{3}$}
] {};
\end{tikzpicture}.
$$
The linear combination matrix $\Avet$ and structural-like matrix $\Svet$ are given by respectively,  
$$
\Avet_{(3\times2)} =  \left[\begin{array}{cc}
-2 & -4\phantom{.0} \\ 
-3 & -2\phantom{.0} \\ 
\phantom{+}0 & -0.5 \\ 
\end{array}\right] \quad \mbox{and}
\quad \Svet_{(5\times2)} = \begin{tikzpicture}[baseline]
\matrix[
    matrix of math nodes, 
    row sep=2pt,
    column sep=8pt,
    left delimiter={[},
    right delimiter={]},
    inner xsep=0pt
    ] (m) {
-2 & -4\phantom{.0} \\ 
-3 & -2\phantom{.0} \\ 
\phantom{+}0 & -0.5 \\ 
\phantom{+}1 & \phantom{+}0\phantom{.0}\\
\phantom{+}0 & \phantom{+}1\phantom{.0}\\
    };
\draw[dashed] ($0.5*(m-3-1.south west)+0.5*(m-4-1.north west)$) -- ($0.5*(m-3-2.south east)+0.5*(m-4-2.north east)$);

\node[
  fit=(m-1-2)(m-3-2),
  inner xsep=10pt,inner ysep=0,
  right delimiter=\},
  label={[label distance=0.25cm]right:$\;\Avet_{(3\times 2)}$}
] {};

\node[
  fit=(m-4-2)(m-5-2),
  inner xsep=10pt,inner ysep=0,
  right delimiter=\},
  label={[label distance=0.25cm]right:$\;\Ivet_{2}$}
] {};
\end{tikzpicture}.
$$
These matrices are such that
$$
\cvet_t = \Avet_{(3\times2)}\uvet_t, \qquad \yvet_t = \Svet_{(5\times2)} \uvet_t, \qquad \mbox{for } t = 1,\dots, T,
$$
where 
$$
\cvet_t = \left[\begin{array}{c}
	x_{1,t}\\
	x_{2,t}\\
	x_{4,t}\\
\end{array}\right], 
\qquad \uvet_t = \left[\begin{array}{c}
	x_{3,t}\\
	x_{5,t}\\
\end{array}\right], \quad \mbox{and}
\quad \yvet_t = \left[\begin{array}{c}
	\cvet_t\\
	\uvet_t\\
\end{array}\right].
$$

\subsection{Linearly constrained multiple time series, $n = 4$ and $p = 3$}
First, consider a system of $n=4$ linearly constrained time series such that for $t = 1,\dots, T$
$$
\left\{\begin{array}{ccccccccccc}
	x_{1,t} &-& 2x_{2,t} &-& x_{3,t} &+& 3x_{4,t}&=&0\\ 
	2x_{1,t} &-& 4x_{2,t} &-& 3x_{3,t} &+& 2x_{4,t}&=&0\\ 
	4x_{1,t} &-& 8x_{2,t} &-& 6x_{3,t} &+& 4x_{4,t}&=&0\\ 
\end{array}\right.
$$
and
$$
\Gammavet_{(3\times4)}\xvet_t = \Zerovet_{(3\times1)}, \qquad t = 1, \dots, T,
$$
where $\xvet_t = \left[x_{1,t} \; x_{2,t} \; x_{3,t} \; x_{4,t}\right]'$ and 
$$
\Gammavet = \begin{tikzpicture}[baseline]
\matrix[
    matrix of math nodes, 
    row sep=3pt,
    column sep=8pt,
    left delimiter={[},
    right delimiter={]},
    inner xsep=3pt
    ] (m) {
1 & -2 & -1 & 3 \\ 
2 & -4 & -3 & 2 \\ 
4 & -8 & -6 & 4 \\ 
    };
    \node[fit=(m-1-1)(m-3-1), inner sep = 0, outer sep = 0, fill = red, opacity = 0.25] {};
    \node[fit=(m-1-2)(m-3-2), inner sep = 0, outer sep = 0, fill = blue, opacity = 0.25] {};
    \node[fit=(m-1-4)(m-3-4), inner sep = 0, outer sep = 0, fill = blue, opacity = 0.25] {};
    \node[fit=(m-1-3)(m-3-3), inner sep = 0, outer sep = 0, fill = red, opacity = 0.25] {};
\end{tikzpicture} \xrightarrow{\quad\mbox{rref}\quad}  \begin{tikzpicture}[baseline]
\matrix[
    matrix of math nodes, 
    row sep=3pt,
    column sep=8pt,
    left delimiter={[},
    right delimiter={]},
    inner xsep=3pt
    ] (m) {
1 & -2 & 0 & 7 \\ 
0 & \phantom{+}0 & 1 & 4 \\ 
0 & \phantom{+}0 & 0 & 0 \\ 
    };
    \node[fit=(m-1-1)(m-3-1), inner sep = 0, outer sep = 0, fill = red, opacity = 0.25] {};
    \node[fit=(m-1-2)(m-3-2), inner sep = 0, outer sep = 0, fill = blue, opacity = 0.25] {};
    \node[fit=(m-1-4)(m-3-4), inner sep = 0, outer sep = 0, fill = blue, opacity = 0.25] {};
    \node[fit=(m-1-3)(m-3-3), inner sep = 0, outer sep = 0, fill = red, opacity = 0.25] {};
    \node[fit=(m-3-1)(m-3-4), dashed, draw, inner sep = 0.5mm, outer sep = 0, line width=0.6mm] (zero){};
\end{tikzpicture}.
$$
In this case, it clearly appears that there is a redundant relationship (e.g., the third equation is equal to the second one multiplied by 2). This is confirmed by the fact that, reducing the coefficient matrix $\Gammavet$, we obtain $n_c = 2$ constrained variables ({\color{red}red} background) in position $\{1,4\}$, $n_u = 2$ free variables ({\color{blue}blue} background) in position $\{2,3\}$, and the last row is null. Then, 
$$
\Pvet_{(4\times 4)} = \left[\begin{array}{cccc}
1 & 0 & 0 & 0 \\ 
0 & 0 & 1 & 0 \\ 
0 & 1 & 0 & 0 \\ 
0 & 0 & 0 & 1 \\ 
\end{array}\right], \qquad \Zvet_{(2 \times 4)} = \begin{tikzpicture}[baseline]
\matrix[
    matrix of math nodes, 
    row sep=3pt,
    column sep=8pt,
    left delimiter={[},
    right delimiter={]},
    inner xsep=3pt
    ] (m) {
1 & -2 & 0 & 7 \\ 
0 & \phantom{+}0 & 1 & 4 \\ 
    };
    \node[fit=(m-1-1)(m-2-1), inner sep = 0, outer sep = 0, fill = red, opacity = 0.25] {};
    \node[fit=(m-1-2)(m-2-2), inner sep = 0, outer sep = 0, fill = blue, opacity = 0.25] {};
    \node[fit=(m-1-4)(m-2-4), inner sep = 0, outer sep = 0, fill = blue, opacity = 0.25] {};
    \node[fit=(m-1-3)(m-2-3), inner sep = 0, outer sep = 0, fill = red, opacity = 0.25] {};
\end{tikzpicture}
$$
and
$$
\Cvet_{(2 \times 4)} = \Zvet\Pvet' =
\begin{tikzpicture}[baseline]
\matrix[
    matrix of math nodes, 
    row sep=3pt,
    column sep=8pt,
    left delimiter={[},
    right delimiter={]},
    inner xsep=3pt
    ] (m) {
1 & 0 & -2 & 7 \\ 
0 & 1 & \phantom{+}0 & 4 \\
    };
\draw[dashed] ($0.5*(m-1-2.north east)+0.5*(m-1-3.north west)$) -- ($0.5*(m-2-2.south east)+0.5*(m-2-3.south west)$);
    \node[fit=(m-1-1)(m-2-1), inner sep = 0, outer sep = 0, fill = red, opacity = 0.25] {};
    \node[fit=(m-1-2)(m-2-2), inner sep = 0, outer sep = 0, fill = red, opacity = 0.25] {};
    \node[fit=(m-1-4)(m-2-4), inner sep = 0, outer sep = 0, fill = blue, opacity = 0.25] {};
    \node[fit=(m-1-3)(m-2-3), inner sep = 0, outer sep = 0, fill = blue, opacity = 0.25] {};
	
\node[
  fit=(m-2-3.south west) (m-2-4.south east),
  yshift = 0.1cm,
  inner xsep=0pt,inner ysep=0cm,
  below delimiter=\},
  label={[label distance=0.3cm]below:$-\Avet_{(2\times 3)}$}
] {};

\node[
  fit=(m-2-1.south west) (m-2-2.south east),
  yshift = 0.1cm,
  inner xsep=0pt,inner ysep=0cm,
  below delimiter=\},
  label={[label distance=0.3cm]below:$\Ivet_{3}$}
] {};
\end{tikzpicture}.
$$
Finally,  
$$
\Avet_{(3\times2)} =  \left[\begin{array}{cc}
-2 & 7 \\ 
\phantom{+}0 & 4 \\
\end{array}\right] \quad \mbox{and}
\quad \Svet_{(5\times2)} = \begin{tikzpicture}[baseline]
\matrix[
    matrix of math nodes, 
    row sep=2pt,
    column sep=8pt,
    left delimiter={[},
    right delimiter={]},
    inner xsep=0pt
    ] (m) {
-2 & 7 \\ 
\phantom{+}0 & 4 \\
\phantom{+}1 & 0\\
\phantom{+}0 & 1\\
    };
\draw[dashed] ($0.5*(m-2-1.south west)+0.5*(m-3-1.north west)$) -- ($0.5*(m-2-2.south east)+0.5*(m-3-2.north east)$);

\node[
  fit=(m-1-2)(m-2-2),
  inner xsep=10pt,inner ysep=0,
  right delimiter=\},
  label={[label distance=0.25cm]right:$\;\Avet_{(2\times 2)}$}
] {};

\node[
  fit=(m-3-2)(m-4-2),
  inner xsep=10pt,inner ysep=0,
  right delimiter=\},
  label={[label distance=0.25cm]right:$\;\Ivet_{2}$}
] {};
\end{tikzpicture}.
$$

\subsection{Linearly constrained multiple time series, $n = 35$ and $p = 15$}\label{ssec:bigex}

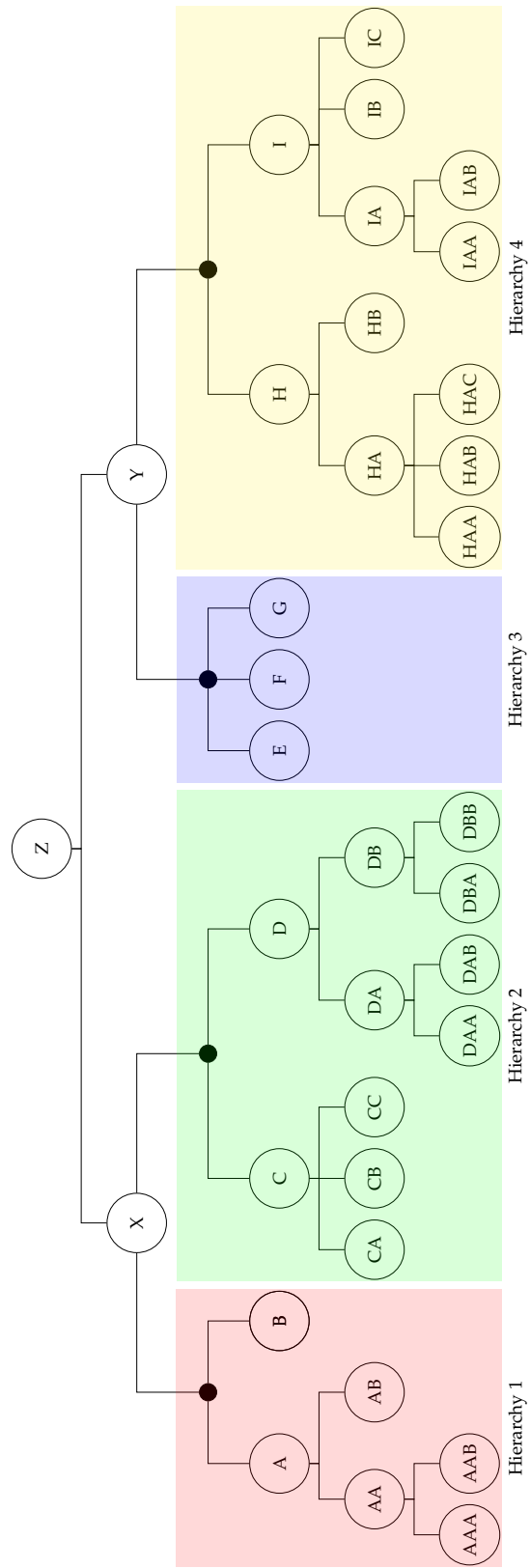
\begin{figure}[p]
\centering
\resizebox{!}{0.9\textheight}{
\rotatebox{90}{\begin{tikzpicture}[baseline=(current  bounding  box.center),
			rel/.append style={shape=circle,
				draw=black,
			minimum width=1.25cm,
			minimum height=1.25cm},
			connection/.style ={inner sep =0, outer sep =0}]
				
			\node[rel] at (0, 0) (AAA){AAA};
			\node[rel] at (1.5, 0) (AAB){AAB};
			\node[rel] at (3, 2) (AB){AB};
			\node[rel] at (4.5, 4) (B){B};
			\node[rel] at (6, 2) (CA){CA};
			\node[rel] at (7.5, 2) (CB){CB};
			\node[rel] at (9, 2) (CC){CC};
			\node[rel] at (10.5, 0) (DAA){DAA};
			\node[rel] at (12, 0) (DAB){DAB};
			\node[rel] at (13.5, 0) (DBA){DBA};
			\node[rel] at (15, 0) (DBB){DBB};
			\node[rel] at (16.5, 4) (E){E};
			\node[rel] at (18, 4) (F){F};
			\node[rel] at (19.5, 4) (G){G};
			\node[rel] at (21, 0) (HAA){HAA};
			\node[rel] at (22.5, 0) (HAB){HAB};
			\node[rel] at (24, 0) (HAC){HAC};
			\node[rel] at (25.5, 2) (HB){HB};
			\node[rel] at (27, 0) (IAA){IAA};
			\node[rel] at (28.5, 0) (IAB){IAB};
			\node[rel] at (30, 2) (IB){IB};
			\node[rel] at (31.5, 2) (IC){IC};
			
			\node[rel] at (0.75, 2) (AA){AA};
			\node[rel] at (11.25, 2) (DA){DA};
			\node[rel] at (14.25, 2) (DB){DB};
			\node[rel] at (22.5, 2) (HA){HA};
			\node[rel] at (27.75, 2) (IA){IA};
			
			\node[rel] at (29.25, 4) (I){I};
			\node[rel] at (24, 4) (H){H};
			\node[rel] at (12.75, 4) (D){D};
			\node[rel] at (7.5, 4) (C){C};
			\node[rel] at (1.5, 4) (A){A};
			
			\node[rel] at (4.5, 4) (B){B};

			\node[shape = circle, draw = black, fill = black] at (3, 5.5) (X1){};
			\node[shape = circle, draw = black, fill = black] at (10.125, 5.5) (X2){};
			\node[shape = circle, draw = black, fill = black] at (18, 5.5) (Y1){};
			\node[shape = circle, draw = black, fill = black] at (26.625, 5.5) (Y2){};
			\node[rel] at (6.5625, 7) (X){X};
			\node[rel] at (22.3125, 7) (Y){Y};
			
			\node[rel] at (14.4375, 9) (T){Z};
			
			\relation{0.2}{AA}{A};
			\relation{0.2}{AB}{A};
			\relation{0.2}{AAB}{AA};
			\relation{0.2}{AAA}{AA};

			\relation{0.2}{CA}{C};
			\relation{0.2}{CB}{C};
			\relation{0.2}{CC}{C};
			
			\relation{0.2}{DA}{D};
			\relation{0.2}{DB}{D};
			\relation{0.2}{DAA}{DA};
			\relation{0.2}{DAB}{DA};
			\relation{0.2}{DBA}{DB};
			\relation{0.2}{DBB}{DB};
			
			\relation{0.2}{HA}{H};
			\relation{0.2}{HB}{H};
			\relation{0.2}{HAA}{HA};
			\relation{0.2}{HAB}{HA};
			\relation{0.2}{HAC}{HA};
			
			\relation{0.2}{IA}{I};
			\relation{0.2}{IB}{I};
			\relation{0.2}{IC}{I};
			\relation{0.2}{IAA}{IA};
			\relation{0.2}{IAB}{IA};
						
			\relationW{A}{X1};
			\relationE{B}{X1};
			
			\relationW{C}{X2};
			\relationE{D}{X2};
			
			\relationW{E}{Y1};
			\relation{0}{F}{Y1};
			\relationE{G}{Y1};
			
			\relationW{H}{Y2};
			\relationE{I}{Y2};
			
			\relationW{X1}{X};
			\relationE{X2}{X};
			
			\relationW{Y1}{Y};
			\relationE{Y2}{Y};
			
			\relation{0.2}{X}{T};
			\relation{0.2}{Y}{T};

			\node[minimum width=1.25cm,
			minimum height=1.25cm] at (4.5, 5.5) (con0){};
			\node[minimum width=1.25cm,
			minimum height=1.25cm] at (6, 0) (con1){};
			\node[minimum width=1.25cm,
			minimum height=1.25cm] at (15, 5.5) (con2){};
			\node[minimum width=1.25cm,
			minimum height=1.25cm] at (16.5, 0) (con3){};
			\node[minimum width=1.25cm,
			minimum height=1.25cm] at (19.5, 5.5) (con4){};
			\node[minimum width=1.25cm,
			minimum height=1.25cm] at (31.5, 5.5) (con5){};

    	\node[fit=(AAA)(con0), rectangle, dashed, inner sep = 1, fill = red, opacity = 0.15, label= below:{Hierarchy 1}] {};
    	\node[fit=(con1)(con2), rectangle, dashed, inner sep = 1, fill = green, opacity = 0.15, label= below:{Hierarchy 2}] {};
    	\node[fit=(con3)(con4), rectangle, dashed, inner sep = 1, fill = blue, opacity = 0.15, label= below:{Hierarchy 3}] {};
    	\node[fit=(HAA)(con5), rectangle, dashed, inner sep = 1, fill = yellow, opacity = 0.15, label= below:{Hierarchy 4}] {};
		\end{tikzpicture}}}
		\vskip0.15cm
		\caption{Representation of a multiple linearly constrained time series formed by four genuine hierarchies in which hierarchies 1 and 2 share the same top-level variable $X$ and hierarchies 3 and 4 share the same top-level variable $Y$. The total variable $Z$ is the sum of $X$ and $Y$.}
\label{fig:hierXY}
\end{figure}

\autoref{fig:hierXY} shows a rather complex linearly constrained system, consisting in four hierarchies sharing only the top-level variables. Hierarchies 1 and 2 share the same top-level variable $X$ and hierarchies 3 and 4 share the same top-level variable $Y$. The total variable $Z$ is the sum of $X$ and $Y$. \autoref{tab:strucXY1} reports the number of bottom ($n_b$), upper ($n_a$) and all ($n$) time series of the four component hierarchies and \autoref{tab:strucXY2}, the number of free ($n_u$), constrained ($n_c$) and all ($n$) time series.

This type of linear constrained structure is not unusual in National Accounts, for example when $GDP$ from different accounting sides is considered. 

\begin{table}[!t]
	\centering
	\begin{subtable}{0.45\linewidth}
	\centering
	\begin{tabular}{c|cccc|c}
	\toprule
	 & \multicolumn{4}{c|}{Hierarchy} & Complete\\
	 & 1 & 2 & 3 & 4 & structure\\
	\midrule
	$n_b$ & 4 & 7 & 3 & 8 & 22\\
	$n_a$ & 3 & 5 & 1 & 6 & 13\\
	$n$ & 7 & 12 & 4 & 14 & 35 \\
	\bottomrule
	\end{tabular}	
	\caption{Bottom and upper variables}
	\label{tab:strucXY1}
	\end{subtable}
	\hfill
	\begin{subtable}{0.45\linewidth}
	\centering
	\begin{tabular}{c|cccc|c}
	\toprule
	 & \multicolumn{4}{c|}{Hierarchy} & Complete\\
	 & 1 & 2 & 3 & 4 & structure\\
	\midrule
	$n_u$ & 4 & 7 & 3 & 8 & 20\\
	$n_c$ & 3 & 5 & 1 & 6 & 15\\
	$n$ & 7 & 12 & 4 & 14 & 35 \\
	\bottomrule
	\end{tabular}	
	\caption{Free (unconstrained) and constrained variables}
	\label{tab:strucXY2}
	\end{subtable}
	\caption{\autoref{tab:strucXY1} reports the number of bottom ($n_b$), upper ($n_a$) and all ($n$) time series of the four component hierarchies. \autoref{tab:strucXY2} reports the number of free ($n_u$), constrained ($n_c$) and all ($n$) time series.}
\label{tab:strucXY}
\end{table}

Moving from the top of the structure in \autoref{fig:hierXY}, the relationships between the 35 variables can be expressed as
\begin{enumerate}[nosep]
	\item $Z = X + Y$
	\item $X = A + B$
	\item $X = C + D$
	\item $Y = E + F + G$
	\item $Y = H + I$
	\item $A = AB + AAA + AAB$
	\item $AA = AAA + AAB$
	\item $C = CA + CB + CC$ 
	\item $D = DAA + DAB + DBA + DBB$
	\item $DA = DAA + DAB$
	\item $DB = DBA + DBB$
	\item $H = HA + HB$
	\item $HA = HAA + HAB + HAC$
	\item $I = IB + IC + IAA + IAB$
	\item $IA = IAA + IAB$
\end{enumerate}	
These relationships can be grouped into a system of 15 equations:
$$\begin{aligned}
x_{{Z},t} - x_{{X},t} - x_{{Y},t} &= 0, \\ 
x_{{X},t} - x_{{A},t} - x_{{B},t} &= 0, \\ 
x_{{X},t} - x_{{C},t} - x_{{D},t} &= 0, \\ 
x_{{Y},t} - x_{{E},t} - x_{{F},t} - x_{{G},t} &= 0, \\ 
x_{{Y},t} - x_{{H},t} - x_{{I},t} &= 0, \\ 
x_{{A},t} - x_{{AB},t} - x_{{AAA},t} - x_{{AAB},t} &= 0, \\ 
x_{{AA},t} - x_{{AAA},t} - x_{{AAB},t} &= 0, \\ 
x_{{C},t} - x_{{CA},t} - x_{{CB},t} - x_{{CC},t} &= 0 \\ 
\end{aligned}\qquad \qquad 
\begin{aligned}
x_{{D},t} - x_{{DAA},t} - x_{{DAB},t} - x_{{DBA},t} - x_{{DBB},t} &= 0, \\ 
x_{{DA},t} - x_{{DAA},t} - x_{{DAB},t} &= 0, \\ 
x_{{DB},t} - x_{{DBA},t} - x_{{DBB},t} &= 0, \\ 
x_{{H},t} - x_{{HA},t} - x_{{HB},t} &= 0, \\ 
x_{{HA},t} - x_{{HAA},t} - x_{{HAB},t} - x_{{HAC},t} &= 0, \\ 
x_{{I},t} - x_{{IB},t} - x_{{IC},t} - x_{{IAA},t} - x_{{IAB},t} &= 0, \\ 
x_{{IA},t} - x_{{IAA},t} - x_{{IAB},t} &= 0, \\
&
\end{aligned}
$$
with coefficient matrix
$$
\Gammavet_{(15\times35)} = \begin{adjustbox}{width=0.8\textwidth}
\begin{tikzpicture}[baseline]
\matrix[matrix of math nodes, 
    every node/.append style={anchor=base,align=center,
    text depth=.5ex,text height=2ex,text width=1.5em},
    row sep=0pt,
    column sep=2pt,
    left delimiter={[},
    right delimiter={]},
    inner xsep=1pt
    ] (m) {
1 & -1 & -1 & 0 & 0 & 0 & 0 & 0 & 0 & 0 & 0 & 0 & 0 & 0 & 0 & 0 & 0 & 0 & 0 & 0 & 0 & 0 & 0 & 0 & 0 & 0 & 0 & 0 & 0 & 0 & 0 & 0 & 0 & 0 & 0 \\ 
0 & 1 & 0 & -1 & -1 & 0 & 0 & 0 & 0 & 0 & 0 & 0 & 0 & 0 & 0 & 0 & 0 & 0 & 0 & 0 & 0 & 0 & 0 & 0 & 0 & 0 & 0 & 0 & 0 & 0 & 0 & 0 & 0 & 0 & 0 \\ 
0 & 1 & 0 & 0 & 0 & -1 & -1 & 0 & 0 & 0 & 0 & 0 & 0 & 0 & 0 & 0 & 0 & 0 & 0 & 0 & 0 & 0 & 0 & 0 & 0 & 0 & 0 & 0 & 0 & 0 & 0 & 0 & 0 & 0 & 0 \\ 
0 & 0 & 1 & 0 & 0 & 0 & 0 & -1 & -1 & -1 & 0 & 0 & 0 & 0 & 0 & 0 & 0 & 0 & 0 & 0 & 0 & 0 & 0 & 0 & 0 & 0 & 0 & 0 & 0 & 0 & 0 & 0 & 0 & 0 & 0 \\ 
0 & 0 & 1 & 0 & 0 & 0 & 0 & 0 & 0 & 0 & -1 & -1 & 0 & 0 & 0 & 0 & 0 & 0 & 0 & 0 & 0 & 0 & 0 & 0 & 0 & 0 & 0 & 0 & 0 & 0 & 0 & 0 & 0 & 0 & 0 \\ 
0 & 0 & 0 & 1 & 0 & 0 & 0 & 0 & 0 & 0 & 0 & 0 & 0 & -1 & 0 & 0 & 0 & 0 & 0 & 0 & 0 & 0 & 0 & 0 & -1 & -1 & 0 & 0 & 0 & 0 & 0 & 0 & 0 & 0 & 0 \\ 
0 & 0 & 0 & 0 & 0 & 0 & 0 & 0 & 0 & 0 & 0 & 0 & 1 & 0 & 0 & 0 & 0 & 0 & 0 & 0 & 0 & 0 & 0 & 0 & -1 & -1 & 0 & 0 & 0 & 0 & 0 & 0 & 0 & 0 & 0 \\ 
0 & 0 & 0 & 0 & 0 & 1 & 0 & 0 & 0 & 0 & 0 & 0 & 0 & 0 & -1 & -1 & -1 & 0 & 0 & 0 & 0 & 0 & 0 & 0 & 0 & 0 & 0 & 0 & 0 & 0 & 0 & 0 & 0 & 0 & 0 \\ 
0 & 0 & 0 & 0 & 0 & 0 & 1 & 0 & 0 & 0 & 0 & 0 & 0 & 0 & 0 & 0 & 0 & 0 & 0 & 0 & 0 & 0 & 0 & 0 & 0 & 0 & -1 & -1 & -1 & -1 & 0 & 0 & 0 & 0 & 0 \\ 
0 & 0 & 0 & 0 & 0 & 0 & 0 & 0 & 0 & 0 & 0 & 0 & 0 & 0 & 0 & 0 & 0 & 1 & 0 & 0 & 0 & 0 & 0 & 0 & 0 & 0 & -1 & -1 & 0 & 0 & 0 & 0 & 0 & 0 & 0 \\ 
0 & 0 & 0 & 0 & 0 & 0 & 0 & 0 & 0 & 0 & 0 & 0 & 0 & 0 & 0 & 0 & 0 & 0 & 1 & 0 & 0 & 0 & 0 & 0 & 0 & 0 & 0 & 0 & -1 & -1 & 0 & 0 & 0 & 0 & 0 \\ 
0 & 0 & 0 & 0 & 0 & 0 & 0 & 0 & 0 & 0 & 1 & 0 & 0 & 0 & 0 & 0 & 0 & 0 & 0 & -1 & -1 & 0 & 0 & 0 & 0 & 0 & 0 & 0 & 0 & 0 & 0 & 0 & 0 & 0 & 0 \\ 
0 & 0 & 0 & 0 & 0 & 0 & 0 & 0 & 0 & 0 & 0 & 0 & 0 & 0 & 0 & 0 & 0 & 0 & 0 & 1 & 0 & 0 & 0 & 0 & 0 & 0 & 0 & 0 & 0 & 0 & -1 & -1 & -1 & 0 & 0 \\ 
0 & 0 & 0 & 0 & 0 & 0 & 0 & 0 & 0 & 0 & 0 & 1 & 0 & 0 & 0 & 0 & 0 & 0 & 0 & 0 & 0 & 0 & -1 & -1 & 0 & 0 & 0 & 0 & 0 & 0 & 0 & 0 & 0 & -1 & -1 \\ 
0 & 0 & 0 & 0 & 0 & 0 & 0 & 0 & 0 & 0 & 0 & 0 & 0 & 0 & 0 & 0 & 0 & 0 & 0 & 0 & 0 & 1 & 0 & 0 & 0 & 0 & 0 & 0 & 0 & 0 & 0 & 0 & 0 & -1 & -1 \\
    };
\matrix[matrix of math nodes, below = 1.1ex of m,
    every node/.append style={anchor=base,align=center,
    text depth=.5ex,text height=0.75ex,text width=1.5em, font=\tiny},
    row sep=0pt,
    column sep=2pt,
    inner xsep=1pt,
    above delimiter = |
    ] (m1) {
{Z} & {X} & {Y} & {A} & {B} & {C} & {D} & {E} & {F} & {G} & {H} & {I} & {AA} & {AB} & {CA} & {CB} & {CC} & {DA} & {DB} & {HA} & {HB} & {IA} & {IB} & {IC} & {AAA} & {AAB} & {DAA} & {DAB} & {DBA} & {DBB} & {HAA} & {HAB} & {HAC} & {IAA} & {IAB} \\
    };
\end{tikzpicture}
\end{adjustbox}.
$$

Reducing this matrix through rref yields  
$$
\Zvet_{(15\times35)} = \begin{adjustbox}{width=0.8\textwidth}
\begin{tikzpicture}[baseline]
\matrix[matrix of math nodes, 
    every node/.append style={anchor=base,align=center,
    text depth=.5ex,text height=2ex,text width=1.5em},
    row sep=0pt,
    column sep=2pt,
    left delimiter={[},
    right delimiter={]},
    inner xsep=1pt
    ] (m) {
1 & 0 & 0 & 0 & 0 & 0 & 0 & 0 & 0 & 0 & 0 & 0 & 0 & 0 & -1 & -1 & -1 & 0 & 0 & 0 & -1 & 0 & -1 & -1 & 0 & 0 & -1 & -1 & -1 & -1 & -1 & -1 & -1 & -1 & -1 \\ 
0 & 1 & 0 & 0 & 0 & 0 & 0 & 0 & 0 & 0 & 0 & 0 & 0 & 0 & -1 & -1 & -1 & 0 & 0 & 0 & 0 & 0 & 0 & 0 & 0 & 0 & -1 & -1 & -1 & -1 & 0 & 0 & 0 & 0 & 0 \\ 
0 & 0 & 1 & 0 & 0 & 0 & 0 & 0 & 0 & 0 & 0 & 0 & 0 & 0 & 0 & 0 & 0 & 0 & 0 & 0 & -1 & 0 & -1 & -1 & 0 & 0 & 0 & 0 & 0 & 0 & -1 & -1 & -1 & -1 & -1 \\ 
0 & 0 & 0 & 1 & 0 & 0 & 0 & 0 & 0 & 0 & 0 & 0 & 0 & -1 & 0 & 0 & 0 & 0 & 0 & 0 & 0 & 0 & 0 & 0 & -1 & -1 & 0 & 0 & 0 & 0 & 0 & 0 & 0 & 0 & 0 \\ 
0 & 0 & 0 & 0 & 1 & 0 & 0 & 0 & 0 & 0 & 0 & 0 & 0 & 1 & -1 & -1 & -1 & 0 & 0 & 0 & 0 & 0 & 0 & 0 & 1 & 1 & -1 & -1 & -1 & -1 & 0 & 0 & 0 & 0 & 0 \\ 
0 & 0 & 0 & 0 & 0 & 1 & 0 & 0 & 0 & 0 & 0 & 0 & 0 & 0 & -1 & -1 & -1 & 0 & 0 & 0 & 0 & 0 & 0 & 0 & 0 & 0 & 0 & 0 & 0 & 0 & 0 & 0 & 0 & 0 & 0 \\ 
0 & 0 & 0 & 0 & 0 & 0 & 1 & 0 & 0 & 0 & 0 & 0 & 0 & 0 & 0 & 0 & 0 & 0 & 0 & 0 & 0 & 0 & 0 & 0 & 0 & 0 & -1 & -1 & -1 & -1 & 0 & 0 & 0 & 0 & 0 \\ 
0 & 0 & 0 & 0 & 0 & 0 & 0 & 1 & 1 & 1 & 0 & 0 & 0 & 0 & 0 & 0 & 0 & 0 & 0 & 0 & -1 & 0 & -1 & -1 & 0 & 0 & 0 & 0 & 0 & 0 & -1 & -1 & -1 & -1 & -1 \\ 
0 & 0 & 0 & 0 & 0 & 0 & 0 & 0 & 0 & 0 & 1 & 0 & 0 & 0 & 0 & 0 & 0 & 0 & 0 & 0 & -1 & 0 & 0 & 0 & 0 & 0 & 0 & 0 & 0 & 0 & -1 & -1 & -1 & 0 & 0 \\ 
0 & 0 & 0 & 0 & 0 & 0 & 0 & 0 & 0 & 0 & 0 & 1 & 0 & 0 & 0 & 0 & 0 & 0 & 0 & 0 & 0 & 0 & -1 & -1 & 0 & 0 & 0 & 0 & 0 & 0 & 0 & 0 & 0 & -1 & -1 \\ 
0 & 0 & 0 & 0 & 0 & 0 & 0 & 0 & 0 & 0 & 0 & 0 & 1 & 0 & 0 & 0 & 0 & 0 & 0 & 0 & 0 & 0 & 0 & 0 & -1 & -1 & 0 & 0 & 0 & 0 & 0 & 0 & 0 & 0 & 0 \\ 
0 & 0 & 0 & 0 & 0 & 0 & 0 & 0 & 0 & 0 & 0 & 0 & 0 & 0 & 0 & 0 & 0 & 1 & 0 & 0 & 0 & 0 & 0 & 0 & 0 & 0 & -1 & -1 & 0 & 0 & 0 & 0 & 0 & 0 & 0 \\ 
0 & 0 & 0 & 0 & 0 & 0 & 0 & 0 & 0 & 0 & 0 & 0 & 0 & 0 & 0 & 0 & 0 & 0 & 1 & 0 & 0 & 0 & 0 & 0 & 0 & 0 & 0 & 0 & -1 & -1 & 0 & 0 & 0 & 0 & 0 \\ 
0 & 0 & 0 & 0 & 0 & 0 & 0 & 0 & 0 & 0 & 0 & 0 & 0 & 0 & 0 & 0 & 0 & 0 & 0 & 1 & 0 & 0 & 0 & 0 & 0 & 0 & 0 & 0 & 0 & 0 & -1 & -1 & -1 & 0 & 0 \\ 
0 & 0 & 0 & 0 & 0 & 0 & 0 & 0 & 0 & 0 & 0 & 0 & 0 & 0 & 0 & 0 & 0 & 0 & 0 & 0 & 0 & 1 & 0 & 0 & 0 & 0 & 0 & 0 & 0 & 0 & 0 & 0 & 0 & -1 & -1 \\ 
    };
 \matrix[matrix of math nodes, below = 1.1ex of m,
    every node/.append style={anchor=base,align=center,
    text depth=.5ex,text height=0.75ex,text width=1.5em, font=\tiny},
    row sep=0pt,
    column sep=2pt,
    inner xsep=1pt,
    above delimiter = |
    ] (m1) {
{Z} & {X} & {Y} & {A} & {B} & {C} & {D} & {E} & {F} & {G} & {H} & {I} & {AA} & {AB} & {CA} & {CB} & {CC} & {DA} & {DB} & {HA} & {HB} & {IA} & {IB} & {IC} & {AAA} & {AAB} & {DAA} & {DAB} & {DBA} & {DBB} & {HAA} & {HAB} & {HAC} & {IAA} & {IAB} \\
    };
    \foreach \x in {1, 2, 3, 4, 5, 6, 7, 8, 11, 12, 13, 18, 19, 20, 22}
    	\node[fit=(m-1-\x)(m-15-\x), inner sep = 0, outer sep = 0, fill = red, opacity = 0.15] {};
    
    \foreach \x in {9, 10, 14, 15, 16, 17, 21, 23, 24, 25, 26, 27, 28, 29, 30, 31, 32, 33, 34, 35}
    	\node[fit=(m-1-\x)(m-15-\x), inner sep = 0, outer sep = 0, fill = blue, opacity = 0.15] {};
\end{tikzpicture}
\end{adjustbox}
$$
and
$$
\begin{aligned}
\Cvet_{(15\times35)} &= \Zvet\Pvet' \\
&= \begin{adjustbox}{width=0.8\textwidth}
\begin{tikzpicture}[baseline]
\matrix[matrix of math nodes, 
    every node/.append style={anchor=base,align=center,
    text depth=.5ex,text height=2ex,text width=1.5em},
    row sep=0pt,
    column sep=2pt,
    left delimiter={[},
    right delimiter={]},
    inner xsep=1pt
    ] (m) {
1 & 0 & 0 & 0 & 0 & 0 & 0 & 0 & 0 & 0 & 0 & 0 & 0 & 0 & 0 & 0 & 0 & 0 & -1 & -1 & -1 & -1 & -1 & -1 & 0 & 0 & -1 & -1 & -1 & -1 & -1 & -1 & -1 & -1 & -1 \\ 
0 & 1 & 0 & 0 & 0 & 0 & 0 & 0 & 0 & 0 & 0 & 0 & 0 & 0 & 0 & 0 & 0 & 0 & -1 & -1 & -1 & 0 & 0 & 0 & 0 & 0 & -1 & -1 & -1 & -1 & 0 & 0 & 0 & 0 & 0 \\ 
0 & 0 & 1 & 0 & 0 & 0 & 0 & 0 & 0 & 0 & 0 & 0 & 0 & 0 & 0 & 0 & 0 & 0 & 0 & 0 & 0 & -1 & -1 & -1 & 0 & 0 & 0 & 0 & 0 & 0 & -1 & -1 & -1 & -1 & -1 \\ 
0 & 0 & 0 & 1 & 0 & 0 & 0 & 0 & 0 & 0 & 0 & 0 & 0 & 0 & 0 & 0 & 0 & -1 & 0 & 0 & 0 & 0 & 0 & 0 & -1 & -1 & 0 & 0 & 0 & 0 & 0 & 0 & 0 & 0 & 0 \\ 
0 & 0 & 0 & 0 & 1 & 0 & 0 & 0 & 0 & 0 & 0 & 0 & 0 & 0 & 0 & 0 & 0 & 1 & -1 & -1 & -1 & 0 & 0 & 0 & 1 & 1 & -1 & -1 & -1 & -1 & 0 & 0 & 0 & 0 & 0 \\ 
0 & 0 & 0 & 0 & 0 & 1 & 0 & 0 & 0 & 0 & 0 & 0 & 0 & 0 & 0 & 0 & 0 & 0 & -1 & -1 & -1 & 0 & 0 & 0 & 0 & 0 & 0 & 0 & 0 & 0 & 0 & 0 & 0 & 0 & 0 \\ 
0 & 0 & 0 & 0 & 0 & 0 & 1 & 0 & 0 & 0 & 0 & 0 & 0 & 0 & 0 & 0 & 0 & 0 & 0 & 0 & 0 & 0 & 0 & 0 & 0 & 0 & -1 & -1 & -1 & -1 & 0 & 0 & 0 & 0 & 0 \\ 
0 & 0 & 0 & 0 & 0 & 0 & 0 & 1 & 0 & 0 & 0 & 0 & 0 & 0 & 0 & 1 & 1 & 0 & 0 & 0 & 0 & -1 & -1 & -1 & 0 & 0 & 0 & 0 & 0 & 0 & -1 & -1 & -1 & -1 & -1 \\ 
0 & 0 & 0 & 0 & 0 & 0 & 0 & 0 & 1 & 0 & 0 & 0 & 0 & 0 & 0 & 0 & 0 & 0 & 0 & 0 & 0 & -1 & 0 & 0 & 0 & 0 & 0 & 0 & 0 & 0 & -1 & -1 & -1 & 0 & 0 \\ 
0 & 0 & 0 & 0 & 0 & 0 & 0 & 0 & 0 & 1 & 0 & 0 & 0 & 0 & 0 & 0 & 0 & 0 & 0 & 0 & 0 & 0 & -1 & -1 & 0 & 0 & 0 & 0 & 0 & 0 & 0 & 0 & 0 & -1 & -1 \\ 
0 & 0 & 0 & 0 & 0 & 0 & 0 & 0 & 0 & 0 & 1 & 0 & 0 & 0 & 0 & 0 & 0 & 0 & 0 & 0 & 0 & 0 & 0 & 0 & -1 & -1 & 0 & 0 & 0 & 0 & 0 & 0 & 0 & 0 & 0 \\ 
0 & 0 & 0 & 0 & 0 & 0 & 0 & 0 & 0 & 0 & 0 & 1 & 0 & 0 & 0 & 0 & 0 & 0 & 0 & 0 & 0 & 0 & 0 & 0 & 0 & 0 & -1 & -1 & 0 & 0 & 0 & 0 & 0 & 0 & 0 \\ 
0 & 0 & 0 & 0 & 0 & 0 & 0 & 0 & 0 & 0 & 0 & 0 & 1 & 0 & 0 & 0 & 0 & 0 & 0 & 0 & 0 & 0 & 0 & 0 & 0 & 0 & 0 & 0 & -1 & -1 & 0 & 0 & 0 & 0 & 0 \\ 
0 & 0 & 0 & 0 & 0 & 0 & 0 & 0 & 0 & 0 & 0 & 0 & 0 & 1 & 0 & 0 & 0 & 0 & 0 & 0 & 0 & 0 & 0 & 0 & 0 & 0 & 0 & 0 & 0 & 0 & -1 & -1 & -1 & 0 & 0 \\ 
0 & 0 & 0 & 0 & 0 & 0 & 0 & 0 & 0 & 0 & 0 & 0 & 0 & 0 & 1 & 0 & 0 & 0 & 0 & 0 & 0 & 0 & 0 & 0 & 0 & 0 & 0 & 0 & 0 & 0 & 0 & 0 & 0 & -1 & -1 \\ 
    };

 \matrix[matrix of math nodes, below = 1.1ex of m,
    every node/.append style={anchor=base,align=center,
    text depth=.5ex,text height=0.75ex,text width=1.5em, font=\tiny},
    row sep=0pt,
    column sep=2pt,
    inner xsep=1pt,
    above delimiter = |
    ] (m1) {
{Z} & {X} & {Y} & {A} & {B} & {C} & {D} & {E} & {H} & {I} & {AA} & {DA} & {DB} & {HA} & {IA} & {F} & {G} & {AB} & {CA} & {CB} & {CC} & {HB} & {IB} & {IC} & {AAA} & {AAB} & {DAA} & {DAB} & {DBA} & {DBB} & {HAA} & {HAB} & {HAC} & {IAA} & {IAB} \\
    };
    
    \foreach \x in {1,...,15}
    	\node[fit=(m-1-\x)(m-15-\x), inner sep = 0, outer sep = 0, fill = red, opacity = 0.15] {};
    
    \foreach \x in {16,...,35}
    	\node[fit=(m-1-\x)(m-15-\x), inner sep = 0, outer sep = 0, fill = blue, opacity = 0.15] {};
\end{tikzpicture}
\end{adjustbox},
\end{aligned}
$$
where $\Pvet$ is a ($35 \times 35$) permutation matrix moving the pivot columns (red background) in $\Zvet$ to the left side of matrix $\Cvet$:
$$
\Pvet_{(35\times35)} = \begin{adjustbox}{width=0.85\textwidth}
\begin{tikzpicture}[baseline]
\matrix[matrix of math nodes, 
    every node/.append style={anchor=base,align=center,
    text depth=.5ex,text height=2ex,text width=1.5em},
    row sep=0pt,
    column sep=2pt,
    left delimiter={[},
    right delimiter={]},
    inner xsep=1pt
    ] (m) {
1 & 0 & 0 & 0 & 0 & 0 & 0 & 0 & 0 & 0 & 0 & 0 & 0 & 0 & 0 & 0 & 0 & 0 & 0 & 0 & 0 & 0 & 0 & 0 & 0 & 0 & 0 & 0 & 0 & 0 & 0 & 0 & 0 & 0 & 0 \\ 
0 & 1 & 0 & 0 & 0 & 0 & 0 & 0 & 0 & 0 & 0 & 0 & 0 & 0 & 0 & 0 & 0 & 0 & 0 & 0 & 0 & 0 & 0 & 0 & 0 & 0 & 0 & 0 & 0 & 0 & 0 & 0 & 0 & 0 & 0 \\ 
0 & 0 & 1 & 0 & 0 & 0 & 0 & 0 & 0 & 0 & 0 & 0 & 0 & 0 & 0 & 0 & 0 & 0 & 0 & 0 & 0 & 0 & 0 & 0 & 0 & 0 & 0 & 0 & 0 & 0 & 0 & 0 & 0 & 0 & 0 \\ 
0 & 0 & 0 & 1 & 0 & 0 & 0 & 0 & 0 & 0 & 0 & 0 & 0 & 0 & 0 & 0 & 0 & 0 & 0 & 0 & 0 & 0 & 0 & 0 & 0 & 0 & 0 & 0 & 0 & 0 & 0 & 0 & 0 & 0 & 0 \\ 
0 & 0 & 0 & 0 & 1 & 0 & 0 & 0 & 0 & 0 & 0 & 0 & 0 & 0 & 0 & 0 & 0 & 0 & 0 & 0 & 0 & 0 & 0 & 0 & 0 & 0 & 0 & 0 & 0 & 0 & 0 & 0 & 0 & 0 & 0 \\ 
0 & 0 & 0 & 0 & 0 & 1 & 0 & 0 & 0 & 0 & 0 & 0 & 0 & 0 & 0 & 0 & 0 & 0 & 0 & 0 & 0 & 0 & 0 & 0 & 0 & 0 & 0 & 0 & 0 & 0 & 0 & 0 & 0 & 0 & 0 \\ 
0 & 0 & 0 & 0 & 0 & 0 & 1 & 0 & 0 & 0 & 0 & 0 & 0 & 0 & 0 & 0 & 0 & 0 & 0 & 0 & 0 & 0 & 0 & 0 & 0 & 0 & 0 & 0 & 0 & 0 & 0 & 0 & 0 & 0 & 0 \\ 
0 & 0 & 0 & 0 & 0 & 0 & 0 & 1 & 0 & 0 & 0 & 0 & 0 & 0 & 0 & 0 & 0 & 0 & 0 & 0 & 0 & 0 & 0 & 0 & 0 & 0 & 0 & 0 & 0 & 0 & 0 & 0 & 0 & 0 & 0 \\ 
0 & 0 & 0 & 0 & 0 & 0 & 0 & 0 & 0 & 0 & 1 & 0 & 0 & 0 & 0 & 0 & 0 & 0 & 0 & 0 & 0 & 0 & 0 & 0 & 0 & 0 & 0 & 0 & 0 & 0 & 0 & 0 & 0 & 0 & 0 \\ 
0 & 0 & 0 & 0 & 0 & 0 & 0 & 0 & 0 & 0 & 0 & 1 & 0 & 0 & 0 & 0 & 0 & 0 & 0 & 0 & 0 & 0 & 0 & 0 & 0 & 0 & 0 & 0 & 0 & 0 & 0 & 0 & 0 & 0 & 0 \\ 
0 & 0 & 0 & 0 & 0 & 0 & 0 & 0 & 0 & 0 & 0 & 0 & 1 & 0 & 0 & 0 & 0 & 0 & 0 & 0 & 0 & 0 & 0 & 0 & 0 & 0 & 0 & 0 & 0 & 0 & 0 & 0 & 0 & 0 & 0 \\ 
0 & 0 & 0 & 0 & 0 & 0 & 0 & 0 & 0 & 0 & 0 & 0 & 0 & 0 & 0 & 0 & 0 & 1 & 0 & 0 & 0 & 0 & 0 & 0 & 0 & 0 & 0 & 0 & 0 & 0 & 0 & 0 & 0 & 0 & 0 \\ 
0 & 0 & 0 & 0 & 0 & 0 & 0 & 0 & 0 & 0 & 0 & 0 & 0 & 0 & 0 & 0 & 0 & 0 & 1 & 0 & 0 & 0 & 0 & 0 & 0 & 0 & 0 & 0 & 0 & 0 & 0 & 0 & 0 & 0 & 0 \\ 
0 & 0 & 0 & 0 & 0 & 0 & 0 & 0 & 0 & 0 & 0 & 0 & 0 & 0 & 0 & 0 & 0 & 0 & 0 & 1 & 0 & 0 & 0 & 0 & 0 & 0 & 0 & 0 & 0 & 0 & 0 & 0 & 0 & 0 & 0 \\ 
0 & 0 & 0 & 0 & 0 & 0 & 0 & 0 & 0 & 0 & 0 & 0 & 0 & 0 & 0 & 0 & 0 & 0 & 0 & 0 & 0 & 1 & 0 & 0 & 0 & 0 & 0 & 0 & 0 & 0 & 0 & 0 & 0 & 0 & 0 \\ 
0 & 0 & 0 & 0 & 0 & 0 & 0 & 0 & 1 & 0 & 0 & 0 & 0 & 0 & 0 & 0 & 0 & 0 & 0 & 0 & 0 & 0 & 0 & 0 & 0 & 0 & 0 & 0 & 0 & 0 & 0 & 0 & 0 & 0 & 0 \\ 
0 & 0 & 0 & 0 & 0 & 0 & 0 & 0 & 0 & 1 & 0 & 0 & 0 & 0 & 0 & 0 & 0 & 0 & 0 & 0 & 0 & 0 & 0 & 0 & 0 & 0 & 0 & 0 & 0 & 0 & 0 & 0 & 0 & 0 & 0 \\ 
0 & 0 & 0 & 0 & 0 & 0 & 0 & 0 & 0 & 0 & 0 & 0 & 0 & 1 & 0 & 0 & 0 & 0 & 0 & 0 & 0 & 0 & 0 & 0 & 0 & 0 & 0 & 0 & 0 & 0 & 0 & 0 & 0 & 0 & 0 \\ 
0 & 0 & 0 & 0 & 0 & 0 & 0 & 0 & 0 & 0 & 0 & 0 & 0 & 0 & 1 & 0 & 0 & 0 & 0 & 0 & 0 & 0 & 0 & 0 & 0 & 0 & 0 & 0 & 0 & 0 & 0 & 0 & 0 & 0 & 0 \\ 
0 & 0 & 0 & 0 & 0 & 0 & 0 & 0 & 0 & 0 & 0 & 0 & 0 & 0 & 0 & 1 & 0 & 0 & 0 & 0 & 0 & 0 & 0 & 0 & 0 & 0 & 0 & 0 & 0 & 0 & 0 & 0 & 0 & 0 & 0 \\ 
0 & 0 & 0 & 0 & 0 & 0 & 0 & 0 & 0 & 0 & 0 & 0 & 0 & 0 & 0 & 0 & 1 & 0 & 0 & 0 & 0 & 0 & 0 & 0 & 0 & 0 & 0 & 0 & 0 & 0 & 0 & 0 & 0 & 0 & 0 \\ 
0 & 0 & 0 & 0 & 0 & 0 & 0 & 0 & 0 & 0 & 0 & 0 & 0 & 0 & 0 & 0 & 0 & 0 & 0 & 0 & 1 & 0 & 0 & 0 & 0 & 0 & 0 & 0 & 0 & 0 & 0 & 0 & 0 & 0 & 0 \\ 
0 & 0 & 0 & 0 & 0 & 0 & 0 & 0 & 0 & 0 & 0 & 0 & 0 & 0 & 0 & 0 & 0 & 0 & 0 & 0 & 0 & 0 & 1 & 0 & 0 & 0 & 0 & 0 & 0 & 0 & 0 & 0 & 0 & 0 & 0 \\ 
0 & 0 & 0 & 0 & 0 & 0 & 0 & 0 & 0 & 0 & 0 & 0 & 0 & 0 & 0 & 0 & 0 & 0 & 0 & 0 & 0 & 0 & 0 & 1 & 0 & 0 & 0 & 0 & 0 & 0 & 0 & 0 & 0 & 0 & 0 \\ 
0 & 0 & 0 & 0 & 0 & 0 & 0 & 0 & 0 & 0 & 0 & 0 & 0 & 0 & 0 & 0 & 0 & 0 & 0 & 0 & 0 & 0 & 0 & 0 & 1 & 0 & 0 & 0 & 0 & 0 & 0 & 0 & 0 & 0 & 0 \\ 
0 & 0 & 0 & 0 & 0 & 0 & 0 & 0 & 0 & 0 & 0 & 0 & 0 & 0 & 0 & 0 & 0 & 0 & 0 & 0 & 0 & 0 & 0 & 0 & 0 & 1 & 0 & 0 & 0 & 0 & 0 & 0 & 0 & 0 & 0 \\ 
0 & 0 & 0 & 0 & 0 & 0 & 0 & 0 & 0 & 0 & 0 & 0 & 0 & 0 & 0 & 0 & 0 & 0 & 0 & 0 & 0 & 0 & 0 & 0 & 0 & 0 & 1 & 0 & 0 & 0 & 0 & 0 & 0 & 0 & 0 \\ 
0 & 0 & 0 & 0 & 0 & 0 & 0 & 0 & 0 & 0 & 0 & 0 & 0 & 0 & 0 & 0 & 0 & 0 & 0 & 0 & 0 & 0 & 0 & 0 & 0 & 0 & 0 & 1 & 0 & 0 & 0 & 0 & 0 & 0 & 0 \\ 
0 & 0 & 0 & 0 & 0 & 0 & 0 & 0 & 0 & 0 & 0 & 0 & 0 & 0 & 0 & 0 & 0 & 0 & 0 & 0 & 0 & 0 & 0 & 0 & 0 & 0 & 0 & 0 & 1 & 0 & 0 & 0 & 0 & 0 & 0 \\ 
0 & 0 & 0 & 0 & 0 & 0 & 0 & 0 & 0 & 0 & 0 & 0 & 0 & 0 & 0 & 0 & 0 & 0 & 0 & 0 & 0 & 0 & 0 & 0 & 0 & 0 & 0 & 0 & 0 & 1 & 0 & 0 & 0 & 0 & 0 \\ 
0 & 0 & 0 & 0 & 0 & 0 & 0 & 0 & 0 & 0 & 0 & 0 & 0 & 0 & 0 & 0 & 0 & 0 & 0 & 0 & 0 & 0 & 0 & 0 & 0 & 0 & 0 & 0 & 0 & 0 & 1 & 0 & 0 & 0 & 0 \\ 
0 & 0 & 0 & 0 & 0 & 0 & 0 & 0 & 0 & 0 & 0 & 0 & 0 & 0 & 0 & 0 & 0 & 0 & 0 & 0 & 0 & 0 & 0 & 0 & 0 & 0 & 0 & 0 & 0 & 0 & 0 & 1 & 0 & 0 & 0 \\ 
0 & 0 & 0 & 0 & 0 & 0 & 0 & 0 & 0 & 0 & 0 & 0 & 0 & 0 & 0 & 0 & 0 & 0 & 0 & 0 & 0 & 0 & 0 & 0 & 0 & 0 & 0 & 0 & 0 & 0 & 0 & 0 & 1 & 0 & 0 \\ 
0 & 0 & 0 & 0 & 0 & 0 & 0 & 0 & 0 & 0 & 0 & 0 & 0 & 0 & 0 & 0 & 0 & 0 & 0 & 0 & 0 & 0 & 0 & 0 & 0 & 0 & 0 & 0 & 0 & 0 & 0 & 0 & 0 & 1 & 0 \\ 
0 & 0 & 0 & 0 & 0 & 0 & 0 & 0 & 0 & 0 & 0 & 0 & 0 & 0 & 0 & 0 & 0 & 0 & 0 & 0 & 0 & 0 & 0 & 0 & 0 & 0 & 0 & 0 & 0 & 0 & 0 & 0 & 0 & 0 & 1 \\ 
    };

 \matrix[matrix of math nodes, below = 1.1ex of m,
    every node/.append style={anchor=base,align=center,
    text depth=.5ex,text height=0.75ex,text width=1.5em, font=\tiny},
    row sep=0pt,
    column sep=2pt,
    inner xsep=1pt,
    above delimiter = |
    ] (m1) {
{Z} & {X} & {Y} & {A} & {B} & {C} & {D} & {E} & {F} & {G} & {H} & {I} & {AA} & {AB} & {CA} & {CB} & {CC} & {DA} & {DB} & {HA} & {HB} & {IA} & {IB} & {IC} & {AAA} & {AAB} & {DAA} & {DAB} & {DBA} & {DBB} & {HAA} & {HAB} & {HAC} & {IAA} & {IAB}\\
    };

\matrix[matrix of math nodes, right = 2ex of m,
    every node/.append style={anchor=base,align=left,
    text depth=.5ex,text height=2ex,text width=1.5em, font=\tiny},
    row sep=0pt,
    column sep=2pt,
    inner xsep=1pt
    ] (m2) {
{Z} \\ {X} \\ {Y} \\ {A} \\ {B} \\ {C} \\ {D} \\ {E} \\ {H} \\ {I} \\ {AA} \\ {DA} \\ {DB} \\ {HA} \\ {IA} \\ {F} \\ {G} \\ {AB} \\ {CA} \\ {CB} \\ {CC} \\ {HB} \\ {IB} \\ {IC} \\ {AAA} \\ {AAB} \\ {DAA} \\ {DAB} \\ {DBA} \\ {DBB} \\ {HAA} \\ {HAB} \\ {HAC} \\ {IAA} \\ {IAB} \\
    };

\end{tikzpicture}
\end{adjustbox}.
$$
Finally, the linear combination matrix $\Avet$ is given by:
$$
\Avet= \begin{adjustbox}{width=0.8\textwidth}
\begin{tikzpicture}[baseline]
\matrix[matrix of math nodes, 
    every node/.append style={anchor=base,align=center,
    text depth=.5ex,text height=2ex,text width=1.5em},
    row sep=0pt,
    column sep=2pt,
    left delimiter={[},
    right delimiter={]},
    inner xsep=1pt
    ] (m) {
0 & 0 & 0 & 1 & 1 & 1 & 1 & 1 & 1 & 0 & 0 & 1 & 1 & 1 & 1 & 1 & 1 & 1 & 1 & 1 \\ 
0 & 0 & 0 & 1 & 1 & 1 & 0 & 0 & 0 & 0 & 0 & 1 & 1 & 1 & 1 & 0 & 0 & 0 & 0 & 0 \\ 
0 & 0 & 0 & 0 & 0 & 0 & 1 & 1 & 1 & 0 & 0 & 0 & 0 & 0 & 0 & 1 & 1 & 1 & 1 & 1 \\ 
0 & 0 & 1 & 0 & 0 & 0 & 0 & 0 & 0 & 1 & 1 & 0 & 0 & 0 & 0 & 0 & 0 & 0 & 0 & 0 \\ 
0 & 0 & -1 & 1 & 1 & 1 & 0 & 0 & 0 & -1 & -1 & 1 & 1 & 1 & 1 & 0 & 0 & 0 & 0 & 0 \\ 
0 & 0 & 0 & 1 & 1 & 1 & 0 & 0 & 0 & 0 & 0 & 0 & 0 & 0 & 0 & 0 & 0 & 0 & 0 & 0 \\ 
0 & 0 & 0 & 0 & 0 & 0 & 0 & 0 & 0 & 0 & 0 & 1 & 1 & 1 & 1 & 0 & 0 & 0 & 0 & 0 \\ 
-1 & -1 & 0 & 0 & 0 & 0 & 1 & 1 & 1 & 0 & 0 & 0 & 0 & 0 & 0 & 1 & 1 & 1 & 1 & 1 \\ 
0 & 0 & 0 & 0 & 0 & 0 & 1 & 0 & 0 & 0 & 0 & 0 & 0 & 0 & 0 & 1 & 1 & 1 & 0 & 0 \\ 
0 & 0 & 0 & 0 & 0 & 0 & 0 & 1 & 1 & 0 & 0 & 0 & 0 & 0 & 0 & 0 & 0 & 0 & 1 & 1 \\ 
0 & 0 & 0 & 0 & 0 & 0 & 0 & 0 & 0 & 1 & 1 & 0 & 0 & 0 & 0 & 0 & 0 & 0 & 0 & 0 \\ 
0 & 0 & 0 & 0 & 0 & 0 & 0 & 0 & 0 & 0 & 0 & 1 & 1 & 0 & 0 & 0 & 0 & 0 & 0 & 0 \\ 
0 & 0 & 0 & 0 & 0 & 0 & 0 & 0 & 0 & 0 & 0 & 0 & 0 & 1 & 1 & 0 & 0 & 0 & 0 & 0 \\ 
0 & 0 & 0 & 0 & 0 & 0 & 0 & 0 & 0 & 0 & 0 & 0 & 0 & 0 & 0 & 1 & 1 & 1 & 0 & 0 \\ 
0 & 0 & 0 & 0 & 0 & 0 & 0 & 0 & 0 & 0 & 0 & 0 & 0 & 0 & 0 & 0 & 0 & 0 & 1 & 1 \\ 
    };

 \matrix[matrix of math nodes, below = 1.1ex of m,
    every node/.append style={anchor=base,align=center,
    text depth=.5ex,text height=0.75ex,text width=1.5em, font=\tiny},
    row sep=0pt,
    column sep=2pt,
    inner xsep=1pt,
    above delimiter = |
    ] (m1) {
{F} & {G} & {AB} & {CA} & {CB} & {CC} & {HB} & {IB} & {IC} & {AAA} & {AAB} & {DAA} & {DAB} & {DBA} & {DBB} & {HAA} & {HAB} & {HAC} & {IAA} & {IAB} \\
    };
 
 \matrix[matrix of math nodes, right = 2ex of m,
    every node/.append style={anchor=base,align=left,
    text depth=.5ex,text height=2ex,text width=1.5em, font=\tiny},
    row sep=0pt,
    column sep=2pt,
    inner xsep=1pt
    ] (m2) {
{Z} \\ {X} \\ {Y} \\ {A} \\ {B} \\ {C} \\ {D} \\ {E} \\ {H} \\ {I} \\ {AA} \\ {DA} \\ {DB} \\ {HA} \\ {IA} \\
    };
\end{tikzpicture}
\end{adjustbox}.
$$
It's worth noting that $B$ and $E$, originally considered as bottom variables in hierarchies 1 and 3, respectively, in the general linear constrained representation are treated as constrained variables.

Consider now the situation in which the user works with a number of relationships larger than the minimum needed to define the set of linearly independent constraints. For example, the following 3 (redundant) relationships are added to previous ones:
\begin{itemize}[nosep]
	\item[6.1] $A = AA + AB$
	\item[9.1] $D = DA + DB$
	\item[14.1] $I = IA + IB + IC$
\end{itemize}
In this case, matrix $\Gammavet$ has 18 rows instead of 15, and 
$$
\Gammavet_{(18\times35)} = \begin{adjustbox}{width=0.8\textwidth}
\begin{tikzpicture}[baseline]
\matrix[matrix of math nodes, 
    every node/.append style={anchor=base,align=center,
    text depth=.5ex,text height=2ex,text width=1.5em},
    row sep=0pt,
    column sep=2pt,
    left delimiter={[},
    right delimiter={]},
    inner xsep=1pt
    ] (m) {
1 & -1 & -1 & 0 & 0 & 0 & 0 & 0 & 0 & 0 & 0 & 0 & 0 & 0 & 0 & 0 & 0 & 0 & 0 & 0 & 0 & 0 & 0 & 0 & 0 & 0 & 0 & 0 & 0 & 0 & 0 & 0 & 0 & 0 & 0 \\ 
0 & 1 & 0 & -1 & -1 & 0 & 0 & 0 & 0 & 0 & 0 & 0 & 0 & 0 & 0 & 0 & 0 & 0 & 0 & 0 & 0 & 0 & 0 & 0 & 0 & 0 & 0 & 0 & 0 & 0 & 0 & 0 & 0 & 0 & 0 \\ 
0 & 1 & 0 & 0 & 0 & -1 & -1 & 0 & 0 & 0 & 0 & 0 & 0 & 0 & 0 & 0 & 0 & 0 & 0 & 0 & 0 & 0 & 0 & 0 & 0 & 0 & 0 & 0 & 0 & 0 & 0 & 0 & 0 & 0 & 0 \\ 
0 & 0 & 1 & 0 & 0 & 0 & 0 & -1 & -1 & -1 & 0 & 0 & 0 & 0 & 0 & 0 & 0 & 0 & 0 & 0 & 0 & 0 & 0 & 0 & 0 & 0 & 0 & 0 & 0 & 0 & 0 & 0 & 0 & 0 & 0 \\ 
0 & 0 & 1 & 0 & 0 & 0 & 0 & 0 & 0 & 0 & -1 & -1 & 0 & 0 & 0 & 0 & 0 & 0 & 0 & 0 & 0 & 0 & 0 & 0 & 0 & 0 & 0 & 0 & 0 & 0 & 0 & 0 & 0 & 0 & 0 \\ 
0 & 0 & 0 & 1 & 0 & 0 & 0 & 0 & 0 & 0 & 0 & 0 & 0 & -1 & 0 & 0 & 0 & 0 & 0 & 0 & 0 & 0 & 0 & 0 & -1 & -1 & 0 & 0 & 0 & 0 & 0 & 0 & 0 & 0 & 0 \\ 
0 & 0 & 0 & 0 & 0 & 0 & 0 & 0 & 0 & 0 & 0 & 0 & 1 & 0 & 0 & 0 & 0 & 0 & 0 & 0 & 0 & 0 & 0 & 0 & -1 & -1 & 0 & 0 & 0 & 0 & 0 & 0 & 0 & 0 & 0 \\ 
0 & 0 & 0 & 0 & 0 & 1 & 0 & 0 & 0 & 0 & 0 & 0 & 0 & 0 & -1 & -1 & -1 & 0 & 0 & 0 & 0 & 0 & 0 & 0 & 0 & 0 & 0 & 0 & 0 & 0 & 0 & 0 & 0 & 0 & 0 \\ 
0 & 0 & 0 & 0 & 0 & 0 & 1 & 0 & 0 & 0 & 0 & 0 & 0 & 0 & 0 & 0 & 0 & 0 & 0 & 0 & 0 & 0 & 0 & 0 & 0 & 0 & -1 & -1 & -1 & -1 & 0 & 0 & 0 & 0 & 0 \\ 
0 & 0 & 0 & 0 & 0 & 0 & 0 & 0 & 0 & 0 & 0 & 0 & 0 & 0 & 0 & 0 & 0 & 1 & 0 & 0 & 0 & 0 & 0 & 0 & 0 & 0 & -1 & -1 & 0 & 0 & 0 & 0 & 0 & 0 & 0 \\ 
0 & 0 & 0 & 0 & 0 & 0 & 0 & 0 & 0 & 0 & 0 & 0 & 0 & 0 & 0 & 0 & 0 & 0 & 1 & 0 & 0 & 0 & 0 & 0 & 0 & 0 & 0 & 0 & -1 & -1 & 0 & 0 & 0 & 0 & 0 \\ 
0 & 0 & 0 & 0 & 0 & 0 & 0 & 0 & 0 & 0 & 1 & 0 & 0 & 0 & 0 & 0 & 0 & 0 & 0 & -1 & -1 & 0 & 0 & 0 & 0 & 0 & 0 & 0 & 0 & 0 & 0 & 0 & 0 & 0 & 0 \\ 
0 & 0 & 0 & 0 & 0 & 0 & 0 & 0 & 0 & 0 & 0 & 0 & 0 & 0 & 0 & 0 & 0 & 0 & 0 & 1 & 0 & 0 & 0 & 0 & 0 & 0 & 0 & 0 & 0 & 0 & -1 & -1 & -1 & 0 & 0 \\ 
0 & 0 & 0 & 0 & 0 & 0 & 0 & 0 & 0 & 0 & 0 & 1 & 0 & 0 & 0 & 0 & 0 & 0 & 0 & 0 & 0 & 0 & -1 & -1 & 0 & 0 & 0 & 0 & 0 & 0 & 0 & 0 & 0 & -1 & -1 \\ 
0 & 0 & 0 & 0 & 0 & 0 & 0 & 0 & 0 & 0 & 0 & 0 & 0 & 0 & 0 & 0 & 0 & 0 & 0 & 0 & 0 & 1 & 0 & 0 & 0 & 0 & 0 & 0 & 0 & 0 & 0 & 0 & 0 & -1 & -1 \\ 
0 & 0 & 0 & 1 & 0 & 0 & 0 & 0 & 0 & 0 & 0 & 0 & -1 & -1 & 0 & 0 & 0 & 0 & 0 & 0 & 0 & 0 & 0 & 0 & 0 & 0 & 0 & 0 & 0 & 0 & 0 & 0 & 0 & 0 & 0 \\ 
0 & 0 & 0 & 0 & 0 & 0 & 1 & 0 & 0 & 0 & 0 & 0 & 0 & 0 & 0 & 0 & 0 & -1 & -1 & 0 & 0 & 0 & 0 & 0 & 0 & 0 & 0 & 0 & 0 & 0 & 0 & 0 & 0 & 0 & 0 \\ 
0 & 0 & 0 & 0 & 0 & 0 & 0 & 0 & 0 & 0 & 0 & 1 & 0 & 0 & 0 & 0 & 0 & 0 & 0 & 0 & 0 & -1 & -1 & -1 & 0 & 0 & 0 & 0 & 0 & 0 & 0 & 0 & 0 & 0 & 0 \\ 
    };
\matrix[matrix of math nodes, below = 1.1ex of m,
    every node/.append style={anchor=base,align=center,
    text depth=.5ex,text height=0.75ex,text width=1.5em, font=\tiny},
    row sep=0pt,
    column sep=2pt,
    inner xsep=1pt,
    above delimiter = |
    ] (m1) {
{Z} & {X} & {Y} & {A} & {B} & {C} & {D} & {E} & {F} & {G} & {H} & {I} & {AA} & {AB} & {CA} & {CB} & {CC} & {DA} & {DB} & {HA} & {HB} & {IA} & {IB} & {IC} & {AAA} & {AAB} & {DAA} & {DAB} & {DBA} & {DBB} & {HAA} & {HAB} & {HAC} & {IAA} & {IAB} \\
    };
\end{tikzpicture}
\end{adjustbox}
$$
$$
\downarrow \mathrm{rref}
$$
$$
\phantom{\Gammavet_{(18\times35)} = }\begin{adjustbox}{width=0.8\textwidth}
\begin{tikzpicture}[baseline]
\matrix[matrix of math nodes, 
    every node/.append style={anchor=base,align=center,
    text depth=.5ex,text height=2ex,text width=1.5em},
    row sep=0pt,
    column sep=2pt,
    left delimiter={[},
    right delimiter={]},
    inner xsep=1pt
    ] (m) {
1 & 0 & 0 & 0 & 0 & 0 & 0 & 0 & 0 & 0 & 0 & 0 & 0 & 0 & -1 & -1 & -1 & 0 & 0 & 0 & -1 & 0 & -1 & -1 & 0 & 0 & -1 & -1 & -1 & -1 & -1 & -1 & -1 & -1 & -1 \\ 
0 & 1 & 0 & 0 & 0 & 0 & 0 & 0 & 0 & 0 & 0 & 0 & 0 & 0 & -1 & -1 & -1 & 0 & 0 & 0 & 0 & 0 & 0 & 0 & 0 & 0 & -1 & -1 & -1 & -1 & 0 & 0 & 0 & 0 & 0 \\ 
0 & 0 & 1 & 0 & 0 & 0 & 0 & 0 & 0 & 0 & 0 & 0 & 0 & 0 & 0 & 0 & 0 & 0 & 0 & 0 & -1 & 0 & -1 & -1 & 0 & 0 & 0 & 0 & 0 & 0 & -1 & -1 & -1 & -1 & -1 \\ 
0 & 0 & 0 & 1 & 0 & 0 & 0 & 0 & 0 & 0 & 0 & 0 & 0 & -1 & 0 & 0 & 0 & 0 & 0 & 0 & 0 & 0 & 0 & 0 & -1 & -1 & 0 & 0 & 0 & 0 & 0 & 0 & 0 & 0 & 0 \\ 
0 & 0 & 0 & 0 & 1 & 0 & 0 & 0 & 0 & 0 & 0 & 0 & 0 & 1 & -1 & -1 & -1 & 0 & 0 & 0 & 0 & 0 & 0 & 0 & 1 & 1 & -1 & -1 & -1 & -1 & 0 & 0 & 0 & 0 & 0 \\ 
0 & 0 & 0 & 0 & 0 & 1 & 0 & 0 & 0 & 0 & 0 & 0 & 0 & 0 & -1 & -1 & -1 & 0 & 0 & 0 & 0 & 0 & 0 & 0 & 0 & 0 & 0 & 0 & 0 & 0 & 0 & 0 & 0 & 0 & 0 \\ 
0 & 0 & 0 & 0 & 0 & 0 & 1 & 0 & 0 & 0 & 0 & 0 & 0 & 0 & 0 & 0 & 0 & 0 & 0 & 0 & 0 & 0 & 0 & 0 & 0 & 0 & -1 & -1 & -1 & -1 & 0 & 0 & 0 & 0 & 0 \\ 
0 & 0 & 0 & 0 & 0 & 0 & 0 & 1 & 1 & 1 & 0 & 0 & 0 & 0 & 0 & 0 & 0 & 0 & 0 & 0 & -1 & 0 & -1 & -1 & 0 & 0 & 0 & 0 & 0 & 0 & -1 & -1 & -1 & -1 & -1 \\ 
0 & 0 & 0 & 0 & 0 & 0 & 0 & 0 & 0 & 0 & 1 & 0 & 0 & 0 & 0 & 0 & 0 & 0 & 0 & 0 & -1 & 0 & 0 & 0 & 0 & 0 & 0 & 0 & 0 & 0 & -1 & -1 & -1 & 0 & 0 \\ 
0 & 0 & 0 & 0 & 0 & 0 & 0 & 0 & 0 & 0 & 0 & 1 & 0 & 0 & 0 & 0 & 0 & 0 & 0 & 0 & 0 & 0 & -1 & -1 & 0 & 0 & 0 & 0 & 0 & 0 & 0 & 0 & 0 & -1 & -1 \\ 
0 & 0 & 0 & 0 & 0 & 0 & 0 & 0 & 0 & 0 & 0 & 0 & 1 & 0 & 0 & 0 & 0 & 0 & 0 & 0 & 0 & 0 & 0 & 0 & -1 & -1 & 0 & 0 & 0 & 0 & 0 & 0 & 0 & 0 & 0 \\ 
0 & 0 & 0 & 0 & 0 & 0 & 0 & 0 & 0 & 0 & 0 & 0 & 0 & 0 & 0 & 0 & 0 & 1 & 0 & 0 & 0 & 0 & 0 & 0 & 0 & 0 & -1 & -1 & 0 & 0 & 0 & 0 & 0 & 0 & 0 \\ 
0 & 0 & 0 & 0 & 0 & 0 & 0 & 0 & 0 & 0 & 0 & 0 & 0 & 0 & 0 & 0 & 0 & 0 & 1 & 0 & 0 & 0 & 0 & 0 & 0 & 0 & 0 & 0 & -1 & -1 & 0 & 0 & 0 & 0 & 0 \\ 
0 & 0 & 0 & 0 & 0 & 0 & 0 & 0 & 0 & 0 & 0 & 0 & 0 & 0 & 0 & 0 & 0 & 0 & 0 & 1 & 0 & 0 & 0 & 0 & 0 & 0 & 0 & 0 & 0 & 0 & -1 & -1 & -1 & 0 & 0 \\ 
0 & 0 & 0 & 0 & 0 & 0 & 0 & 0 & 0 & 0 & 0 & 0 & 0 & 0 & 0 & 0 & 0 & 0 & 0 & 0 & 0 & 1 & 0 & 0 & 0 & 0 & 0 & 0 & 0 & 0 & 0 & 0 & 0 & -1 & -1 \\ 
0 & 0 & 0 & 0 & 0 & 0 & 0 & 0 & 0 & 0 & 0 & 0 & 0 & 0 & 0 & 0 & 0 & 0 & 0 & 0 & 0 & 0 & 0 & 0 & 0 & 0 & 0 & 0 & 0 & 0 & 0 & 0 & 0 & 0 & 0 \\ 
0 & 0 & 0 & 0 & 0 & 0 & 0 & 0 & 0 & 0 & 0 & 0 & 0 & 0 & 0 & 0 & 0 & 0 & 0 & 0 & 0 & 0 & 0 & 0 & 0 & 0 & 0 & 0 & 0 & 0 & 0 & 0 & 0 & 0 & 0 \\ 
0 & 0 & 0 & 0 & 0 & 0 & 0 & 0 & 0 & 0 & 0 & 0 & 0 & 0 & 0 & 0 & 0 & 0 & 0 & 0 & 0 & 0 & 0 & 0 & 0 & 0 & 0 & 0 & 0 & 0 & 0 & 0 & 0 & 0 & 0 \\ 
    };
\matrix[matrix of math nodes, below = 1.1ex of m,
    every node/.append style={anchor=base,align=center,
    text depth=.5ex,text height=0.75ex,text width=1.5em, font=\tiny},
    row sep=0pt,
    column sep=2pt,
    inner xsep=1pt
    ] (m1) {
{Z} & {X} & {Y} & {A} & {B} & {C} & {D} & {E} & {F} & {G} & {H} & {I} & {AA} & {AB} & {CA} & {CB} & {CC} & {DA} & {DB} & {HA} & {HB} & {IA} & {IB} & {IC} & {AAA} & {AAB} & {DAA} & {DAB} & {DBA} & {DBB} & {HAA} & {HAB} & {HAC} & {IAA} & {IAB} \\
    };
    \foreach \x in {1, 2, 3, 4, 5, 6, 7, 8, 11, 12, 13, 18, 19, 20, 22}
    	\node[fit=(m-1-\x)(m-15-\x), inner sep = 0, outer sep = 0, fill = red, opacity = 0.15] {};
    
    \foreach \x in {9, 10, 14, 15, 16, 17, 21, 23, 24, 25, 26, 27, 28, 29, 30, 31, 32, 33, 34, 35}
    	\node[fit=(m-1-\x)(m-15-\x), inner sep = 0, outer sep = 0, fill = blue, opacity = 0.15] {};
    
       \node[fit=(m-16-1)(m-18-35), dashed, draw, yshift = 0cm, inner sep = 0mm, outer sep = 0, line width=0.6mm] (zero){};
\end{tikzpicture}
\end{adjustbox}.
$$
In the end, the redundant relationships have the effect of adding three null rows to the rref of matrix $\Gammavet$, but the same free (20) and constrained (15) variables are found.

\clearpage
\section{Australian $GDP$ from income and expenditure sides}

\subsection{Australian National Accounts variables}
\begin{table}[H]
	\centering
	\renewcommand{\arraystretch}{1}
\spacingset{1}\centering\footnotesize
\begin{tabular*}{\columnwidth}[width = \textwidth]{llp{0.55\linewidth}}
		\toprule
		\textbf{Variable} & \textbf{Series ID} & \textbf{Description}\\
		\midrule
	Gdp            & A2302467A & GDP - Gross Domestic Product \\
    Sde                   & A2302566J & Statistical Discrepancy(E)\\
    Exp                   & A2302564C & Exports of goods and services\\
    Imp                   & A2302565F & Imports of goods and services\\
    Gne                   & A2302563A & Gross national exp.\\[1ex]
    GneDfdFceGvtNatDef    & A2302523J & Gen. gov. - National; Final consumption exp. - Defence\\
    GneDfdFceGvtNatNdf    & A2302524K & Gen. gov. - National; Final consumption exp. - Non-defence\\
    GneDfdFceGvtNat       & A2302525L & Gen. gov. - National; Final consumption exp.\\
    GneDfdFceGvtSnl       & A2302526R & Gen. gov. - State and local; Final consumption exp,\\
    GneDfdFceGvt          & A2302527T & Gen. gov.; Final consumption exp.\\[1ex]
    GneDfdFce             & A2302529W & All sectors; Final consumption exp.\\
    GneDfdGfcPvtTdwNnu    & A2302543T & Pvt.; Gross fixed capital formation (GFCF)\\
    GneDfdGfcPvtTdwAna    & A2302544V & Pvt.; GFCF - Dwellings - Alterations and additions\\
    GneDfdGfcPvtTdw       & A2302545W & Pvt.; GFCF - Dwellings - Total\\
    GneDfdGfcPvtOtc       & A2302546X & Pvt.; GFCF - Ownership transfer costs\\[1ex]
    GneDfdGfcPvtPbiNdcNbd & A2302533L & Pvt. GFCF - Non-dwelling construction - New building\\
    GneDfdGfcPvtPbiNdcNec & A2302534R & Pvt.; GFCF - Non-dwelling construction - New engineering construction\\
    GneDfdGfcPvtPbiNdcSha & A2302535T & Pvt.; GFCF - Non-dwelling construction - Net purchase of second hand assets\\[1ex]
    GneDfdGfcPvtPbiNdc    & A2302536V & Pvt.; GFCF - Non-dwelling construction - Total\\
    GneDfdGfcPvtPbiNdmNew & A2302530F & Pvt.; GFCF - Machinery and equipment - New\\
    GneDfdGfcPvtPbiNdmSha & A2302531J & Pvt.; GFCF - Machinery and equipment - Net purchase of second hand assets\\
    GneDfdGfcPvtPbiNdm    & A2302532K & Pvt.; GFCF - Machinery and equipment - Total\\[1ex]
    GneDfdGfcPvtPbiCbr    & A2716219R & Pvt.; GFCF - Cultivated biological resources\\
    GneDfdGfcPvtPbiIprRnd & A2716221A & Pvt.; GFCF - Intellectual property products - Research and development\\
    GneDfdGfcPvtPbiIprMnp & A2302539A & Pvt.; GFCF - Intellectual property products - Mineral and petroleum exploration\\[2ex]
    GneDfdGfcPvtPbiIprCom & A2302538X & Pvt.; GFCF - Intellectual property products - Computer software\\
    GneDfdGfcPvtPbiIprArt & A2302540K & Pvt.; GFCF - Intellectual property products - Artistic originals\\
    GneDfdGfcPvtPbiIpr    & A2716220X & Pvt.; GFCF - Intellectual property products Total\\
    GneDfdGfcPvtPbi       & A2302542R & Pvt.;  GFCF - Total private business investment\\
    GneDfdGfcPvt          & A2302547A & Pvt.; GFCF\\[1ex]
    GneDfdGfcPubPcpCmw    & A2302548C & Plc. corporations - Commonwealth; GFCF\\
    GneDfdGfcPubPcpSnl    & A2302549F & Plc. corporations - State and local; GFCF\\
    GneDfdGfcPubPcp       & A2302550R & Plc. corporations; GFCF Total\\
    GneDfdGfcPubGvtNatDef & A2302551T & Gen. gov. - National; GFCF - Defence\\
    GneDfdGfcPubGvtNatNdf & A2302552V & Gen. gov. - National ; GFCF - Non-defence\\[1ex]
    GneDfdGfcPubGvtNat    & A2302553W & Gen. gov. - National ; GFCF Total\\
    GneDfdGfcPubGvtSnl    & A2302554X & Gen. gov. - State and local; GFCF\\
    GneDfdGfcPubGvt       & A2302555A & Gen. gov.; GFCF\\
    GneDfdGfcPub          & A2302556C & Plc.; GFCF\\
    GneDfdGfc             & A2302557F & All sectors; GFCF\\
		\bottomrule
	\end{tabular*}
	\caption{Variables, series IDs and their descriptions for the expenditure approach. Source: Athanasopoulos, G., P. Gamakumara, A. Panagiotelis, R.J. Hyndman, and M. Affan. 2020. Hierarchical Forecasting, In Macroeconomic Forecasting in the Era of Big Data, ed. Fuleky, P., Volume 52, 689--719. Cham: Springer International Publishing. \doi{10.1007/978-3-030-31150-6 21}.}
	\label{Tab:expenditure-hierarchy-1}
\end{table}

\begin{table}[H]
	\centering
	\renewcommand{\arraystretch}{1}
\spacingset{1}\centering\footnotesize
\begin{tabular*}{\columnwidth}[width = \textwidth]{llp{0.55\linewidth}}
		\toprule
		\textbf{Variable} & \textbf{Series ID} & \textbf{Description}\\
		\midrule
    GneDfdHfc          & A2302254W & Household Final Consumption expenditure\\
    GneDfdFceHfcFud    & A2302237V & Food\\
    GneDfdFceHfcAbt    & A3605816F & Alcoholic beverages and tobacco\\
    GneDfdFceHfcAbtCig & A2302238W & Cigarettes and tobacco\\
    GneDfdFceHfcAbtAlc & A2302239X & Alcoholic beverages\\[1ex]
    GneDfdFceHfcCnf    & A2302240J & Clothing and footwear\\
    GneDfdFceHfcHwe    & A3605680F & Housing, water, electricity, gas and other fuels\\
    GneDfdFceHfcHweRnt & A3605681J & Actual and imputed rent for housing\\
    GneDfdFceHfcHweWsc & A3605682K & Water and sewerage charges\\
    GneDfdFceHfcHweEgf & A2302242L & Electricity, gas and other fuel\\[1ex]
    GneDfdFceHfcFhe    & A2302243R & Furnishings and household equipment\\
    GneDfdFceHfcFheFnt & A3605683L & Furniture, floor coverings and household goods\\
    GneDfdFceHfcFheApp & A3605684R & Household appliances\\
    GneDfdFceHfcFheTls & A3605685T & Household tools\\
    GneDfdFceHfcHlt    & A2302244T & Health\\[2ex]
    GneDfdFceHfcHltMed & A3605686V & Medicines, medical aids and therapeutic appliances\\
    GneDfdFceHfcHltHsv & A3605687W & Total health services\\
    GneDfdFceHfcTpt    & A3605688X & Transport\\
    GneDfdFceHfcTptPvh & A2302245V & Purchase of vehicles\\
    GneDfdFceHfcTptOvh & A2302246W & Operation of vehicles\\[1ex]
    GneDfdFceHfcTptTsv & A2302247X & Transport services\\
    GneDfdFceHfcCom    & A2302248A & Communications\\
    GneDfdFceHfcRnc    & A2302249C & Recreation and culture\\
    GneDfdFceHfcEdc    & A2302250L & Education services\\
    GneDfdFceHfcHcr    & A2302251R & Hotels, cafes and restaurants\\[1ex]
    GneDfdFceHfcHcrCsv & A3605694V & Catering services\\
    GneDfdFceHfcHcrAsv & A3605695W & Accommodation services\\
    GneDfdFceHfcMis    & A3605696X & Miscellaneous goods and services\\
    GneDfdFceHfcMisOgd & A3605697A & Other goods\\
    GneDfdFceHfcMisIfs & A2302252T & Insurance and other financial services\\
    GneDfdFceHfcMisOsv & A3606485T & Other services\\
		\bottomrule
	\end{tabular*}
	\label{Tab:expenditure-hierarchy-3}
	\caption{Variables, series IDs and their descriptions for Household Final Consumption - expenditure approach. Source: Athanasopoulos, G., P. Gamakumara, A. Panagiotelis, R.J. Hyndman, and M. Affan. 2020. Hierarchical Forecasting, In Macroeconomic Forecasting in the Era of Big Data, ed. Fuleky, P., Volume 52, 689--719. Cham: Springer International Publishing. \doi{10.1007/978-3-030-31150-6 21}.}
\end{table}

\begin{table}[H]
	\centering
	\renewcommand{\arraystretch}{1}
\spacingset{1}\centering\footnotesize
\begin{tabular*}{\columnwidth}[width = \textwidth]{llp{0.55\linewidth}}
		\toprule
		\textbf{Variable} & \textbf{Series ID} & \textbf{Description}\\
		\midrule
    GneCii       & A2302562X  & Changes in Inventories\\
    GneCiiPfm    & A2302560V  & Farm\\
    GneCiiPba    & A2302561W  & Public authorities\\
    GneCiiPnf    & A2302559K  & Private; Non-farm Total\\
    GneCiiPnfMin & A83722619L & Private; Mining (B)\\[1ex]
    GneCiiPnfMan & A3348511X  & Private; Manufacturing (C)\\
    GneCiiPnfWht & A3348512A  & Private; Wholesale trade (F)\\
    GneCiiPnfRet & A3348513C  & Private; Retail trade (G)\\
    GneCiiPnfOnf & A2302273C  & Private; Non-farm; Other non-farm industries\\
		\bottomrule
	\end{tabular*}
	\label{Tab:expenditure-hierarchy-2}
	\caption{Variables, series IDs and their descriptions for Changes in Inventories - expenditure approach. Source: Athanasopoulos, G., P. Gamakumara, A. Panagiotelis, R.J. Hyndman, and M. Affan. 2020. Hierarchical Forecasting, In Macroeconomic Forecasting in the Era of Big Data, ed. Fuleky, P., Volume 52, 689--719. Cham: Springer International Publishing. \doi{10.1007/978-3-030-31150-6 21}.}
\end{table}

\begin{table}[H]
	\centering
	\renewcommand{\arraystretch}{1}
\spacingset{1}\centering\footnotesize
\begin{tabular*}{\columnwidth}[width = \textwidth]{llp{0.55\linewidth}}
		\toprule
		\textbf{Variable} & \textbf{Series ID} & \textbf{Description}\\
		\midrule
    Sdi             & A2302413V & Statistical discrepancy (I)\\
    Tsi             & A2302412T & Taxes less subsidies (I)\\
    TfiCoeWns       & A2302399K & Compensation of employees; Wages and salaries\\
    TfiCoeEsc       & A2302400J & Compensation of employees; Employers' social contributions\\[1ex]
    TfiCoe          & A2302401K & Compensation of employees\\
    TfiGosCopNfnPvt & A2323369L & Private non-financial corporations; Gross operating surplus\\
    TfiGosCopNfnPub & A2302403R & Public non-financial corporations; Gross operating surplus\\
    TfiGosCopNfn    & A2302404T & Non-financial corporations; Gross operating surplus\\
    TfiGosCopFin    & A2302405V & Financial corporations; Gross operating surplus\\[1ex]
    TfiGosCop       & A2302406W & Total corporations; Gross operating surplus\\
    TfiGosGvt       & A2298711F & General government; Gross operating surplus\\
    TfiGosDwl       & A2302408A & Dwellings owned by persons; Gross operating surplus\\
    TfiGos          & A2302409C & All sectors; Gross operating surplus\\
    TfiGmi          & A2302410L & Gross mixed income\\
    Tfi             & A2302411R & Total factor income\\
		\bottomrule
	\end{tabular*}
	\label{Tab:income-hierarchy}
	\caption{Variables, series IDs and their descriptions for the income approach. Source: 
	Athanasopoulos, G., P. Gamakumara, A. Panagiotelis, R.J. Hyndman, and M. Affan. 2020. Hierarchical Forecasting, In Macroeconomic Forecasting in the Era of Big Data, ed. Fuleky, P., Volume 52, 689--719. Cham: Springer International Publishing. \doi{10.1007/978-3-030-31150-6 21}.}
\end{table}

\newpage
\subsection{Accuracy indices}

\begin{figure}[H]
	\centering
	\includegraphics[width = \linewidth]{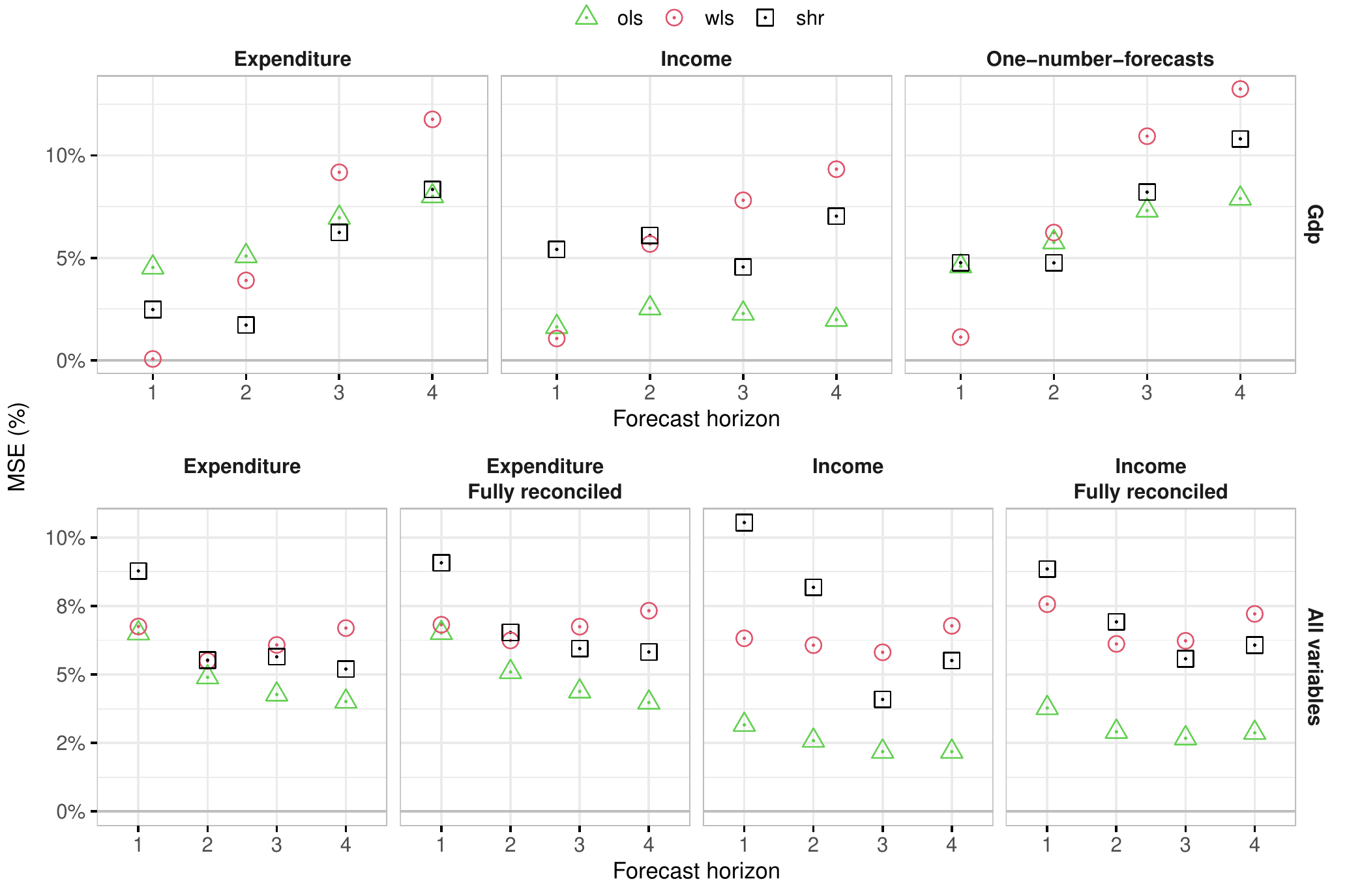}
	\caption{MSE-skill scores (relative to base forecasts) for point forecasts from alternative reconciliation approaches (Australian QNA variables).}
	\label{img:AUS_mse}
\end{figure}

\begin{figure}[H]
	\centering
	\includegraphics[width = \linewidth]{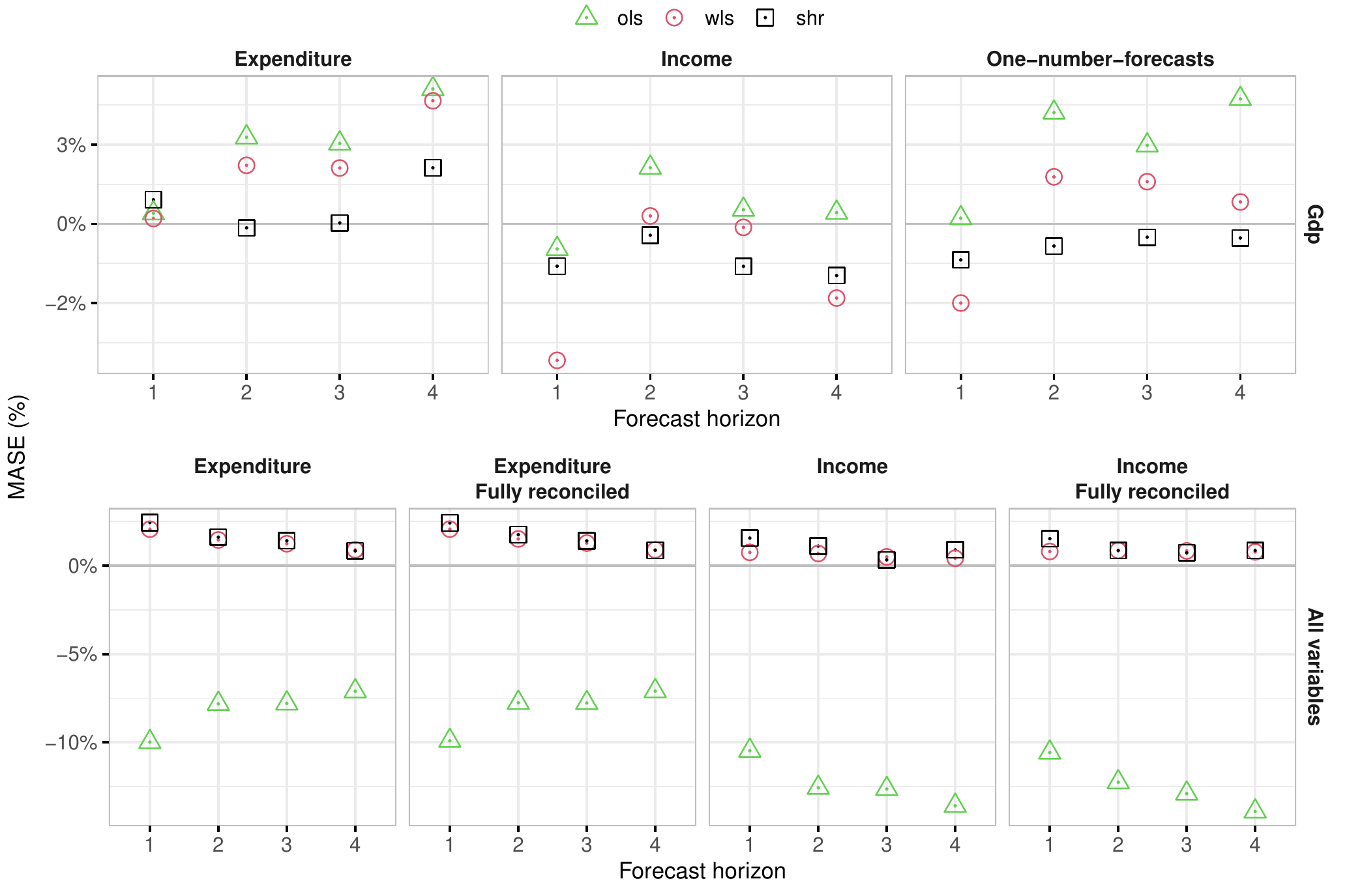}
	\caption{MASE-skill scores (relative to base forecasts) for point forecasts from alternative reconciliation approaches (Australian QNA variables).}
	\label{img:AUS_mase}
\end{figure}

\begin{figure}[p]
	\centering
	\begin{subfigure}[b]{\textwidth}
         \centering
         \includegraphics[width = \linewidth]{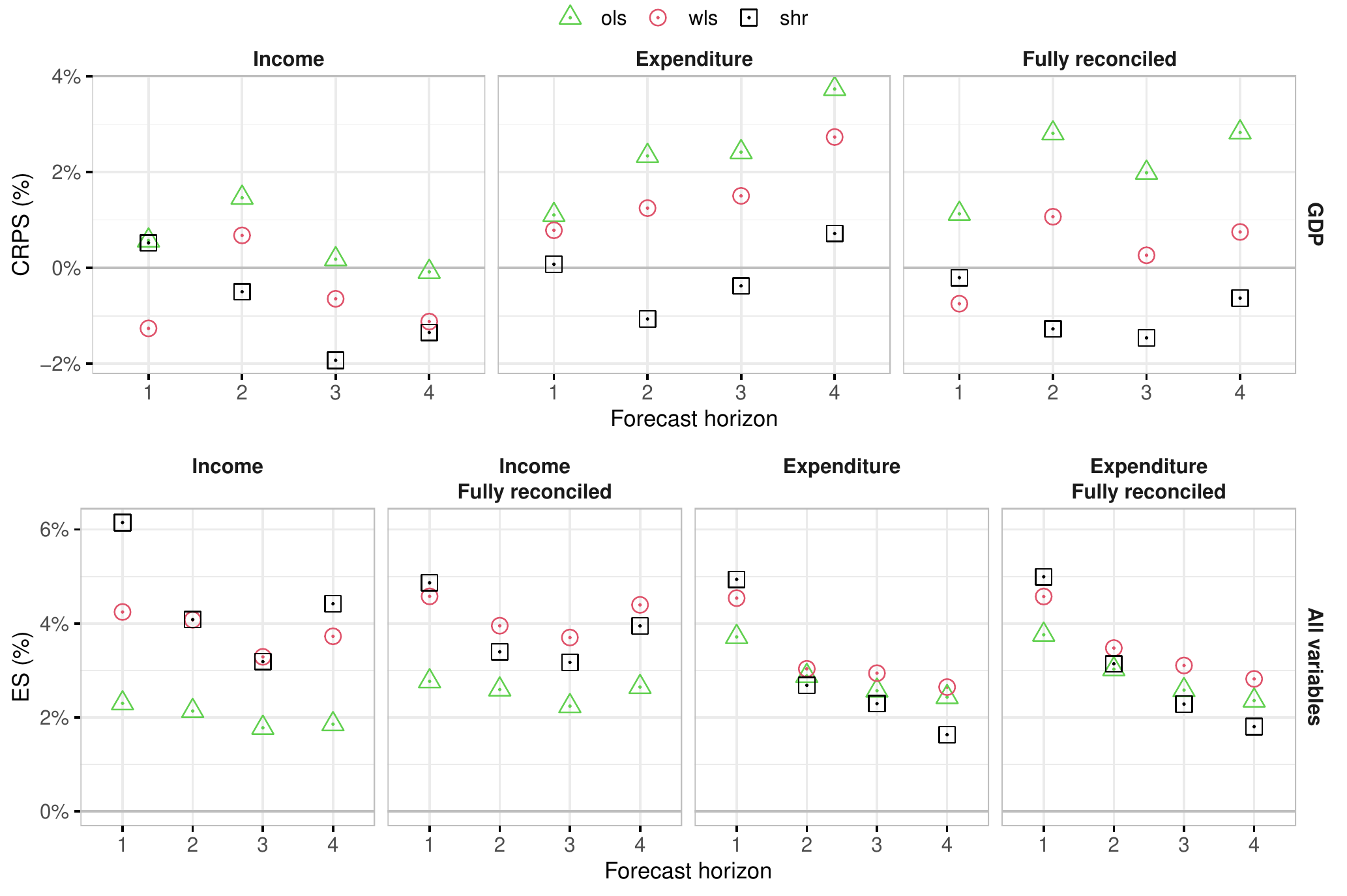}
         \vspace{-0.65cm}
         \caption{Non parametric framework}
         \label{img:AUS_es_boot}
         \vspace{-0.15cm}
     \end{subfigure}
	\begin{subfigure}[b]{\textwidth}
         \centering
         \includegraphics[width = \linewidth]{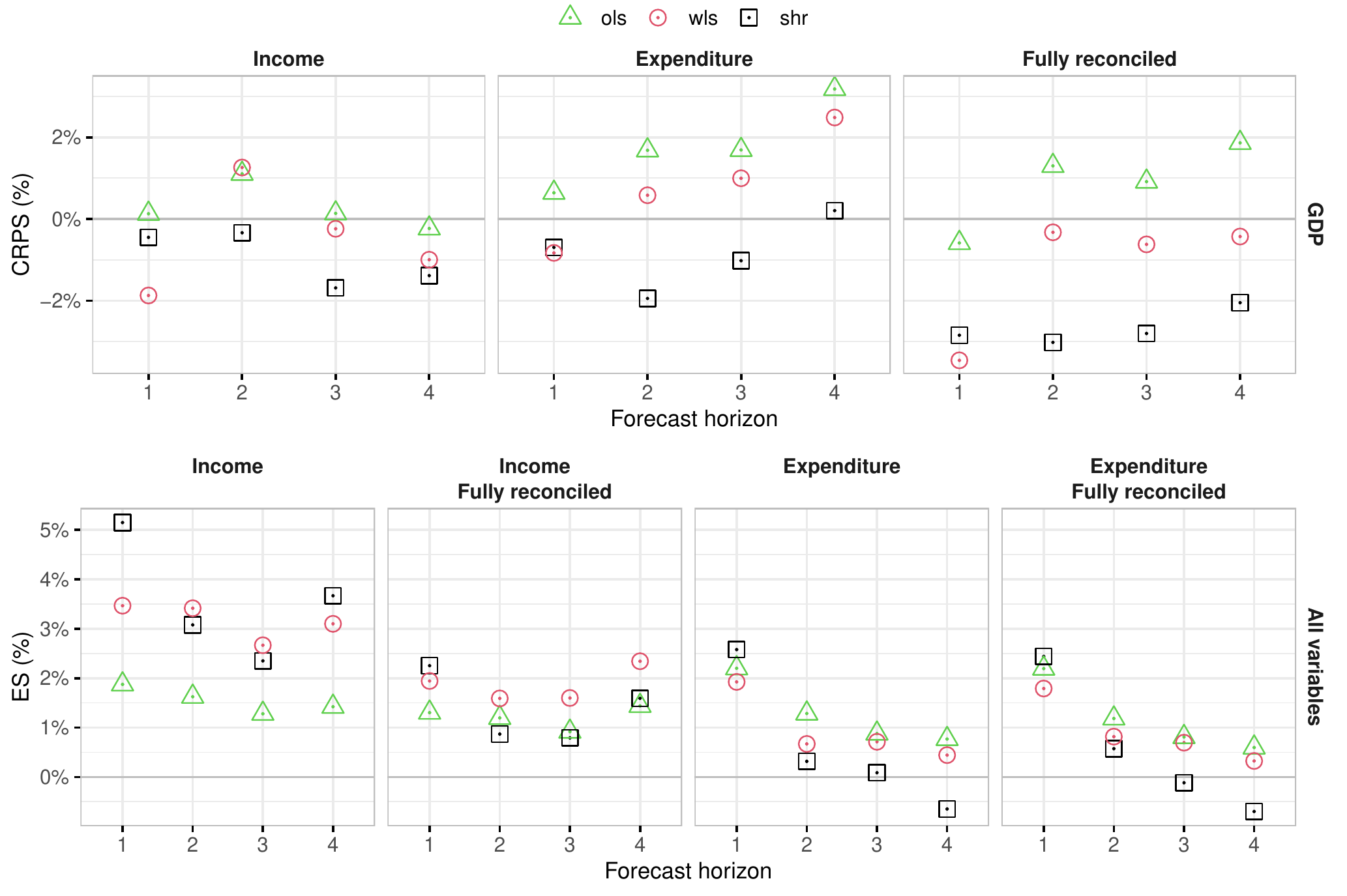}
         \vspace{-0.65cm}
         \caption{Gaussian framework}
         \label{img:AUS_es_gauss}
         \vspace{-0.15cm}
     \end{subfigure}
	\caption{CRPS and ES-skill scores (relative to base forecasts) for the probabilistic forecasts from alternative reconciliation approaches (Australian QNA variables).}
	\label{img:AUS_es}
	\vspace*{-\baselineskip}
\end{figure}

\begin{figure}[H]
	\centering
	\begin{subfigure}[b]{\textwidth}
         \centering
         \includegraphics[width = \linewidth]{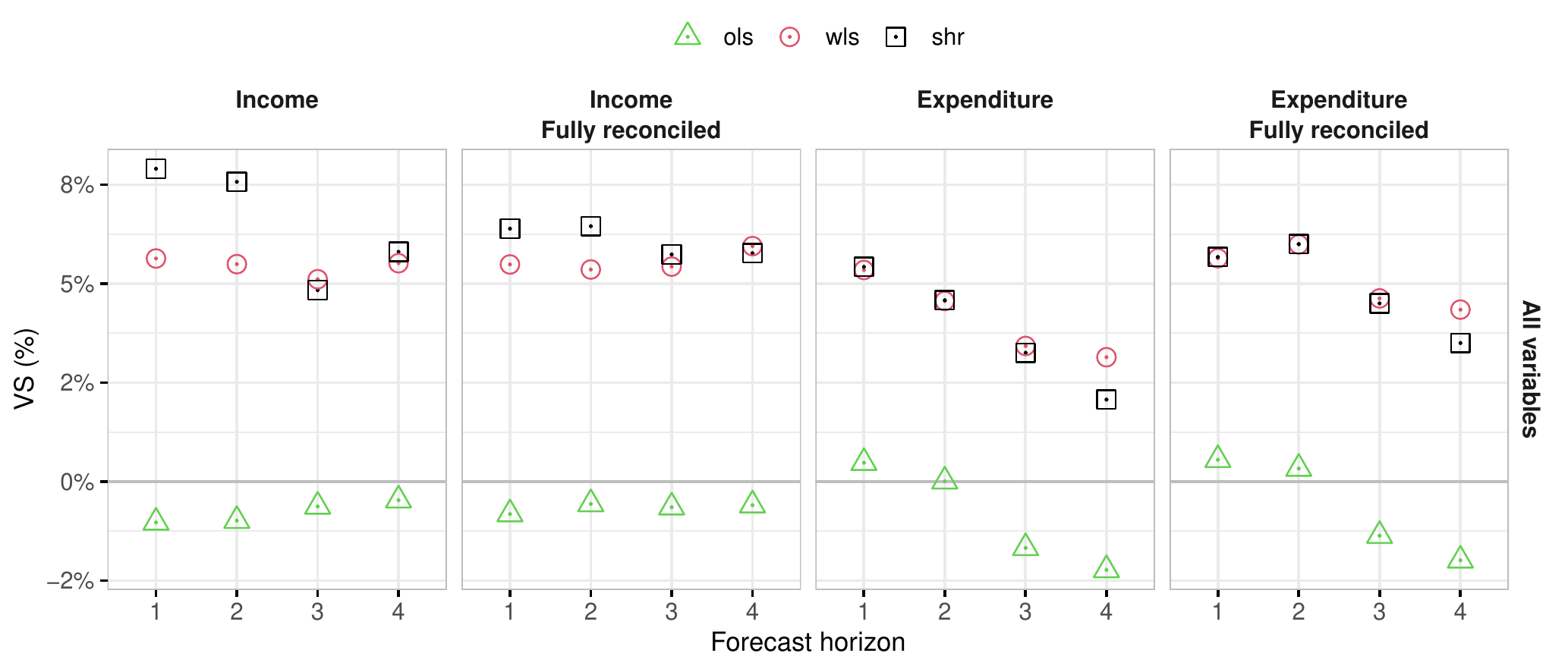}
         \caption{Non parametric joint bootstrap}
         \label{fig:AUS_vs_boot}
     \end{subfigure}
	\begin{subfigure}[b]{\textwidth}
         \centering
         \includegraphics[width = \linewidth]{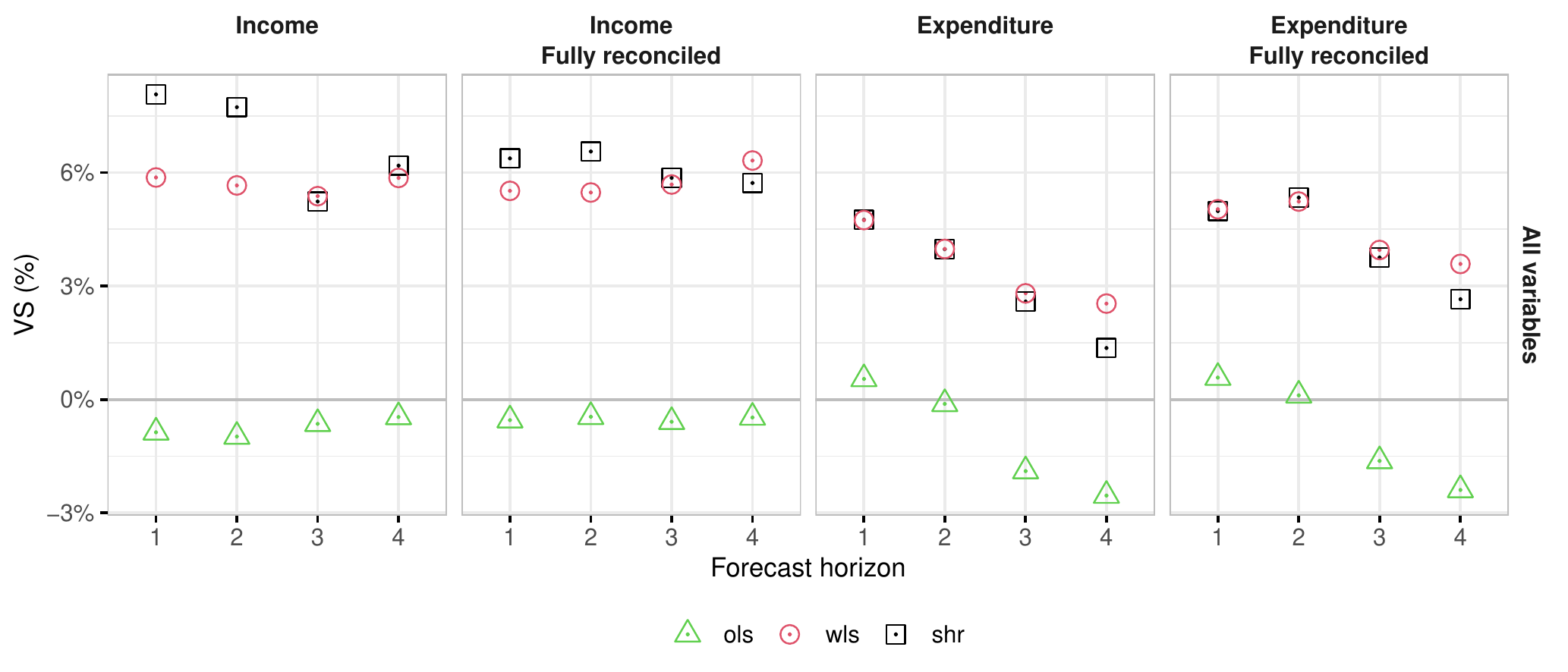}
         \caption{Gaussian distribution}
         \label{fig:AUS_vs_gauss}
     \end{subfigure}
	\caption{VS-skill scores (relative to base forecasts) for the probabilistic forecasts from alternative reconciliation approaches (Australian QNA variables).}
	\label{img:AUS_vs}
\end{figure}

\section{European Area $GDP$ from output, income and expenditure sides}

\subsection{European National Accounts variables}

\begin{table}[H]
\centering
\renewcommand{\arraystretch}{1}
\spacingset{1}\centering\footnotesize
	\begin{tabular}{r|p{0.65\linewidth}}
	\toprule
	\textbf{ID} & \textbf{Description} \\
	\midrule
	GDP & Gross domestic product \\
	\cmidrule{2-2}
	\multicolumn{2}{c}{\textit{expenditure side}}\\
	P3 & Final consumption expenditure \\
 	P3\_S13 & Final consumption expenditure of general government \\
 	P31\_S141\_S15 & Household and NPISH final consumption expenditure \\
 	P31\_S13 & Individual consumption expenditure of general government \\
 	P5G & Gross capital formation \\
 	P521\_P53 & Changes in inventories and acquisitions less disposals of valuables \\
 	P6 & Exports of goods and services \\
 	P7 & Imports of goods and services \\
 	B11 & External balance of goods and services\\
 	B111 & External balance - Goods\\
 	B112 & External balance - Services \\
 	P3\_P5 & Final consumption expenditure and gross capital formation \\
	P3\_P6 & Final consumption expenditure, gross capital formation and exports of goods and services \\
	P32\_S13 & Collective consumption expenditure of general government\\
	P31\_S14 & Final consumption expenditure of households\\
	P31\_S15 & Final consumption expenditure of NPISH\\
	P51G & Gross fixed capital formation\\
	P52 & Changes in inventories\\
	P53 & Acquisitions less disposals of valuables\\
	P61 & Exports of goods\\
	P62 & Exports of services\\
	P71 & Imports of goods\\
	P72 & Imports of services\\
	YA0 & Statistical discrepancy (expenditure approach)\\
	\cmidrule{2-2}
	\multicolumn{2}{c}{\textit{income side}}\\
	D1 & Compensation of employees\\
	D2X3 & Taxes on production and imports less subsidies\\
	D11 & Wages and salaries\\
	D12 & Employers' social contributions\\
	B2A3G & Operating surplus and mixed income, gross\\
	D2 & Taxes on production and imports\\
	D3 & Subsidies\\
	YA2 & Statistical discrepancy (income approach)\\
	\cmidrule{2-2}
	\multicolumn{2}{c}{\textit{output side}}\\
	D21X31 & Taxes less subsidies on products\\
	B1G & Total gross value added\\
	YA1 & Statistical discrepancy (output/production approach)\\
	\bottomrule
	\end{tabular}
\caption{List of the European National Accounts variables (ESA 2010 classification). \\ Source: \url{https://ec.europa.eu/ eurostat/web/national- accounts/data/database}}
\label{tab:all_GDP}
\end{table}

\subsection{Accuracy indices}
\begin{figure}[H]
	\centering
	\begin{subfigure}[b]{\textwidth}
         \centering
         \includegraphics[width = \linewidth]{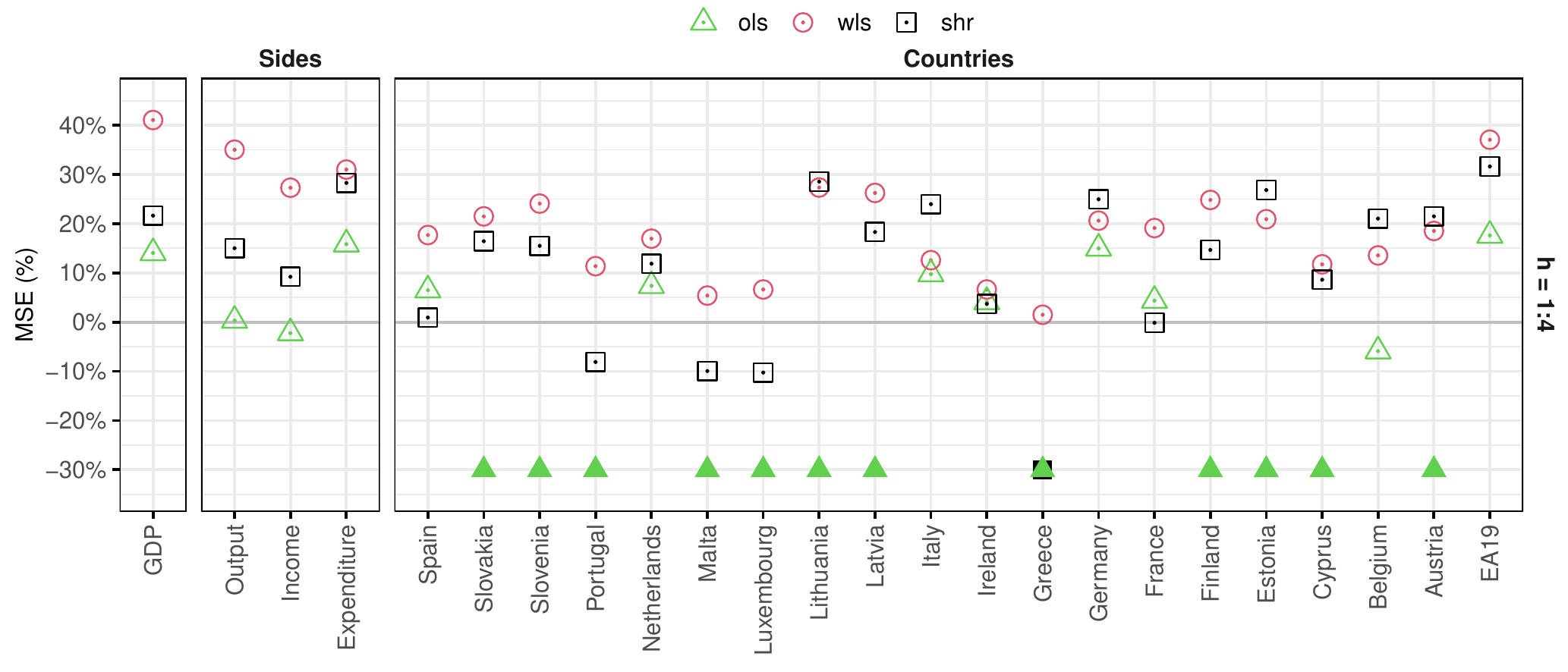}
         \vspace{-0.75cm}
         \caption{Point forecasts}
         \label{img:EA_es_point}
         \vspace{-0.15cm}
     \end{subfigure}
	\begin{subfigure}[b]{\textwidth}
         \centering
         \includegraphics[width = \linewidth]{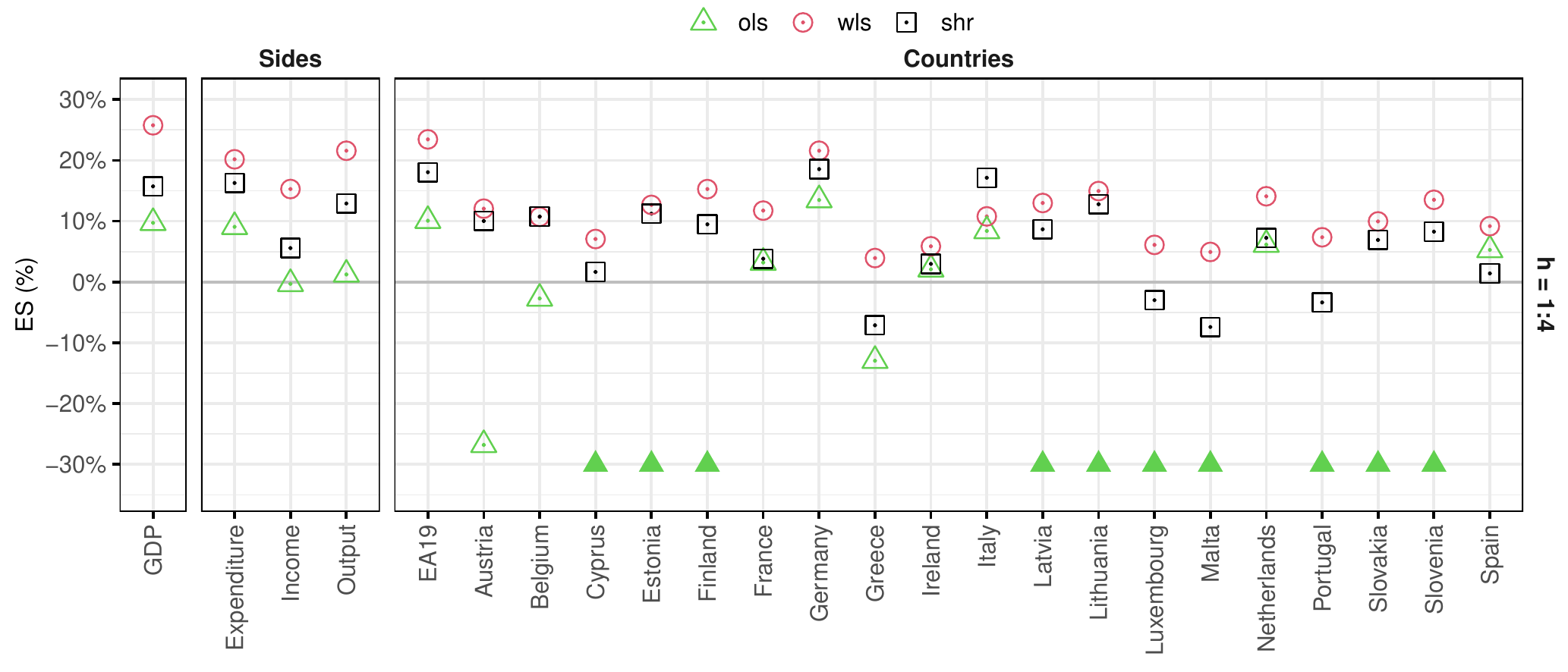}
         \vspace{-0.75cm}
         \caption{Non parametric joint bootstrap probabilistic forecasts}
         \label{img:EA_es_boot}
         \vspace{-0.15cm}
     \end{subfigure}
	\begin{subfigure}[b]{\textwidth}
         \centering
         \includegraphics[width = \linewidth]{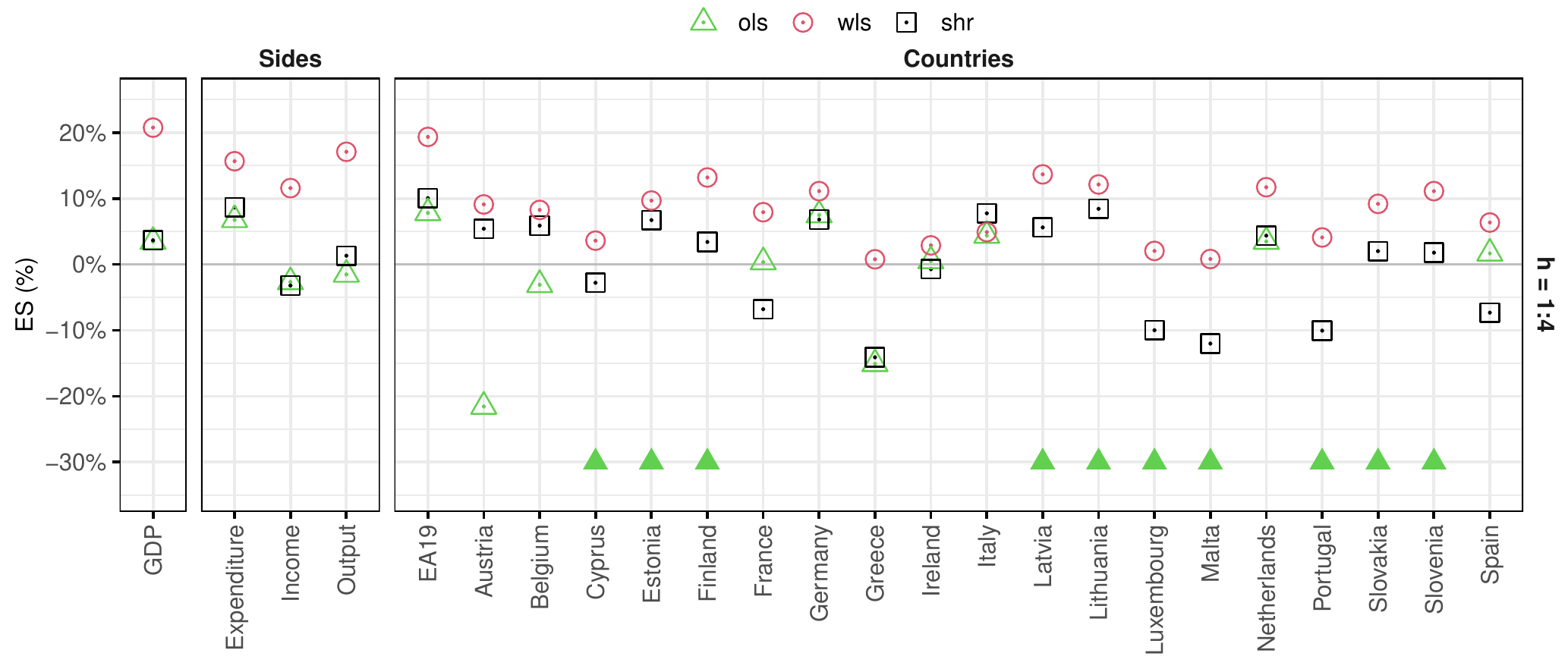}
         \vspace{-0.75cm}
         \caption{Gaussian probabilistic forecasts}
         \label{img:EA_es_gauss}
         \vspace{-0.15cm}
     \end{subfigure}
	\caption{MSE and ES-skill scores (relative to base forecasts) for the point and probabilistic forecasts from alternative reconciliation approaches (European Area QNA). To make the figure more readable, the filled symbols indicate that the skill score is less than $-30$\%.}
	\label{img:EA_es}
\end{figure}

\begin{figure}[H]
	\centering
	\begin{subfigure}[b]{\textwidth}
         \centering
         \includegraphics[width = \linewidth]{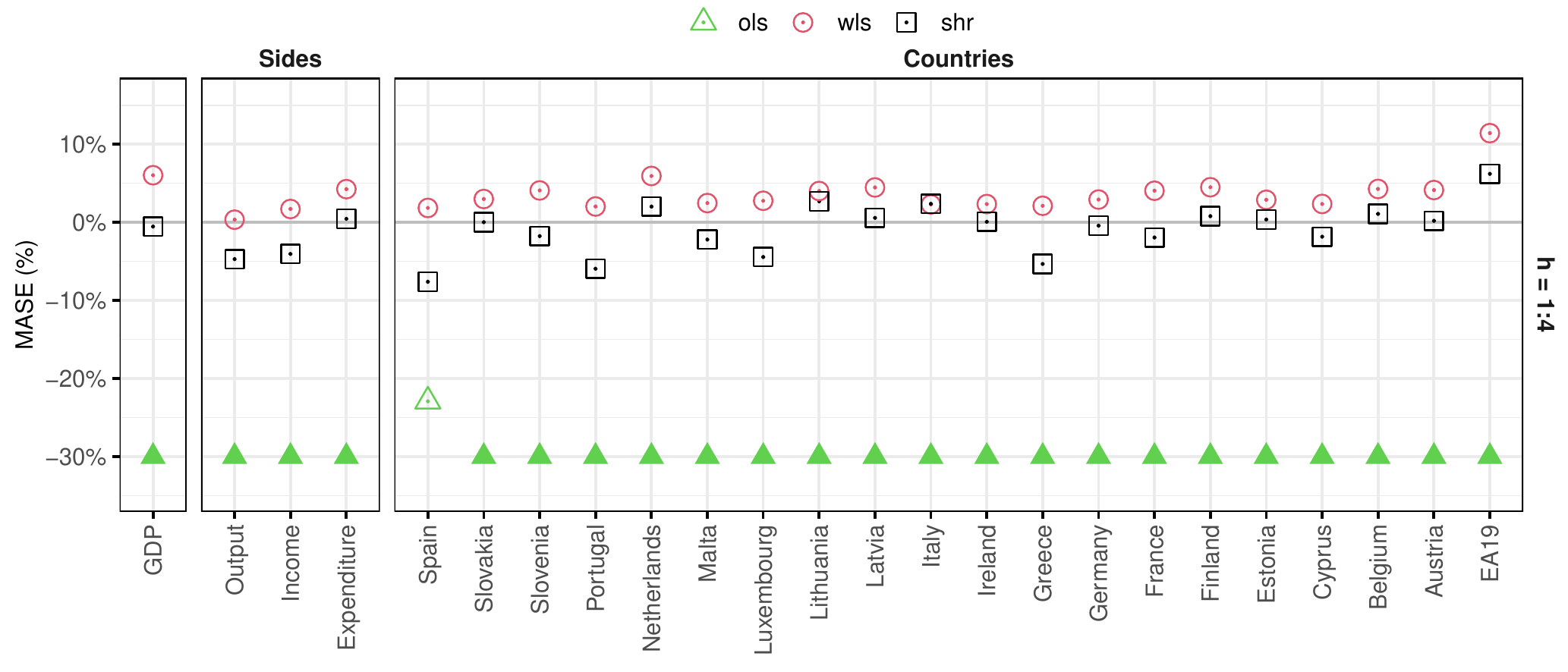}
         \caption{Point forecasts}
         \label{fig:EA_mase_point}
     \end{subfigure}
	\begin{subfigure}[b]{\textwidth}
         \centering
         \includegraphics[width = \linewidth]{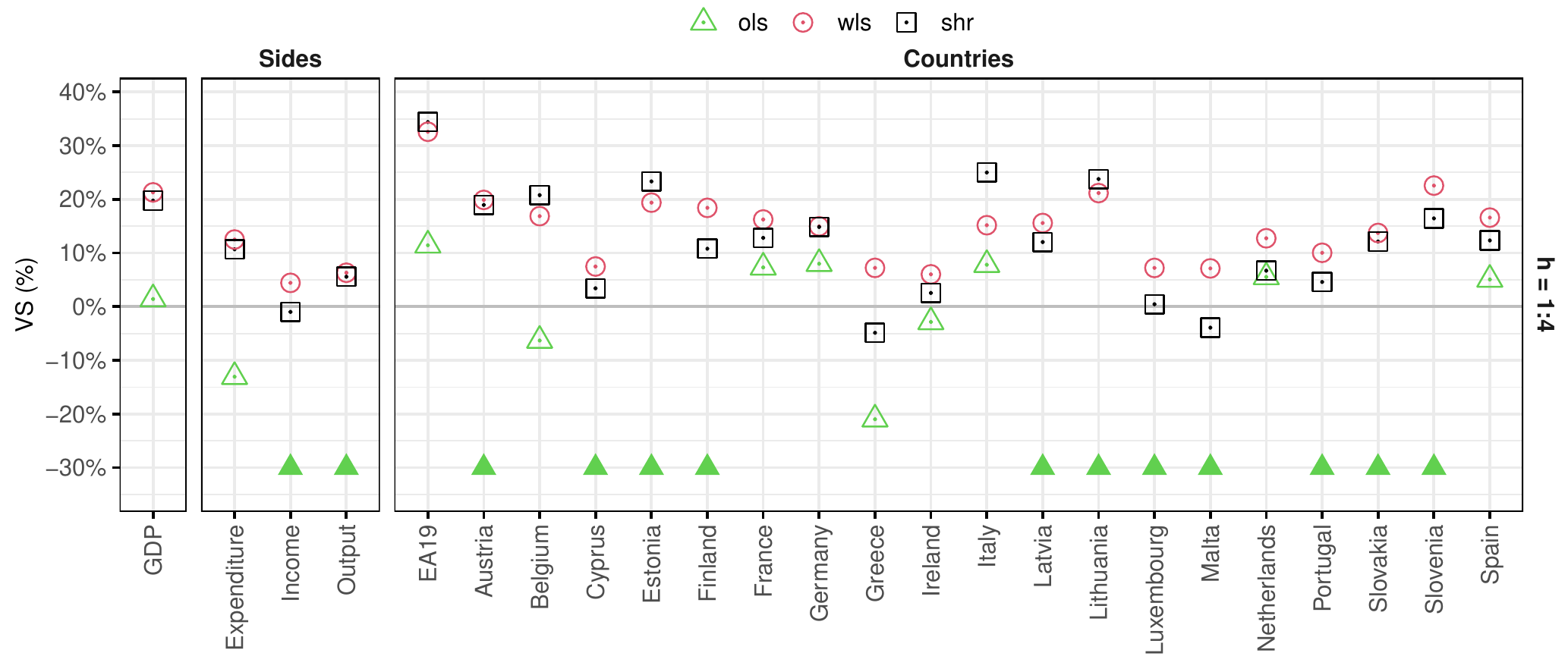}
         \caption{Non parametric joint bootstrap probabilistic forecasts}
         \label{fig:EA_vs_boot}
     \end{subfigure}
	\begin{subfigure}[b]{\textwidth}
         \centering
         \includegraphics[width = \linewidth]{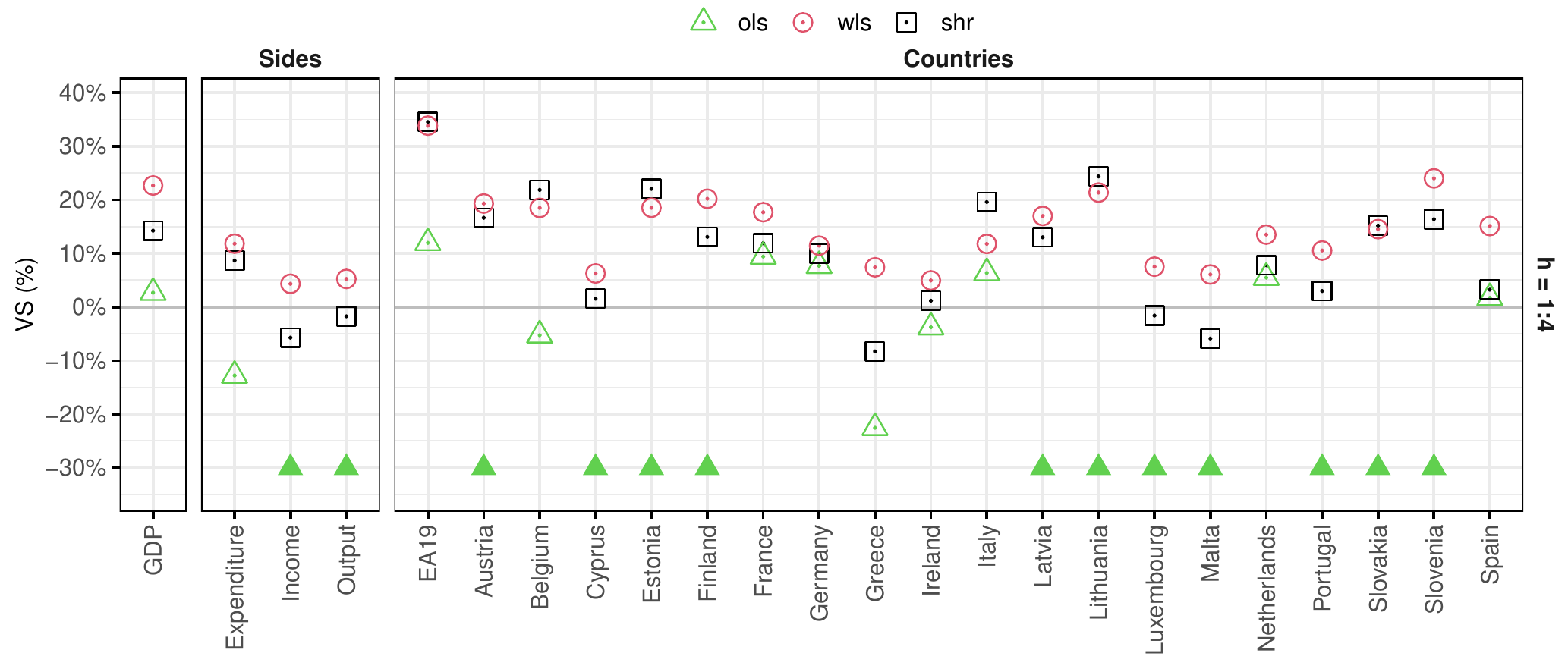}
         \caption{Gaussian probabilistic forecasts}
         \label{fig:EA_vs_gauss}
     \end{subfigure}
	\caption{MASE and VS-skill scores (relative to base forecasts) for the point and probabilistic forecasts from alternative reconciliation approaches (European Area QNA). To make the figure more readable, the filled symbols indicate that the skill score is less than $-30$\%.}
	\label{img:EA_vs}
\end{figure}

\begin{figure}[p]
	\centering
	\includegraphics[width = \linewidth]{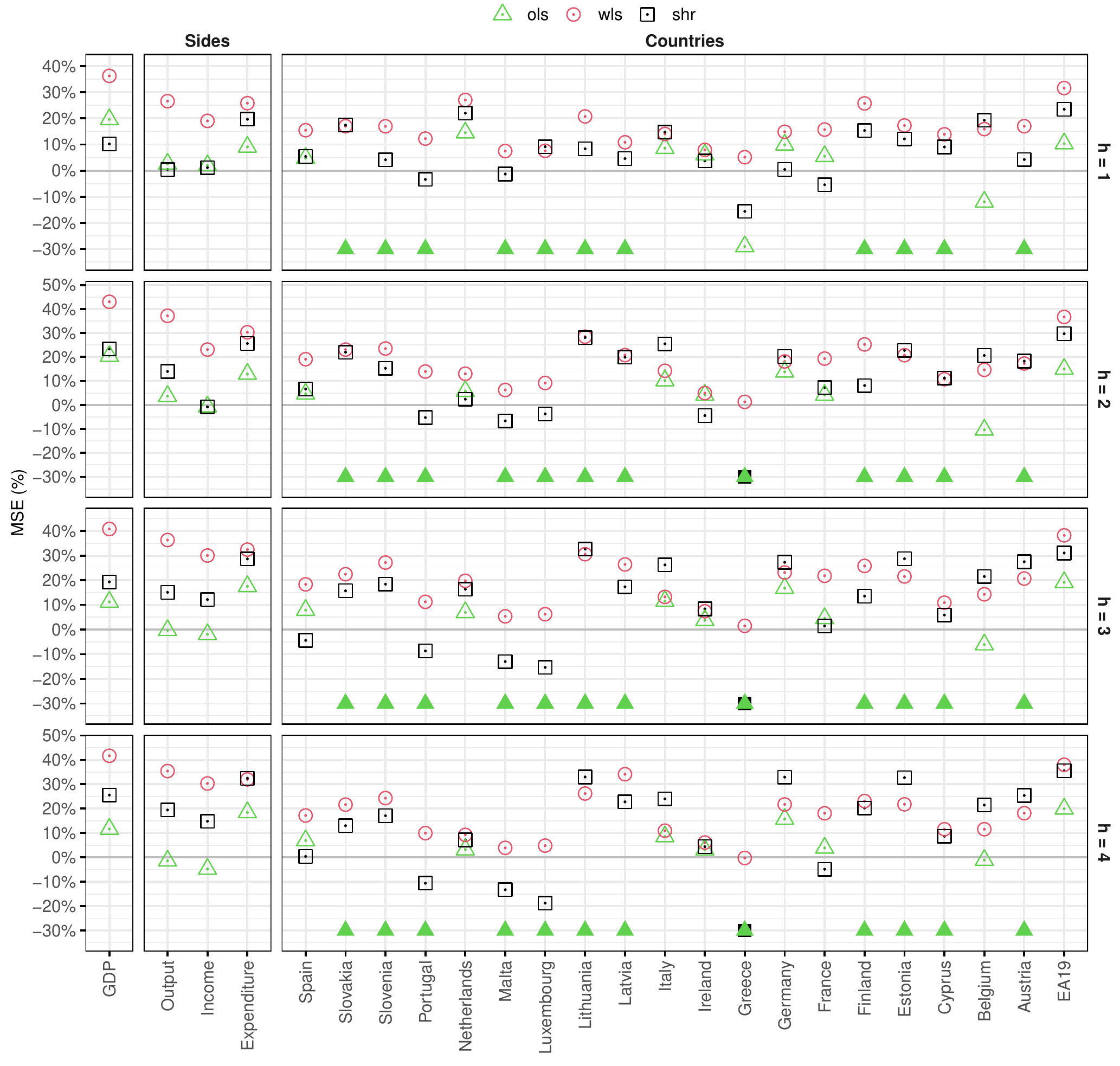}
	\caption{MSE-skill scores for the point forecasts from alternative reconciliation approaches (European Area QNA) for different forecast horizons. To make the figure more readable, the filled symbols indicate that the skill score is less than $-30$\%.}
	\label{img:EA_mseh}
\end{figure}

\begin{figure}[p]
	\centering
	\includegraphics[width = \linewidth]{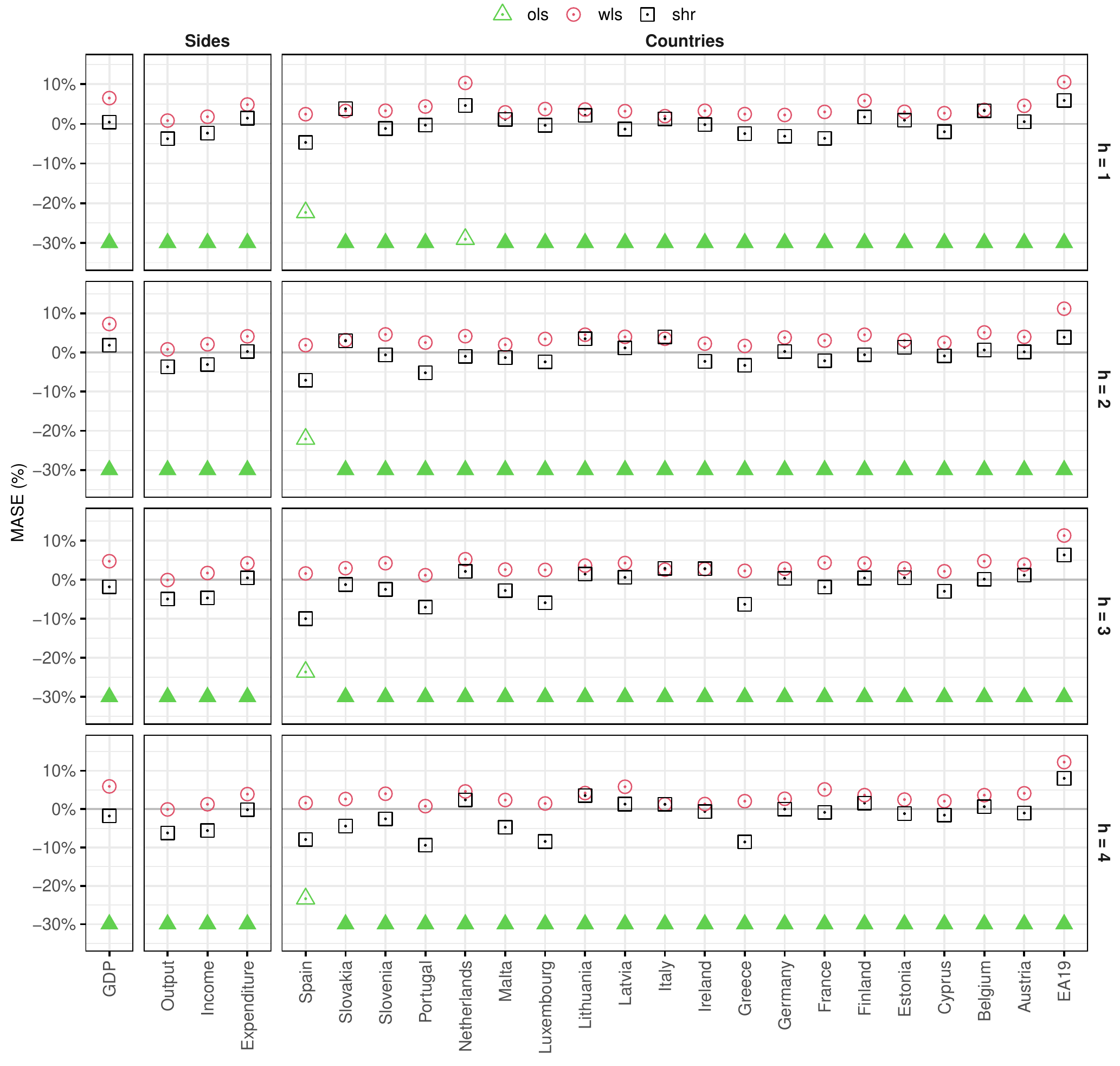}
	\caption{MASE-skill scores for the point forecasts from alternative reconciliation approaches (European Area QNA) for different forecast horizons. To make the figure more readable, the filled symbols indicate that the skill score is less than $-30$\%.}
	\label{img:EA_maseh}
\end{figure}

\begin{figure}[p]
	\centering
	\includegraphics[width = \linewidth]{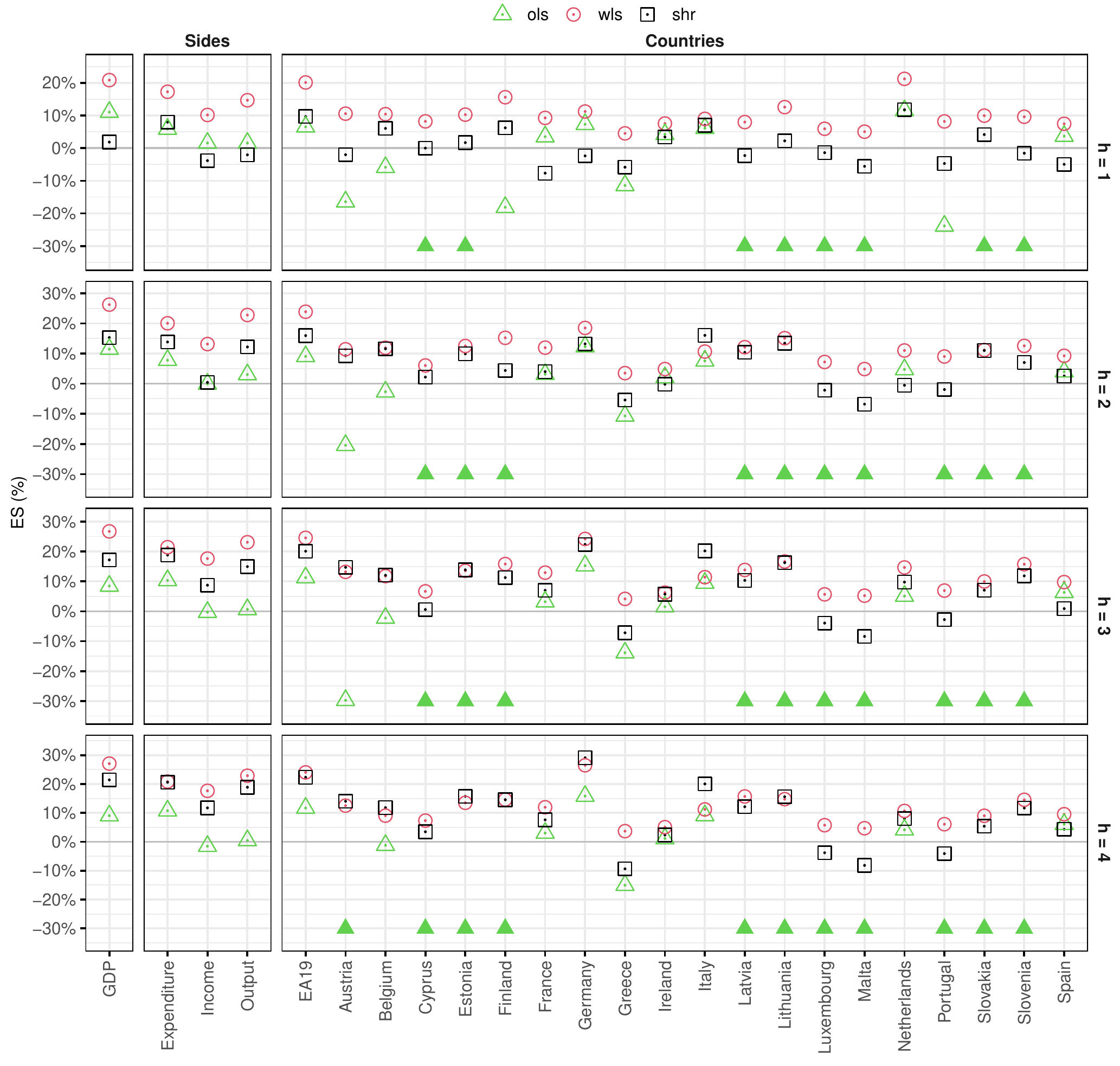}
	\caption{ES-skill scores for the point forecasts from alternative reconciliation approaches (European Area QNA) for different forecast horizons. To make the figure more readable, the filled symbols indicate that the skill score is less than $-30$\%.}
	\label{img:EA_esh}
\end{figure}

\begin{figure}[p]
	\centering
	\includegraphics[width = \linewidth]{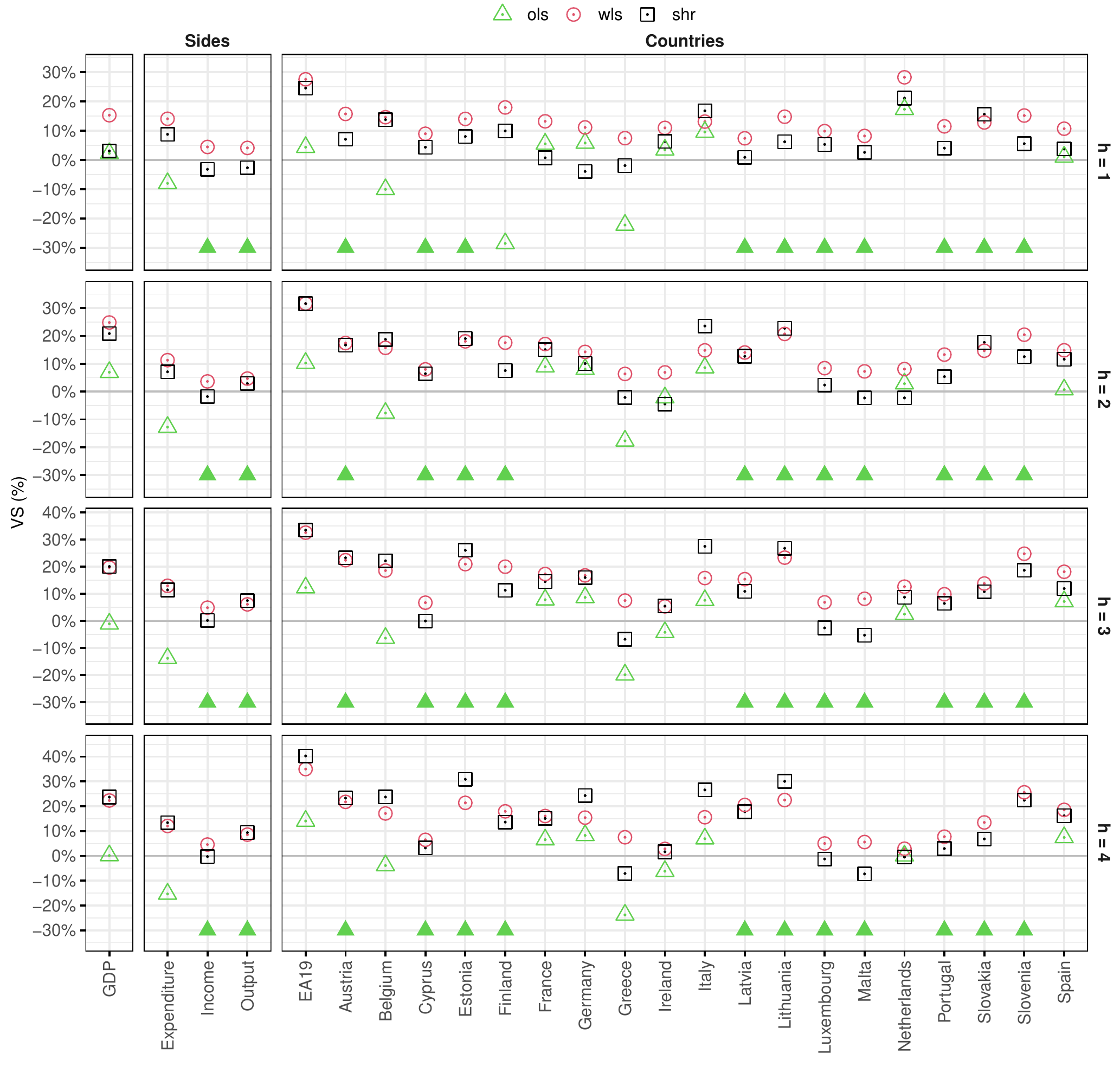}
	\caption{VS-skill scores for the point forecasts from alternative reconciliation approaches (European Area QNA) for different forecast horizons. To make the figure more readable, the filled symbols indicate that the skill score is less than $-30$\%.}
	\label{img:EA_vsh}
\end{figure}